\documentclass[fleqn,usenatbib]{mnras}

\usepackage[T1]{fontenc}
\usepackage{ae,aecompl}

\usepackage{graphicx}	
\usepackage{amsmath}	
\usepackage{amssymb}	
\usepackage{color}
\usepackage{soul}
\usepackage[normalem]{ulem}
\usepackage{mwe,tikz}
\usepackage{float}
\usepackage{subfigure}
\usepackage{multirow}
\usepackage{natbib}
\usepackage{xcolor}
\usepackage{soulpos}

\newcommand{\comments}[1]{}

\defcitealias{vegetti_detection_2010}{V10}
\defcitealias{collett_cosmological_2014}{CA14}
\defcitealias{collett_triple_2020}{CS20}
\defcitealias{minor_unexpected_2021}{M21}

\newcommand{\citetV}{\citetalias{vegetti_detection_2010}}
\newcommand{\citepV}{\citepalias{vegetti_detection_2010}}
\newcommand{\citetM}{\citetalias{minor_unexpected_2021}}

\newcommand{\citetCA}{\citetalias{collett_cosmological_2014}}
\newcommand{\citepCA}{\citepalias{collett_cosmological_2014}}
\newcommand{\citetCS}{\citetalias{collett_triple_2020}}

\title[Gravitational imaging through a compound lens]{Gravitational imaging through a triple source plane lens: revisiting the $\Lambda$CDM--defying dark subhalo in SDSSJ0946+1006}
\author[D. Ballard et al.]{Daniel J. Ballard,$^{1}$\thanks{Contact e-mail: \href{mailto:daniel.ballard@port.ac.uk}{daniel.ballard@port.ac.uk}}
Wolfgang J. R. Enzi,$^{1}$
Thomas E. Collett,$^{1}$
Hannah C. Turner,$^{2}$ \newauthor
and Russell J. Smith$^{2}$
\\
$^{1}$Institute of Cosmology and Gravitation, University of Portsmouth, Burnaby Road, Portsmouth, PO1 3FX, UK\\
$^{2}$Centre for Extragalactic Astronomy, Durham University, South Road, Durham, DH1 3LE, UK
}

\date{Accepted XXX. Received YYY; in original form ZZZ}

\pubyear{2023}

\begin{document}
\label{firstpage}
\pagerange{\pageref{firstpage}--\pageref{lastpage}}
\maketitle

\begin{abstract}
The $\Lambda$CDM paradigm successfully explains the large--scale structure of the Universe, but is less well constrained on sub--galactic scales. Gravitational lens modelling has been used to measure the imprints of dark substructures on lensed arcs, testing the small--scale predictions of $\Lambda$CDM. However, the methods required for these tests are subject to degeneracies among the lens mass model and the source light profile. We present a case study of the unique compound gravitational lens SDSSJ0946+1006, wherein a dark, massive substructure has been detected, whose reported high concentration would be unlikely in a $\Lambda$CDM universe. For the first time, we model the first two background sources in both I-- and U--band HST imaging, as well as VLT--MUSE emission line data for the most distant source. We recover a lensing perturber at a $5.9\sigma$ confidence level with mass $\log_{10}(M_\mathrm{sub}/M_{\odot})=9.2^{+0.4}_{-0.1}$ and concentration $\log_{10}c=2.4^{+0.5}_{-0.3}$. The concentration is more consistent with CDM subhalos than previously reported, and the mass is compatible with that of a dwarf satellite galaxy whose flux is undetectable in the data at the location of the perturber. A wandering black hole with mass $\log_{10}(M_\mathrm{BH}/M_{\odot})=8.9^{+0.2}_{-0.1}$ is a viable alternative model. We systematically investigate alternative assumptions about the complexity of the mass distribution and source reconstruction; in all cases the subhalo is detected at around the $\geq5\sigma$ level. However, the detection significance can be altered substantially (up to $11.3\sigma$) by alternative choices for the source regularisation scheme.
\end{abstract}



%

\section{Introduction} \label{sec:intro}
The standard $\Lambda$CDM model of cosmology describes a dark energy ($\Lambda$) dominated universe whose mass comprises $\sim85\%$ Cold Dark Matter (CDM). In contrast to baryons, this is an exotic type of matter outside of the standard model of particle physics that interacts with electromagnetism very weakly if at all. Assuming that Dark Matter (DM) is a particle, no candidate has been directly observed in a laboratory yet \citep[e.g.][]{roszkowski_wimp_2018,schumann_direct_2019,billard_direct_2022}.

Nonetheless, CDM theory successfully describes observations of the Universe on $\sim$Mpc scales and above \citep[see e.g ][]{bullock_small-scale_2017},  such as the hierarchical formation of large scale structure \citep{anderson_clustering_2014, hildebrandt_kids-450_2017} and the cosmic microwave background \citep{planck_collaboration_planck_2020}. Whilst DM is needed on galactic scales to explain rotation curves \citep{rubin_rotation_1970,rubin_extended_1978,rubin_rotation_1985}, it is entirely possible that the DM is not precisely that of the CDM paradigm; alternative models may be required to explain observed phenomena on smaller, sub--galactic scales \citep{diemand_formation_2007, diemand_clumps_2008}. 
In this lower--mass regime, alternatives to CDM have been proposed to resolve apparent discrepancies between observations and simulations \citep[e.g.][]{del_popolo_small_2017}, though many of these can also be explained by other means than the DM model \cite[see e.g.][]{fairbairn_galactic_2022}.

Alternative DM models make different predictions about the properties of individual halos as well as their populations. For example, higher thermal velocities in Warm Dark Matter \citep[WDM, e.g.][]{schneider_non-linear_2012, lovell_properties_2014} models lead to less concentrated halo mass profiles \citep[e.g.][]{ludlow_mass-concentration-redshift_2016, bose_substructure_2017} and a suppression of small-mass halos \citep[][]{lovell_properties_2014, lovell_spatial_2021}. Deviations from CDM on sub--galactic scales or in dwarf galaxies can, however, be obscured by their tidal interactions with more massive luminous halos \citep[e.g.][]{despali_constraining_2022, moreno_galactic_2022}.

While classical ``hot'' DM models are ruled out by observations of the large--scale Universe \citep[see e.g.][]{primack_hot_2001}, the small scale effects of WDM models are much harder to constrain. The formation of luminous galaxies typically requires a halo mass of around $\gtrsim 5\times10^{9} M_\odot$ \citep{benitez_detailed_2020}, thereby limiting the sample of directly observable satellite galaxies \cite[][]{kim_there_2018, newton_constraints_2021, nadler_dark_2021}. Instead we must rely on observations that are directly sensitive to the gravitational effects of the DM itself, such as strong gravitational lensing. This provides a direct probe of small--mass halos, since the lensing effects of galaxies and halos depend only on their mass, irrespective of their luminosity.

 DM subhalos introduce perturbations on top of the lensing by the main galaxy and its halo. Subhalos -- as well as other small halos projected along the same line--of--sight -- have been revealed primarily by observations of (i) anomalous flux ratios of multiply lensed quasars \citep{mao_evidence_1998, bradac_b1422231_2002, metcalf_flux_2002, mao_anomalous_2004, kochanek_tests_2004, mckean_high_2007, xu_how_2015, gilman_constraints_2019, gilman_warm_2020, hsueh_sharp_2020, nadler_dark_2021};
 (ii) perturbations on the arcs of lensed extended source galaxies \citep{dalal_direct_2002, vegetti_detection_2010, vegetti_quantifying_2010, vegetti_gravitational_2012, vegetti_inference_2014, hezaveh_detection_2016}.
The latter approach, known as gravitational imaging, led to a few detections of DM subhalos in previous studies \citep{vegetti_detection_2010, vegetti_gravitational_2012, nierenberg_detection_2014, hezaveh_detection_2016, nightingale_scanning_2022}, including one notable case in the lens system SDSSJ0946+1006 (henceforth J0946), which is the focus of this work.

J0946 is worthy of further study for two reasons. First, its perturbing subhalo has both an unexpectedly high mass if it is truly a dark matter substructure and not a dwarf satellite assembly of stars \citep[][hereafter \citetV]{vegetti_detection_2010} as well as an unexpectedly high concentration given its mass, making it a substantial outlier with respect to CDM simulations (\citet{nelson_illustris_2015};  \citet{minor_unexpected_2021} -- hereafter \citetM{}).
Second, J0946 is a compound lens system, with a lens at $z_l=0.222$ and three sources at $z_{s1}=0.609$, $z_{s2}=2.035$ and $z_{s3}=5.975$ \citep[][hereafter \citetCS]{collett_triple_2020}. These four galaxies are henceforth referred to as the main deflector, $s1$, $s2$, and $s3$ respectively.

Previous gravitational imaging studies of J0946 have  only considered the lensing of $s1$ as observed in the F814W band by the \textit{Hubble Space Telescope} (HST). In this paper, we extend on previous work in two ways, modelling all three sources in both the near--infrared F814W and the ultraviolet F336W bands simultaneously. Modelling the compound lensing should improve the macro--model of the main deflector, since compound lens modelling is much less affected by degeneracies than the modelling of a single source plane system \citep[see e.g.][]{schneider_source-position_2014}. Furthermore, one of the lensed images of $s3$ happens to fall close to the projected location of the reported dark subhalo, providing additional constraints on its properties. Modelling both HST bands simultaneously will allow us to disentangle source light complexity from mass model complexity, since
 lensing is achromatic whereas star--forming galaxies typically look very different in the ultraviolet and infrared. 

This paper is structured as follows. In Section \ref{sec:compoundlensing}, we describe the data, the geometry of the compound lensing in J0946 and our modelling methodology, and include a test of our sensitivity to a DM substructure. In Section \ref{sec:singleplane}, we present and discuss our results for a single source plane, and compare them to similar literature model setups. In Section \ref{sec:multiplane}, we present and discuss the results of our full triple source plane lens modelling. In Section \ref{sec:systematics}, we then perform systematics tests on various model assumptions. Finally, we conclude our findings in Section \ref{sec:conclusions}.
\section{Methodology}
\label{sec:compoundlensing}
\begin{figure*}
    \centering
    \includegraphics[width=0.9\textwidth]{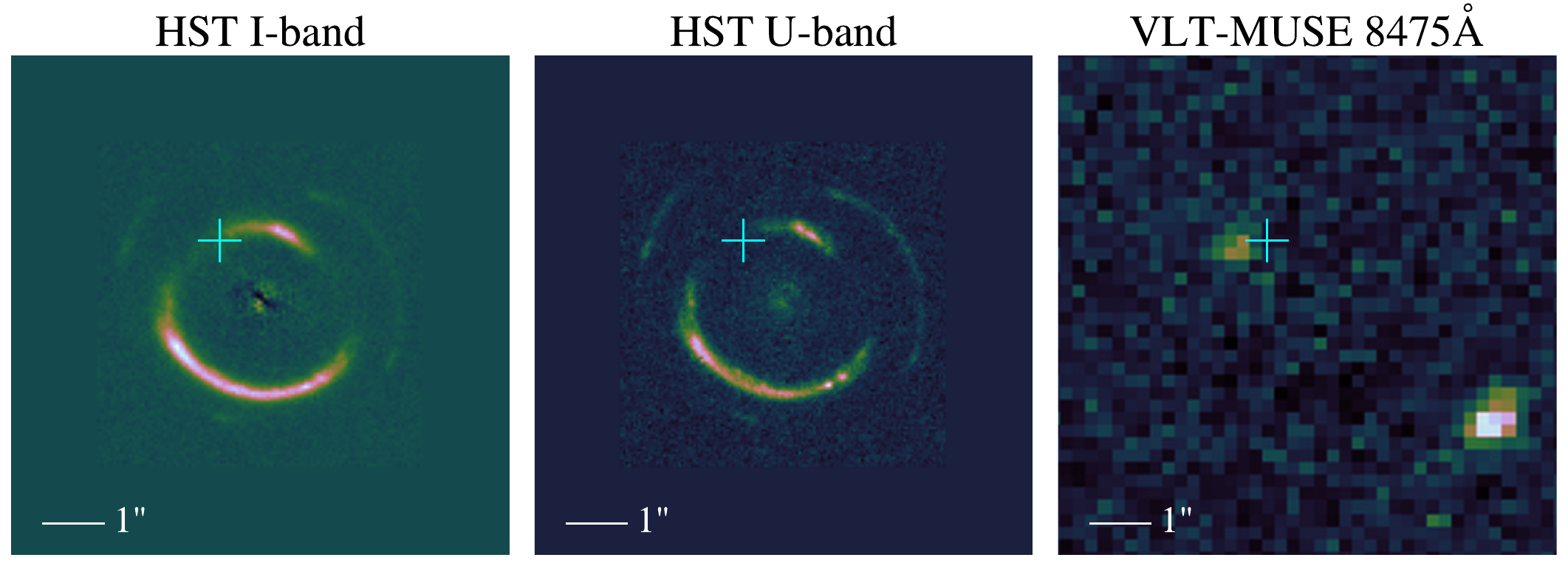}
    \caption{HST imaging of J0946 in the I--band (left) and U--band (middle), and continuum--subtracted VLT--MUSE narrow--band imaging (width 5\,\AA\ centred at $8475$\,\AA) showing the Ly--$\alpha$  
    emission at $z=5.975$ (right). The cyan cross represents the best fit location of the substructure in as reported in \citetV{} (which is visually indistinguishable from the best fit location in \citetM{}).}
    \label{fig:HST+MUSE_sub}
\end{figure*}
\begin{figure}
    \centering
    \includegraphics[width=0.3\textwidth]{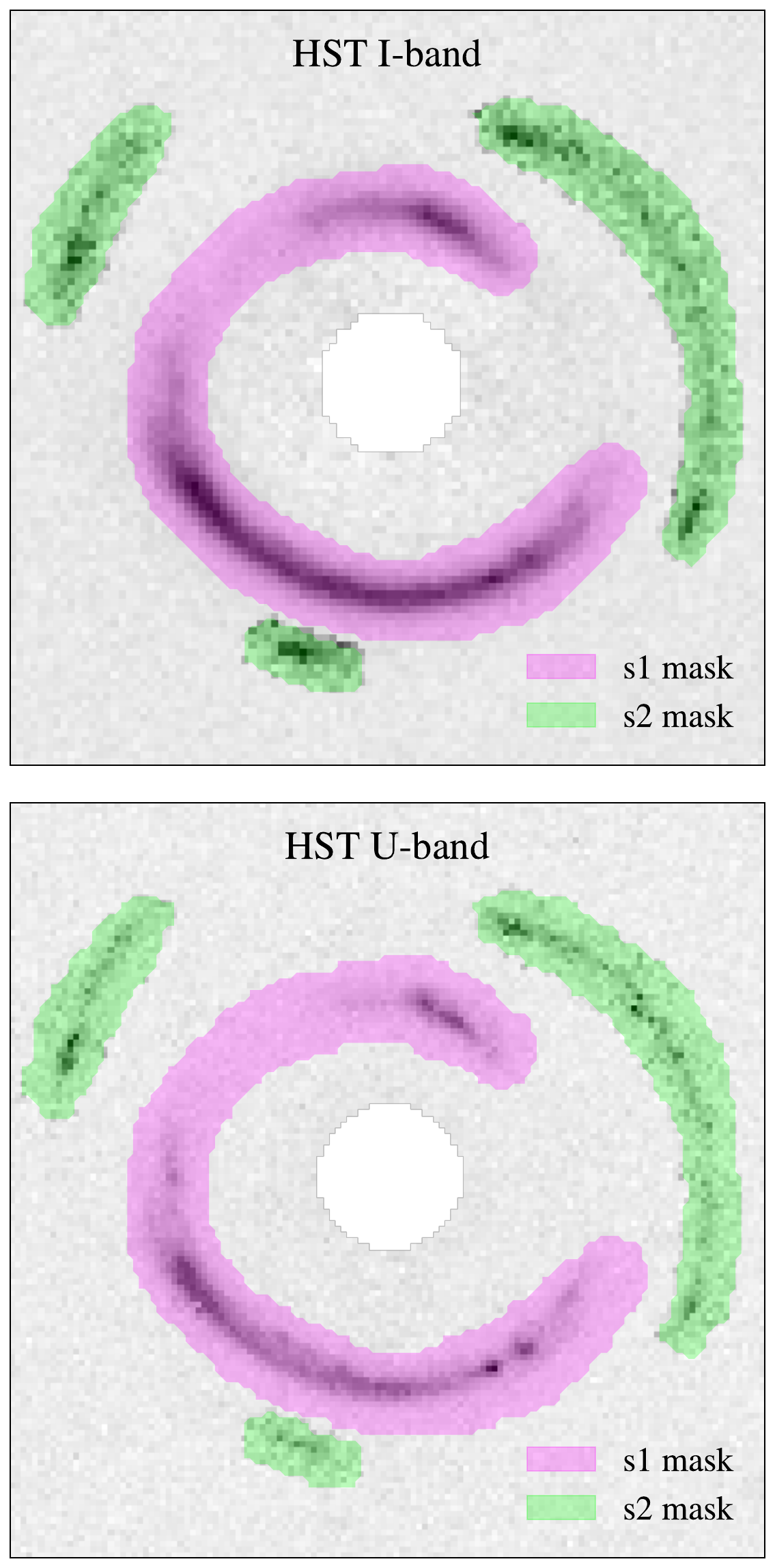}
    \caption{The data pixels used in our modelling of  $s1$ (magenta masked) and $s2$ (green masked) in  I--band (top) and U--band (bottom) HST data. All other pixels are ignored. For illustrative purposes, the image contrast of $s2$ is enhanced and a central region of image pixels is removed.}
    \label{fig:masks}
\end{figure}
\subsection{Data}
\label{sec:data}
We model two HST observations: the 2096\,s ACS image in F814W (I--band) from \cite{gavazzi_sloan_2008} 
and the 5772\,s WFC3/UVIS observation in F336W (U--band) from 
\cite{sonnenfeld_evidence_2012}. The I--band image allows us to compare with previous results in the literature, whilst adding the U--band probes clumpier emission in the source galaxies and gives excellent angular resolution. Though available in the HST archive, we neglect intermediate optical wavelength bands as these are unlikely to capture any qualitatively different structures; the same is true for the longest available wavelength band, WFC3/IR F160W, whose resolution is moreover poorer than the I--band image.
Data in both of our modelled bands are shown in Figure \ref{fig:HST+MUSE_sub}, with the reported location of the substructure from \citetV{} overlaid.

The I--band image as analysed has been drizzled to $0.05\arcsec$/pixel; the U--band image covers the same area but drizzled to $0.04\arcsec$ pixels. We use the same lens light--subtracted I--band image as \citet[][hereafter \citetCA]{collett_cosmological_2014}, but we do not subtract the lens light from the U--band image since it is negligible at this wavelength, at the location of the arcs.
Prior to the lensing analysis, the physical coordinates of the U--band data were aligned to those of the I--band data, to correct for a
small relative offset between the pipeline--processed images.
With the optimised shifts ($\delta x=0.027\arcsec$, $\delta y=-0.023\arcsec$), this correction is smaller than a single pixel.

Figure \ref{fig:HST+MUSE_sub} also shows the VLT--MUSE narrow--band image extracted in a 5\,\AA\ window around 8475\,\AA, capturing Lyman--$\alpha$ emission from the most distant lensed source. This image is not used explicitly in our lens modelling; we instead delens the centroid positions of the two $s3$ image features and their astrometric uncertainties, derived from fitting a Gaussian to each image.
Since the MUSE data have lower angular resolution, the image registration relative to HST is more uncertain than for the HST U--band versus I--band image alignment. To account for this, we artificially blur the I--band image with the MUSE Point Spread Function (PSF) and align this with a simulated HST I--band image of the arcs constructed out of the appropriate wavelength slices of the MUSE data cube. The resultant alignment uncertainty is propagated into the uncertainty of the $s3$ image centroids.

We model image pixels within one of four manually masked regions in the HST imaging of J0946, shown in Figure \ref{fig:masks}. We avoid the computational challenge of modelling both sources simultaneously \citepCA, by reconstructing the two sources and two bands as separate parts of the likelihood, which are simultaneously fit with the same mass model. This is a reasonable approach, since the two rings do not overlap on the sky.

\subsection{Ray Tracing}
For strong gravitational lensing, the source plane position, $\boldsymbol{\beta}$, of a photon is displaced from its observed, lensed, image plane position, $\boldsymbol{\theta}$, by the reduced deflection angle, $\boldsymbol{\alpha}$, according to the lens equation:
\begin{equation}
    \boldsymbol{\beta} = \boldsymbol{\theta} - \boldsymbol{\alpha}(\boldsymbol{\theta}) \, .
    \label{eq:lensequation}
\end{equation}
The deflection angle, $\boldsymbol{\alpha}$, of a lens is related to the lensing potential on its lens plane, $\boldsymbol{\psi}$, such that
\begin{equation}
    \boldsymbol{\alpha}(\boldsymbol{\theta}) = \nabla \boldsymbol{\psi}(\boldsymbol{\theta}) \, ,
    \label{eq:deflectionangle}
\end{equation}
where $\boldsymbol{\psi}$ depends on the 2D projected lens mass distribution, as well as the angular diameter distances between observer, lens and source. 

Equation \ref{eq:lensequation} is for a system with one lens and one source plane, but can be generalised to give the compound lens equation:
\begin{equation}
    \boldsymbol{\theta}_{j}=\boldsymbol{\theta}_{0}-\sum_{i=1}^{j}\eta_{ij}\boldsymbol{\alpha}_{i-1}(\boldsymbol{\theta}_{i-1}) \text{ for j > 0} \, .
    \label{eq:multiplanelensequation}
\end{equation}
Here we have adjusted our notation from Equation \ref{eq:lensequation} to no longer distinguish between lens and source, since in a compound lensing system a single galaxy can be both. In Equation \ref{eq:multiplanelensequation}, $\boldsymbol{\theta}_{i}$ generically denotes an angular position on a redshift plane, {i}, where $i=0$ is the foreground--most lens plane and observed image plane; any $i>0$ refers to the $i^\mathrm{th}$ source (or further lens) plane behind it.

For a lensing plane $l$, the extra parameter $\eta_{ij}$ describes the scaling of the reduced deflection angles from one source plane, $i$, to another, $j$, defined as a ratio of angular diameter distances:
\begin{equation}
    \eta_{ij}=\frac{D_{i}D_{lj}}{D_{li}D_{j}} \, .
    \label{eq:scalefactor}
\end{equation}
Throughout the multi--source plane lensing portions of this work, we define reduced deflection angles of a lens relative to light coming from the plane immediately behind the lens. This is not the convention of \citet{schneider_gravitational_1992}, who define all reduced deflection angles relative to light coming from the furthest plane. Our convention allows easier comparison between our work and other single and double source plane models of J0946.  A detailed explanation of our chosen convention is available in Appendix \ref{sec:scale_factor}.

Throughout this work we fix the angular diameter distances of the system assuming the  $\Lambda$CDM cosmological parameters $\Omega_{\rm m}=0.307$, $\Omega_\Lambda=0.693$, and $h_0=0.6777$ \citep{planck_collaboration_planck_2014}. 

\subsection{Lens Modelling}
\label{sec:lensmodel}
\begin{figure*}
    \centering
    \includegraphics[width=0.9\textwidth]{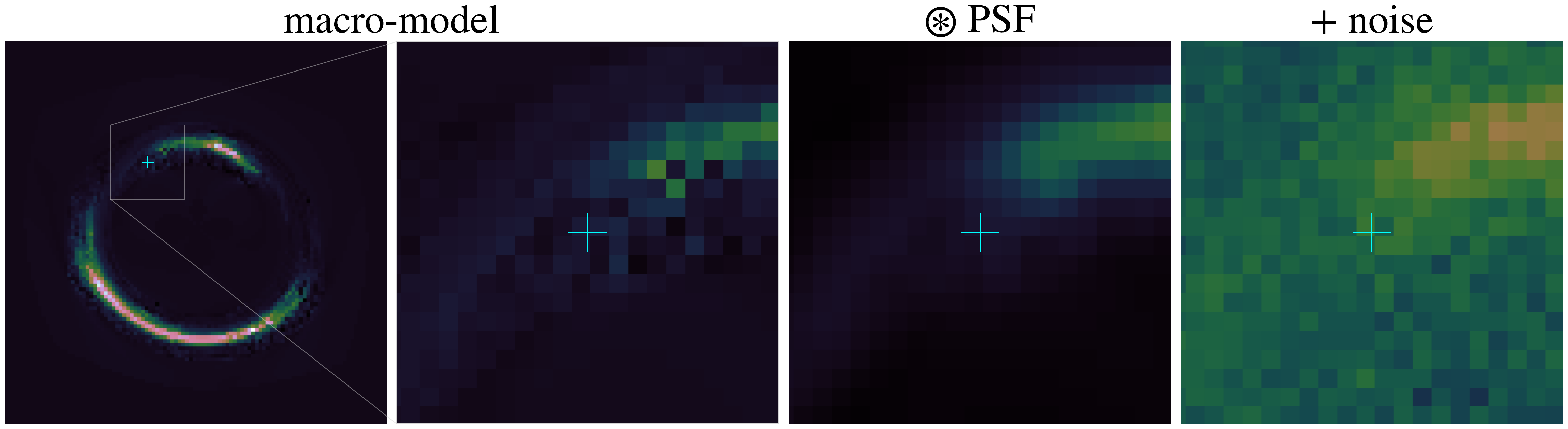}
    \includegraphics[width=0.9\textwidth]{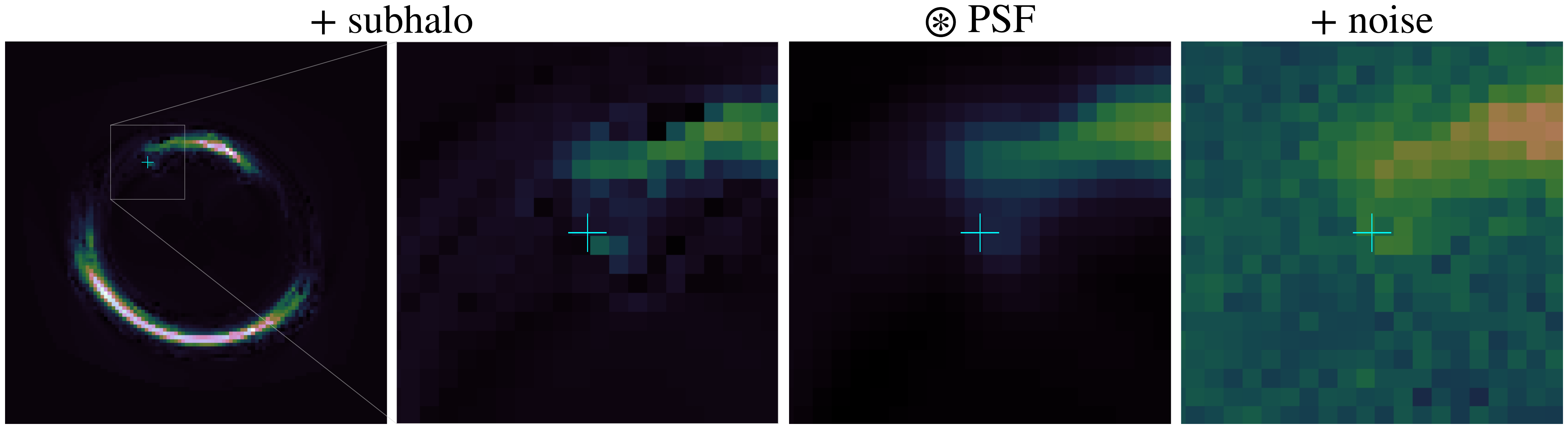}
    \caption{Mock data for our sensitivity test, where panels (left to right) show the initial model image, a zoomed inset around the location of the reported substructure, the effect of blurring by the HST I--band PSF, and the addition of background noise akin to the original HST I--band data. The top row is created from a smooth model for the lens, whilst the bottom row has an injected tNFW subhalo with the parameters of \citetM{} at the cyan cross. The bottom right panel is used as mock data to recover the injected substructure with $\sim5\sigma$ confidence.}
    \label{fig:sensitivity}
\end{figure*}
To model the data, we follow the semi--linear inversion approach of \citet{warren_semilinear_2003}. We define a model for the lensing mass distribution, and for each realisation of the non--linear parameters of that model we linearly solve for the best--fitting source.

\subsubsection{Non--linear Mass Model}
We assume that the main deflector is described by an Elliptical Power Law (EPL) model with external shear. We consider two possible scenarios for evidence comparison: one with and one without a dark subhalo in the form of a truncated Navarro--Frenk--White (tNFW) profile. We refer to these two scenarios as our smooth and tNFW--perturbed models, respectively.
Additionally, in our multi-source plane models in Sections \ref{sec:multiplane} and \ref{sec:systematics}, $s1$ and $s2$ behave as lenses as well as sources; we model their mass distributions as singular isothermal sphere (SIS) profiles.

The EPL profile has six parameters that behave non--linearly in the model: The Einstein radius, $\vartheta_{E}$, the logarithmic slope, $\gamma$, the axis ratio, $q$, the position angle, $\varphi$, and two centroid coordinates $(x, y)$. An SIS is identical to an EPL with $\gamma=2$ and zero ellipticity. The external shear has two non--linear parameters: the shear strength, $\Gamma$, and the shear angle, $\varphi_{\Gamma}$.

The tNFW profile is based upon the profile derived by \citet{navarro_structure_1996}, whose density, $\rho$, at radial distance, $r$, is related to a characteristic density, $\rho_{0}$, by
\begin{equation}
  \rho_{\rm NFW}(r) = \frac{\rho_{0}}{\frac{r}{r_{s}}(1+\frac{r}{r_{s}})^2} \, .
  \label{eq:NFW}
\end{equation}
As in \citetM{}, we do not assume a fixed mass--concentration relation for the substructure, and therefore model both its concentration, $c$, and virial mass, $M_{200}$.
The relation between the scale radius in Equation \ref{eq:NFW}, $r_{s}$, and $c$ is given by:
\begin{equation}
    c = r_{200}/r_{s} \, ,
\end{equation}
where $r_{200}$ is considered the virial radius enclosing $M_{200}$, though is strictly the radius enclosing an average density that is 200 times the critical density of the Universe.

Following \citetM{}, $M_{200}$ is formally defined under the assumption that the subhalo can be considered a field halo, which is then tidally stripped by its massive host. To account for this tidal stripping, we assume that this profile is truncated according to \citet[][]{baltz_analytic_2009}:
\begin{equation}
  \rho_{\rm tNFW}(r) = \frac{r_{t}^{2}}{r_{t}^{2} + (r/r_{s})^{2}}\rho_{\rm NFW}(r) \, .
  \label{eq:tNFW}
\end{equation}
We compute both the virial mass, $M_{200}$, analogous to a non--truncated field halo, as well as the total mass of the substructure, $M_{sub}$, which accounts for the effect of the truncation radius, $r_{t}$. The latter is a finite quantity for the above choice of truncation. It is hence possible that $M_{\mathrm sub} < M_{200}$.
The free parameters of our tNFW profile are $M_{200}$, $c$, $r_{t}$, and centre position $(x, y)$. Throughout this work we assume that the dark perturber is a subhalo at $z=0.222$, the redshift of the main deflector. \citetM{} also find a good fit to the data when the perturber is a line--of--sight halo between the observer and the lens plane, with the mass and concentration marginally decreased but still anomalously high.
\subsubsection{Mass and concentration from simulations}
\label{sec:mass_concentration}
Extrapolating the field halo mass--concentration relation of \citet[][]{shao2023} \citep[based upon the \texttt{CAMELS} suite of hydrodynamic $\Lambda$CDM  simulations,][]{Villaescusa2021} to subhalos of virial mass $M_{200}=10^{10}M_{\odot}$, we expect a mean concentration of $\log_{10}c=1.3$ (with DM only),  $\log_{10}c=1.2$ \citep[with baryonic physics according to \texttt{IllustrisTNG}, see][]{nelson2017,pillepich2018,springel2018,Marinacci2018,Naiman2018,nelson2019}, and $\log_{10}c= 1.4$ \citep[with baryonic physics according to \texttt{SIMBA}, see][]{Dave2019}. 
Taking the mass--concentration relation of \citet{dutton_cold_2014}, we would expect a median 
 value of $\log_{10}c=1.1$. The typical scatter around the mass--concentration relation in simulations is of the order of $\sigma_{\rm scatter} \approx 0.1$
dex \citep[see e.g.][]{dutton_cold_2014}.
We note, however, that the differences that we later quote between these results and our own depend on the assumed parameters describing baryonic physics in the \texttt{IllustrisTNG} and  \texttt{SIMBA} models, i.e. feedback from supernovae and active galactic nuclei.  
\subsubsection{Reconstructing unlensed source brightness distributions}
\label{sec:source}
Since we do not know the morphology of a source a priori, we infer it simultaneously with the lens parameters from the data. It is clear from the clumpiness of the arcs that the sources must be intrinsically irregular. Therefore, we adopt a pixellated free--form reconstruction of the source light. 

Specifically, we evaluate source brightness values defined on an adaptive Voronoi mesh created from a subset of image plane pixels ray--traced onto each source plane. In this work, we cast back all the pixels that fall within the mask of a given source for the band in consideration. The advantage of such an adaptive mesh is that it allows for a higher resolution source at those locations where the magnification through the lens becomes the strongest. We follow \citet[][]{nightingale_pyautolens_2021, nightingale_scanning_2022} and employ a Natural Neighbour Interpolation scheme to determine sub--pixel source brightness values \citep[][]{sibson1981}. We choose this scheme because (i) it yields a smooth Likelihood function which makes sampling the non--linear parameters much easier, and (ii) it forces the gradient of the source to be continuous, which is particularly important for substructure identification. 

To impose the astrophysical prior that sources require a certain degree of smoothness, we additionally introduce a regularisation strength parameter for each source. The brightness values at the vertices follow a Gaussian regularisation prior whose covariance matrix penalises the source brightness gradient or curvature \citep[see][for details]{suyu_bayesian_2006}. Fiducially, we opt for gradient regularisation, in contrast to \citetV{} who use curvature regularisation and \citetM{} who reconstruct their source out of a summation of analytic light profiles. However, since we do not a priori know how smooth our source reconstructions should be, we leave the regularisation strengths for the reconstructions of $s1$ and $s2$ as free parameters to be inferred by the model directly from the data. The centroid position $(x, y)$ of $s3$ is also fit for, but the unlensed light distribution of this source is not reconstructed.

\subsubsection{Posterior and evidence calculation}
For model comparison, we evaluate both the posterior of the non--linear parameters, $\boldsymbol \xi$, and the evidence of our models with and without a substructure. The posterior, $\mathcal{P} (\boldsymbol \xi | \boldsymbol d)$, relates to the likelihood function, $\mathcal{L}_{\rm tot}(\boldsymbol \xi)$, and the prior of model parameters, $\mathcal{P} (\boldsymbol \xi$), according to:
\begin{equation}
\mathcal{P} (\boldsymbol \xi | \boldsymbol d) = \frac{ \mathcal{L}_{\rm tot}(\boldsymbol \xi) \mathcal{P}(\boldsymbol \xi) }{ \mathcal{Z}}\,.
\end{equation}
The full details of $\mathcal{L}_{\rm tot}(\boldsymbol \xi)$ are described in Appendix \ref{sec:likelihood}.
The Bayesian evidence, $\mathcal{Z}$, is an integral of the likelihood multiplied by the prior, which normalizes the posterior,  i.e.:
\begin{equation}
    \mathcal{Z} = \int d \boldsymbol \xi \mathcal{L}_{\rm tot}(\boldsymbol \xi) \mathcal{P}(\boldsymbol \xi) \, .
\end{equation}
We evaluate the posterior and this integral using the preconditioned Monte Carlo package \texttt{pocoMC}  \citep[][]{karamanis_pocomc_2022}. 
 \texttt{pocoMC} generates posterior samples by following a Sequential Monte Carlo scheme combined with a Normalizing Flow, which preconditions the target distribution to remove correlations among its parameters \citep[][]{karamanis_accelerating_2022}\footnote{We choose the default hyper--parameters of \texttt{pocoMC}, i.e. an effective sample size of ESS$ = 0.95$ and correlation coefficient $\gamma = 0.75$, but increase the number of particles to up to 6000. We further set the maximum number of MCMC steps to 10000. We found that these values ensure convergence of the posterior, given the multi--modality of the likelihood.}. Evidences are calculated using the bridge sampling method and consistent with those obtained from the nested sampling algorithm \texttt{MultiNest} \citep[][]{feroz_multinest_2009, feroz_importance_2019}.
When comparing two models, we report the  $N\sigma$ confidence level that one is preferred over the other, i.e. we assume that one of the considered models is true and map the model probability onto the $N\sigma$ probability volume of a Gaussian distribution.

\subsection{Checking the sensitivity of our method for detecting substructures}
\label{sec:sensitivity}
Claiming the detection or non--detection of a substructure requires knowledge of the sensitivity of the data \citep[see e.g.][]{despali_detecting_2022}. To demonstrate that we are, in principle, sensitive to a substructure within the data at the reported location, we create a mock data set based upon our best smooth reconstruction of the I--band image of s1 (see Section \ref{sec:singleplane}) and inject a tNFW profile with the parameters reported in \citetM.  Figure \ref{fig:sensitivity} illustrates how the inclusion of the substructure affects the closest arc, including the effects of the PSF and observational noise. We then remodel this data assuming both a smooth and tNFW--perturbed model, finding that the latter is preferred with a difference in the logarithmic evidence of $\Delta \ln \mathcal{Z} = 15.16\pm0.03$ assuming gradient regularization of the source (corresponding to a $5.2\sigma$ detection significance). Our posteriors are consistent within $1\sigma$ of the input subhalo mass and concentration.
This suggests that we should be able to detect a substructure with similar properties to \citetM{}.
However, since we are reproducing an injected halo whose parameters are exactly known, a more rigorous sensitivity calculation would be required if we were searching for \textit{new} subhalos in J0946.
\section{Single source plane model results and discussion}
\label{sec:singleplane}
In this section, we present the results of our single source plane models for J0946 and compare them with those of previous studies.
\begin{table}
    \caption{The differences in best fit log--likelihood $\Delta\ln\mathcal{L}$ and log--evidence $\Delta\ln\mathcal{Z}$, between smooth and tNFW--perturbed models, shown for our single source plane and triple source plane results. These differences are quoted relative to the smooth case, such that positive values indicate preference for the tNFW--perturbed model. In brackets are the corresponding confidences of the detections.} \centering  
    \begin{tabular}{|l|l|l|}
    \hline
\textbf{Data modelled} & $\boldsymbol{\Delta\ln\mathcal{L}}$ & $\boldsymbol{\Delta\ln\mathcal{Z}}$ (confidence) \\ \hline
    1 source, I--band             & 21.67  & 7.23$\pm$0.03 (3.4$\sigma$) \\
    1 source, I-- \& U--band       & 29.52  & 14.34$\pm$0.04 (5.0$\sigma$) \\
    3 sources, I-- \& U--band      & 38.18  & 19.64$\pm$0.03 (5.9$\sigma$) \\
    \end{tabular}
    \label{tab:dlogL_dlogZ_sigma}
\end{table}
\subsection{I--band Model}
\label{sec:1band_singleplane}
\begin{figure*} \centering
\begin{tikzpicture}[      
        every node/.style={anchor=south west,inner sep=0pt},
        x=1mm, y=1mm,
      ]   
     \node (fig1) at (0,0)
       {\includegraphics[scale=0.175]{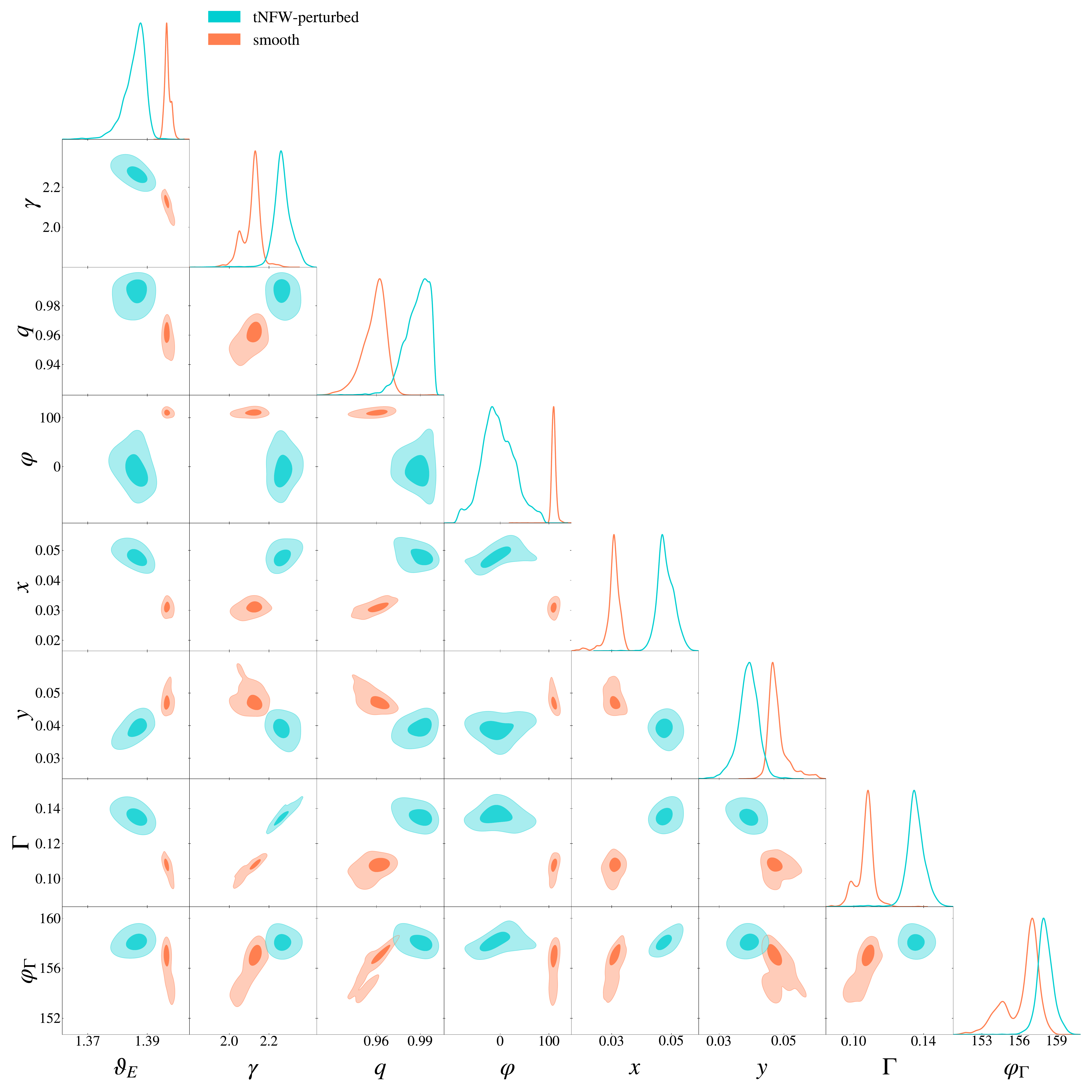}};
    \node (fig3) at (80,120)
       {\includegraphics[scale=0.295]{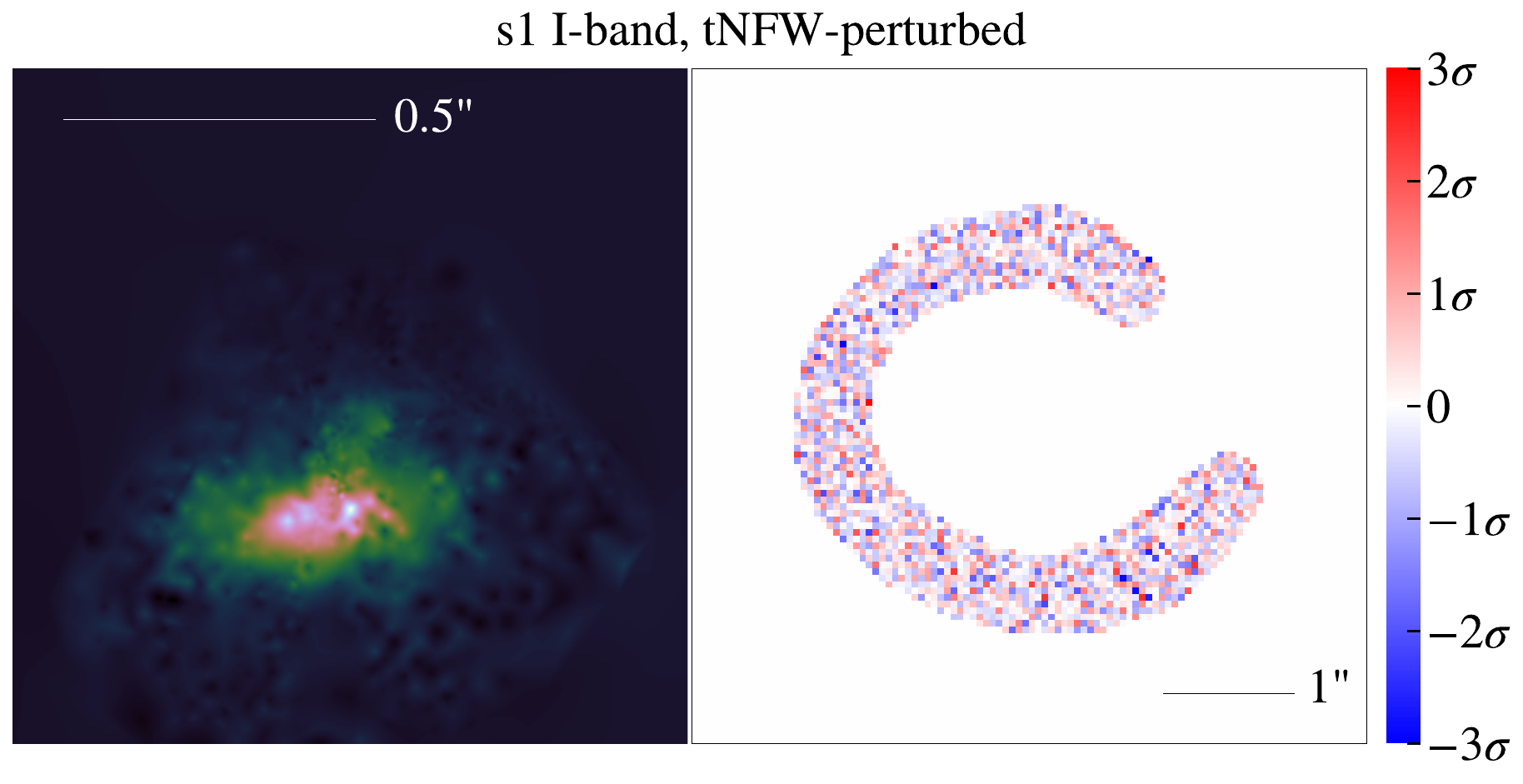}};
\end{tikzpicture}
\caption{$1\sigma$ and $2\sigma$ contours of the posterior distribution for the EPL and external shear parameters for our model of $s1$ in I--band only, with (cyan) and without (orange) the addition of a tNFW substructure. Inset: best fit source reconstruction (left) and residuals between the data and best fit model in units of standard deviation (right). These panels correspond to the tNFW--perturbed models, but are visually indistinguishable to the best fit smooth model results.} 
\label{fig:1band_corner}
\end{figure*}
Modelling the I--band data of the inner arcs alone provides the closest comparison with previous studies of  J0946 (e.g. \citetV{}, \citetM{}). We can reconstruct the data to the noise level assuming our smooth (EPL+Shear) model. Between our smooth and tNFW--perturbed models, we find that the posterior distributions of the macro--model parameters agree within the $\sim3\sigma$ level or closer (with the exception of the $x$ coordinate of the centre of the lens). Posterior distributions for these parameters are shown in Figure \ref{fig:1band_corner}, alongside the best--fit source reconstruction and normalised image--plane residuals, which demonstrate our ability to successfully model these arcs down to the noise level. 

In this single plane, I--band only case, the data prefers the existence of a tNFW substructure ($\Delta\ln\mathcal{Z}=7.23\pm0.03$) with $3.4\sigma$ confidence over the smooth model. Our macro--model parameters are within $4\sigma$ of those reported by \citetV{}. Such differences are most likely due to our prescription of our source model (gradient regularised, versus curvature regularised in \citetV{}) and our wider prior ranges on all parameters. However, it is also noteworthy that comparisons with \citetV{} are non--trivial due to differences in the parameterisation of the EPL mass profile. It is possible that small disagreements found in our results are introduced by such differences in parameter definitions.

The differences in likelihood and evidence between smooth and tNFW--perturbed models are recorded in Table \ref{tab:dlogL_dlogZ_sigma}. All priors and posterior results are documented in Appendix \ref{sec:prior_posterior_tables}.

Regarding the mass and concentration of the substructure, we find $\log_{10}(M_{200} / M_\odot) = 10.8^{+1.3}_{-0.6}$ 
and  $\log_{10}c = 2.0^{+0.3}_{-0.3}$. Our results exceed all of the simulation values with a root--mean--squared difference of 2.7--3.6 $\sigma_{c}$, with $\sigma_{c}$ being the standard deviation of our concentration posterior. 
 Our result is less of an outlier than \citetM{} finds, both because of the greater uncertainty on our inferred parameters and the lower median value of the concentration. The subhalo mass, $\log_{10}(M_\mathrm{sub} / M_\odot) = 10.0^{+0.4}_{-0.3}$, remains perplexing, however, given that such a massive object should host a detectable population of stars \citepV{}.
\subsection{Dual I--band and U--band Model}
\label{sec:2band_singleplane}
Simultaneously modelling the  I-- and U--band data for $s1$ necessitates one additional non--linear parameter (the regularisation strength of the U--band source galaxy) but adds much more data to constrain the lens model. Doing this, the tNFW--perturbed model is preferred over the smooth model with an evidence ratio $\Delta\ln\mathcal{Z}=14.34\pm0.04$, corresponding to a $5.0\sigma$ confidence detection. 

The addition of the U--band yields different posteriors on our macro--model parameters. Comparing with the I--band only case, the mass profile slope for the smooth model is significantly shallower ($\gamma =1.92^{+0.03}_{-0.02}$ versus $2.12^{+0.03}_{-0.07}$). However, when the tNFW perturber is included, both our models prefer a super--isothermal slope ($\gamma =2.27^{+0.05}_{-0.04}$ and $2.23^{+0.02}_{-0.02}$ respectively). The differences in $\gamma$ between smooth and tNFW--perturbed cases are likely caused by a source position transformation, whereby multiple image plane locations that correspond to a single source plane position are invariant under a change in lens model if the source is afforded the flexibility to move \citep{schneider_source-position_2014}. Our multi--plane modelling should not suffer from this effect, as the scalings required for the degeneracy are source redshift dependent. Since we have sources present at multiple very different redshifts, the source position transformation degeneracy is broken, except in extremely contrived scenarios where a transformation of the mass on s1 counterbalances the transformation of the primary lens \citep{schneider_can_2014}.

Despite the significant shifts in the parameters of the macro--model, the substructure mass and concentration are still consistent with the I--band only result within $1\sigma$. Deviations from the predicted mass--concentration relations are on the level of 2.8--3.7 $\sigma_{c}$.
\section{Triple source plane model results and discussion}
\label{sec:multiplane}
\label{sec:tripleplane}
\begin{figure*} \centering
\begin{tikzpicture}[      
        every node/.style={anchor=south west,inner sep=0pt},
        x=1mm, y=1mm,
      ]   
     \node (fig1) at (-115,0)
       {\includegraphics[scale=0.28]{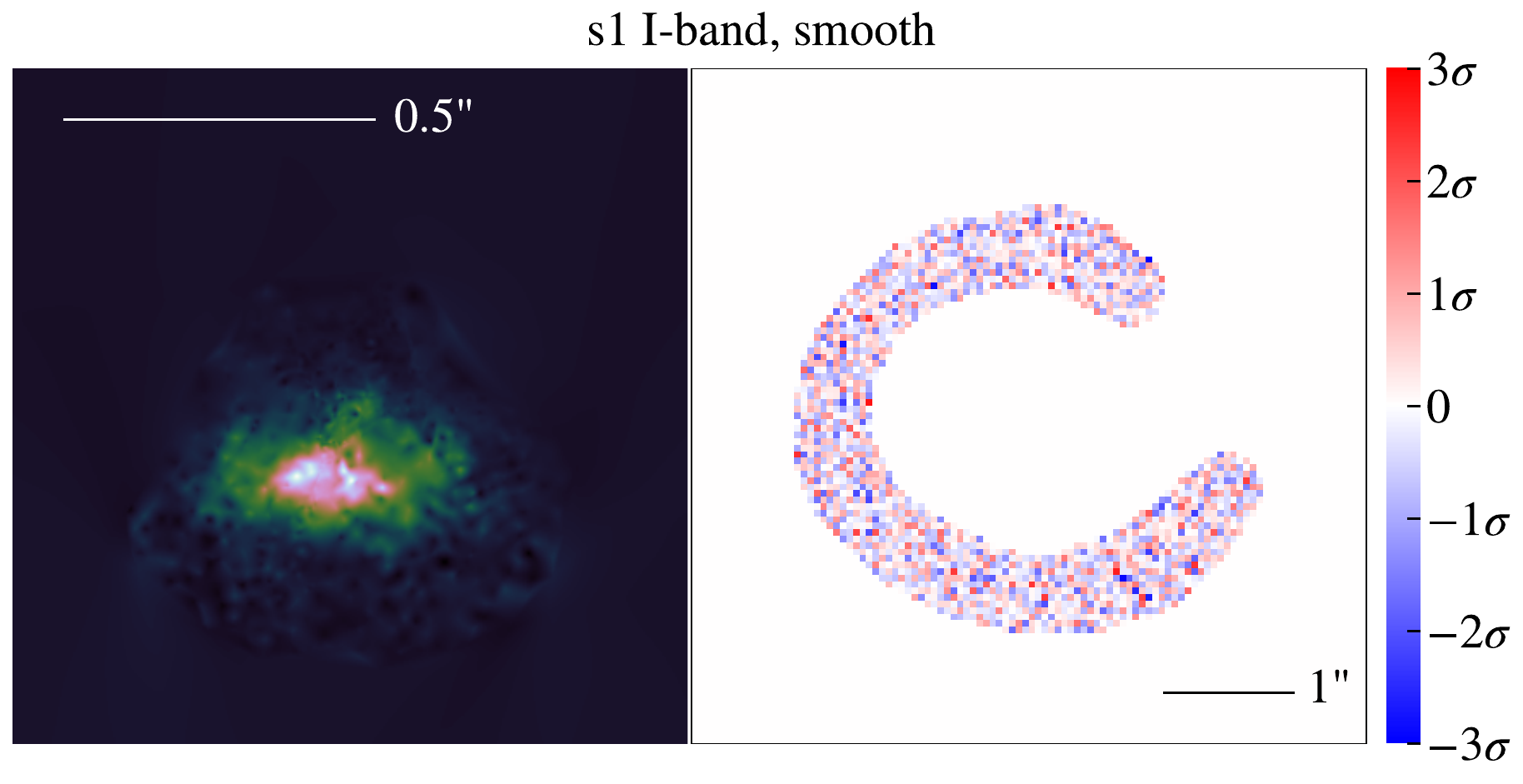}};
     \node (fig3) at (-25,0)
       {\includegraphics[scale=0.28]{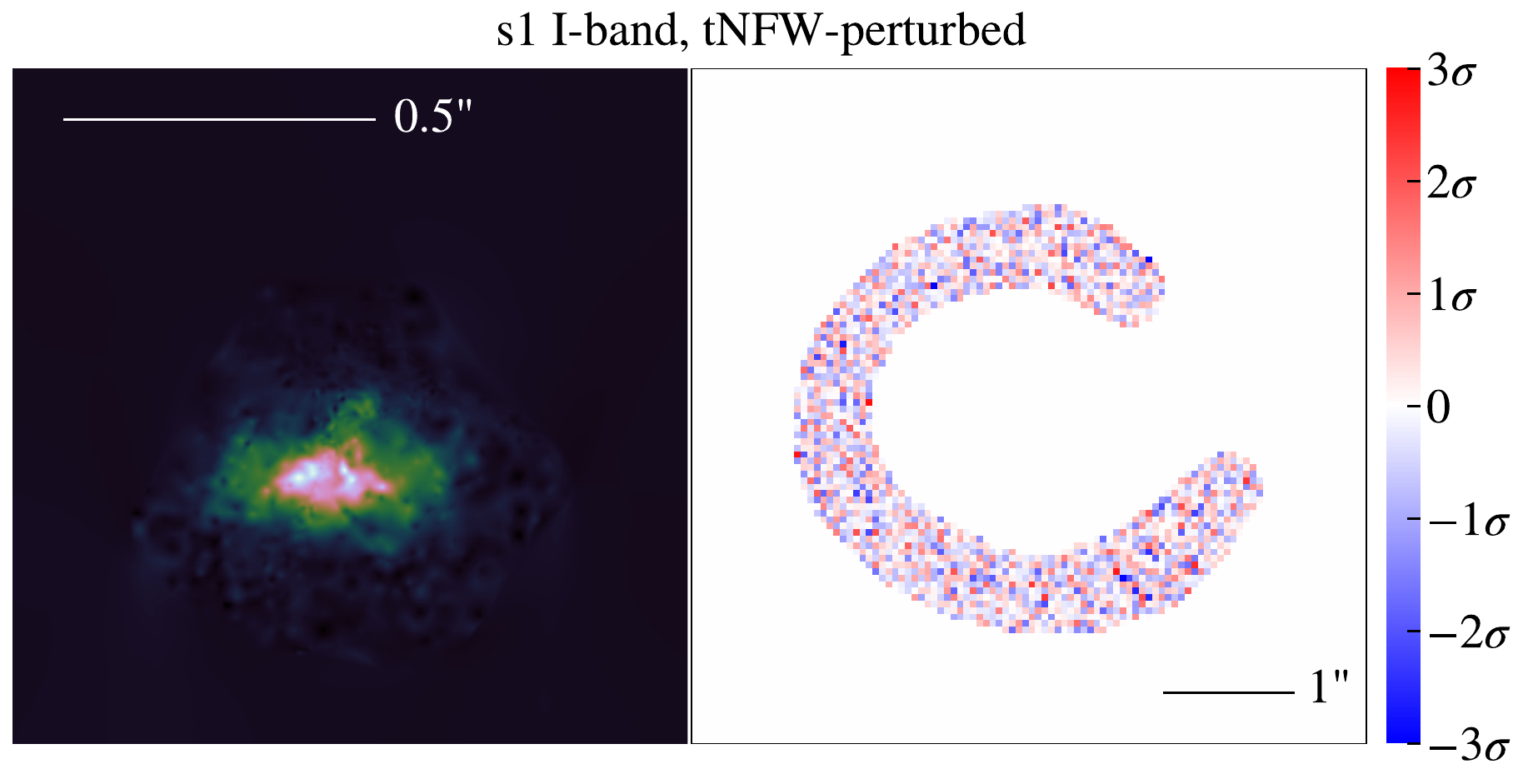}};
     \node (fig5) at (-115,-45)
       {\includegraphics[scale=0.28]{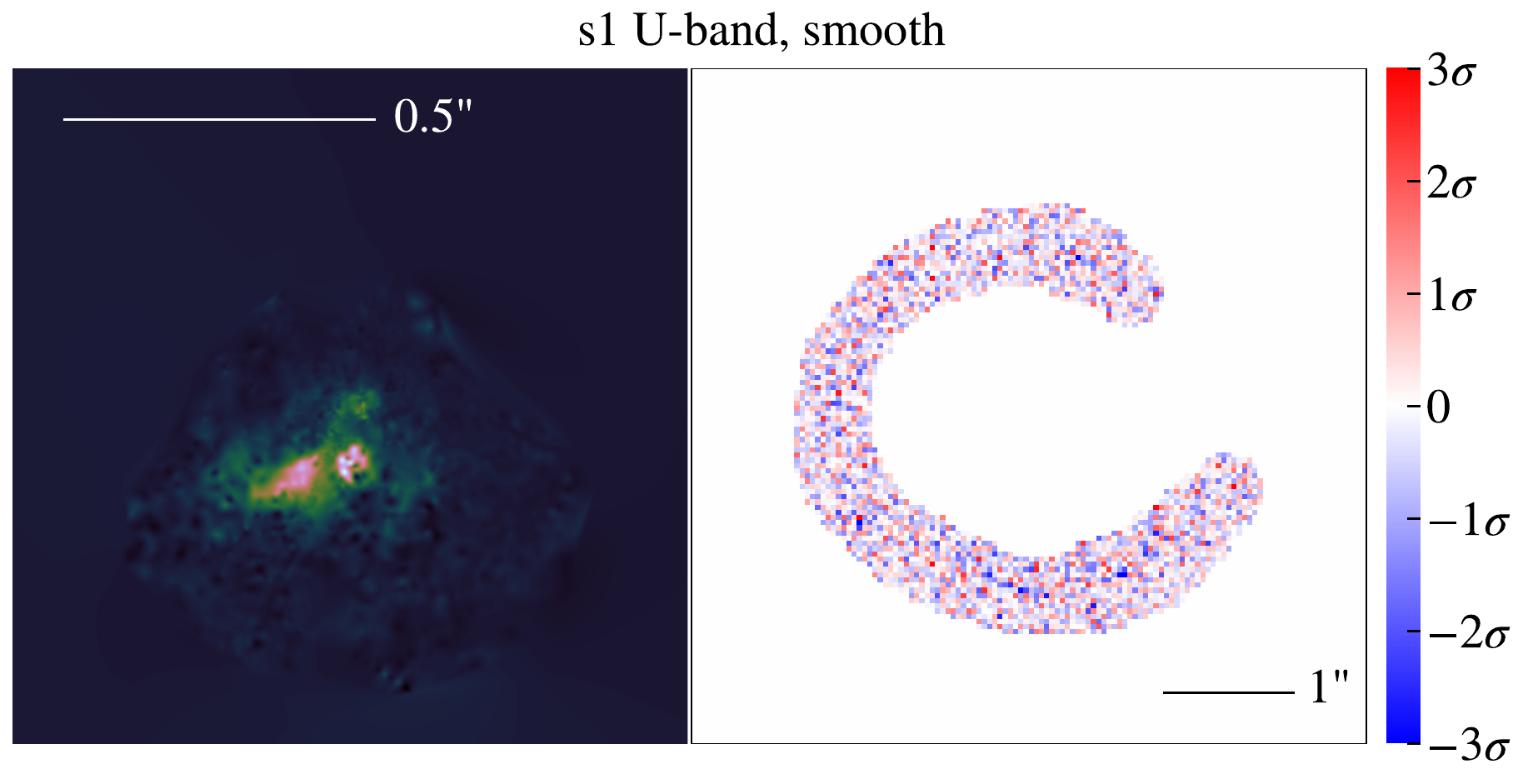}};
     \node (fig7) at (-25,-45)
       {\includegraphics[scale=0.28]{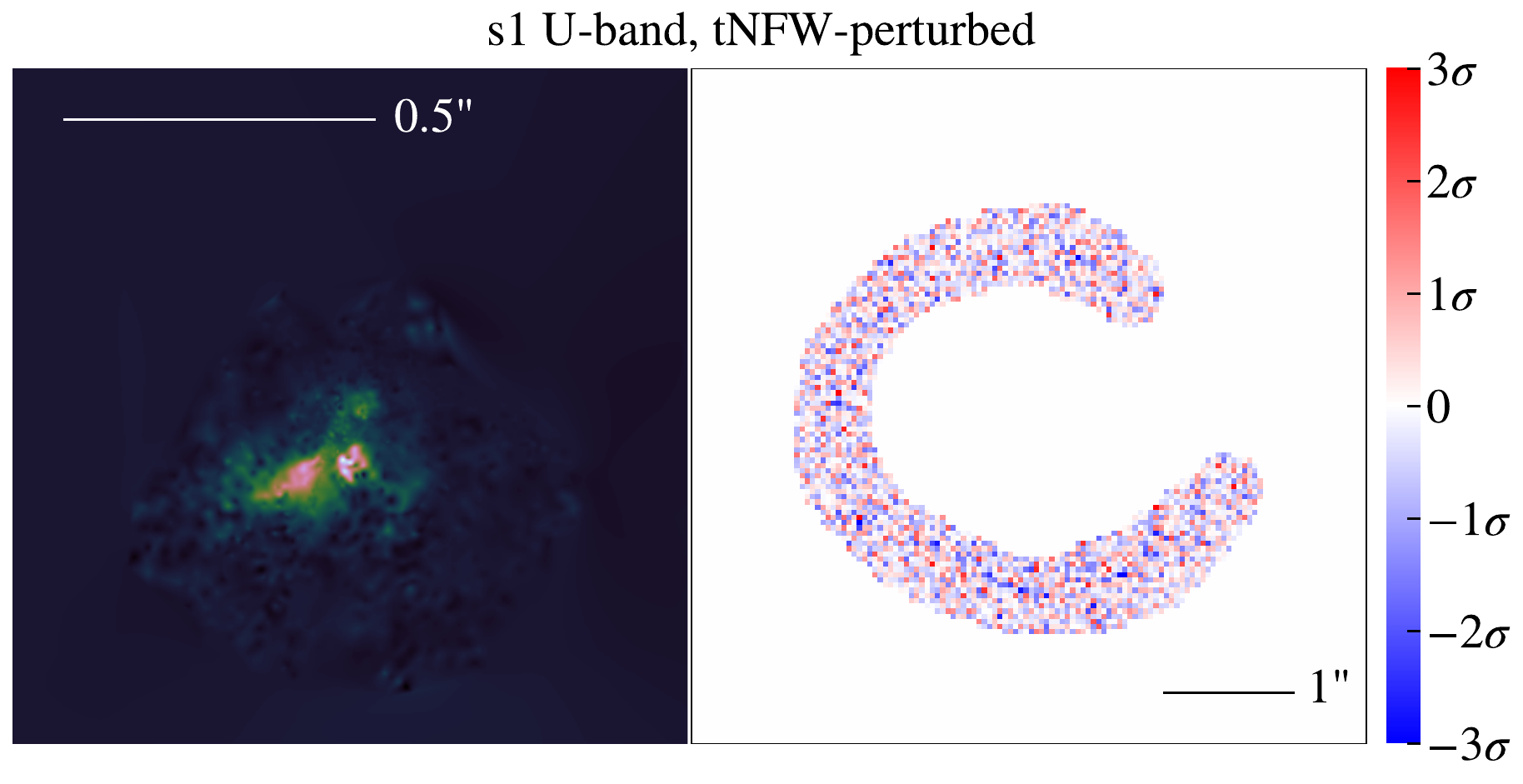}};
     \node (fig9) at (-115,-90)
       {\includegraphics[scale=0.28]{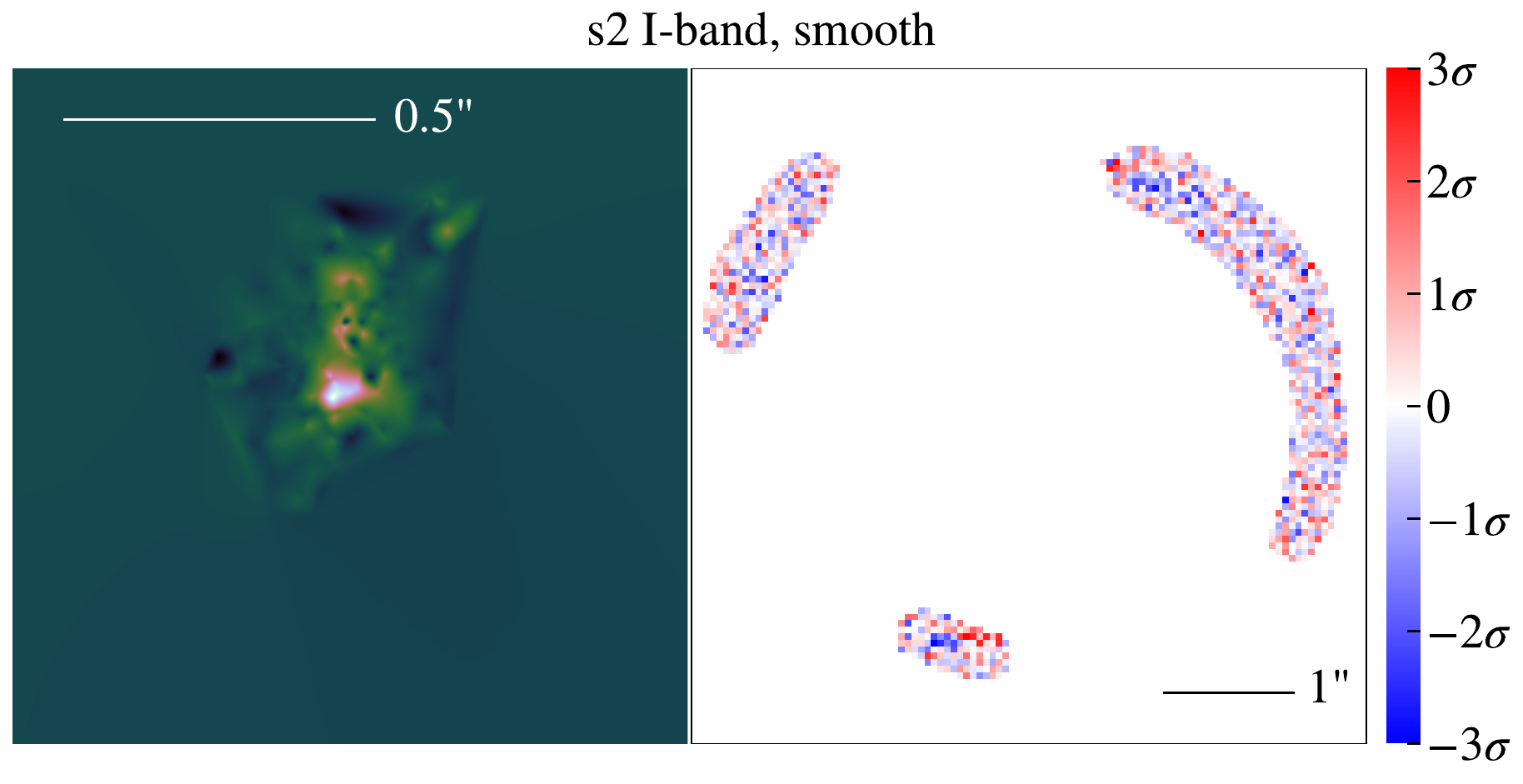}};
     \node (fig11) at (-25,-90)
       {\includegraphics[scale=0.28]{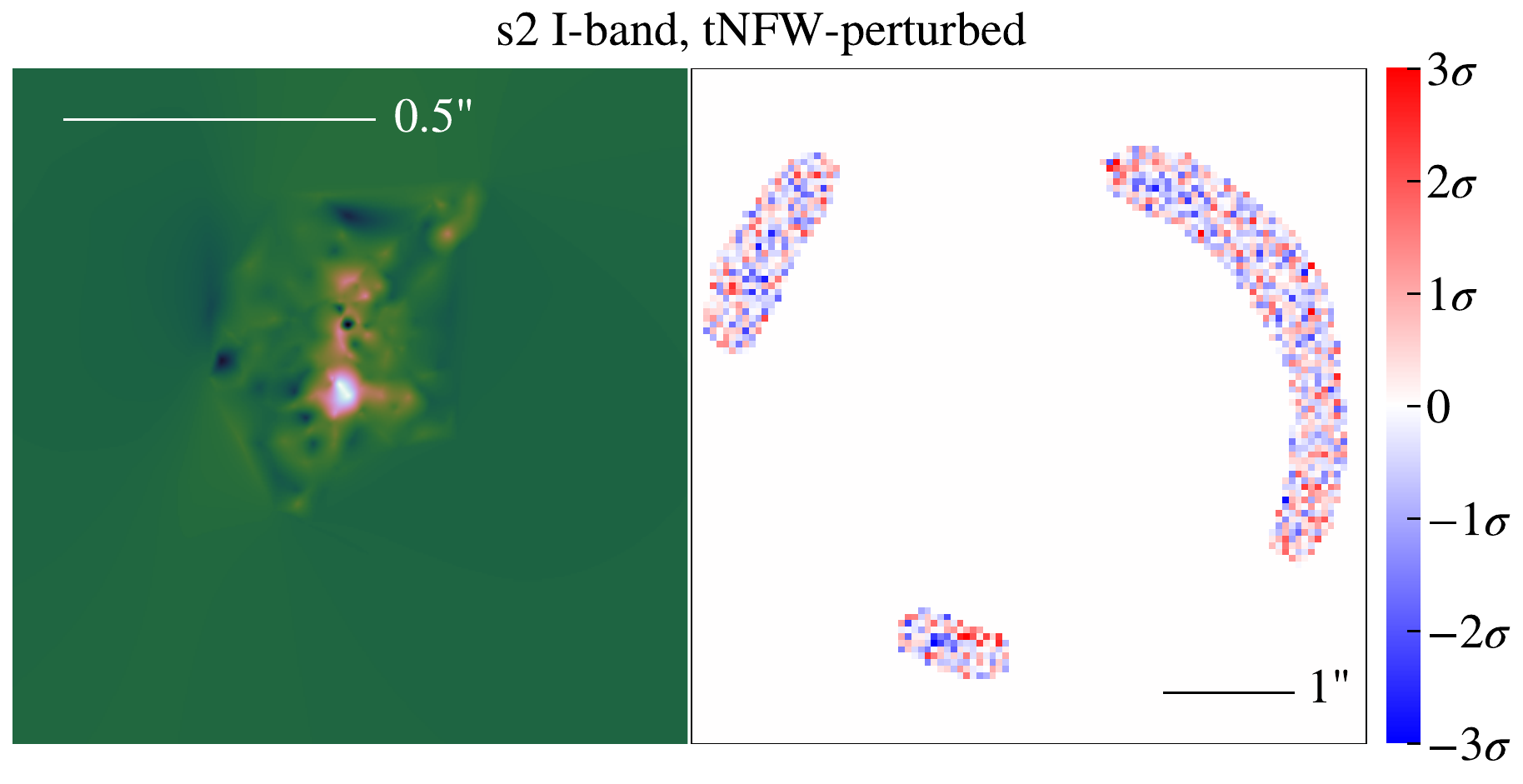}};
     \node (fig13) at (-115,-135)
       {\includegraphics[scale=0.28]{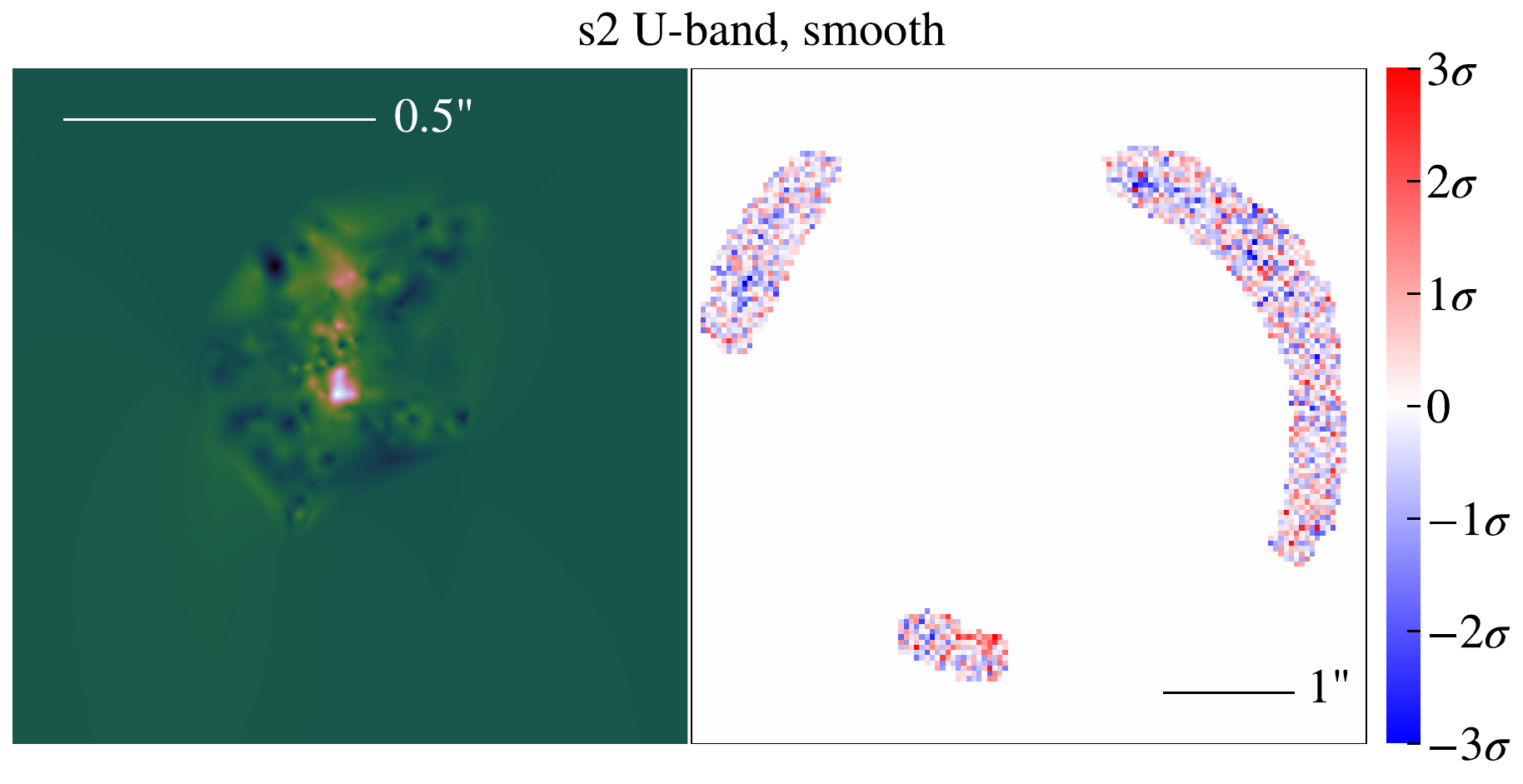}};
     \node (fig15) at (-25,-135)
       {\includegraphics[scale=0.28]{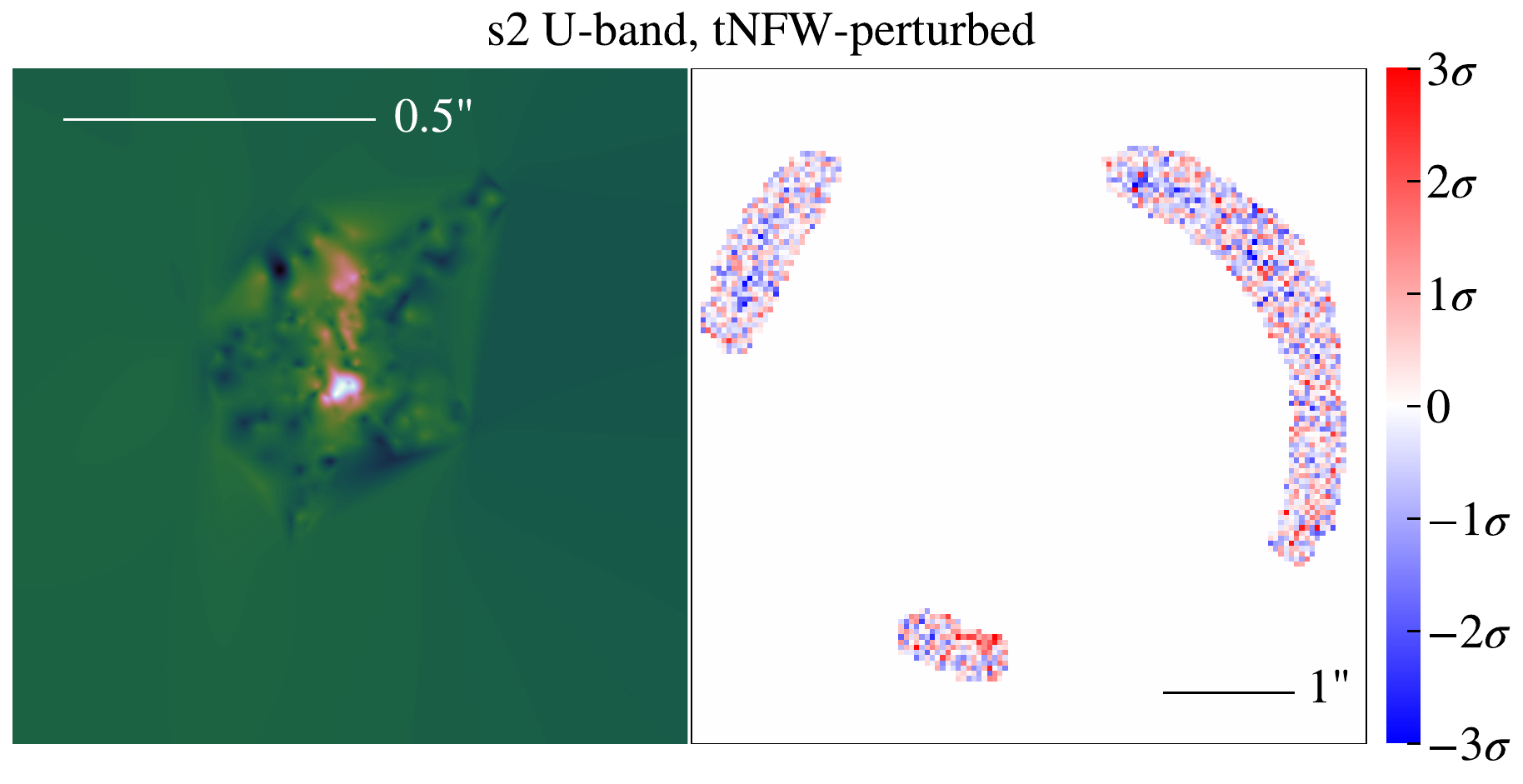}};
\end{tikzpicture}
\caption{Source plane reconstructions and normalised image plane residuals for our best fit smooth (left) and tNFW--perturbed (right) model, for (from top to bottom) $s1$ in I--band, $s1$ in U--band, $s2$ in I--band and $s2$ in U--band.}
\label{fig:sources_residuals_triple_plane}
\end{figure*}
\begin{figure*}
    \centering
    \includegraphics[width=0.999\textwidth]{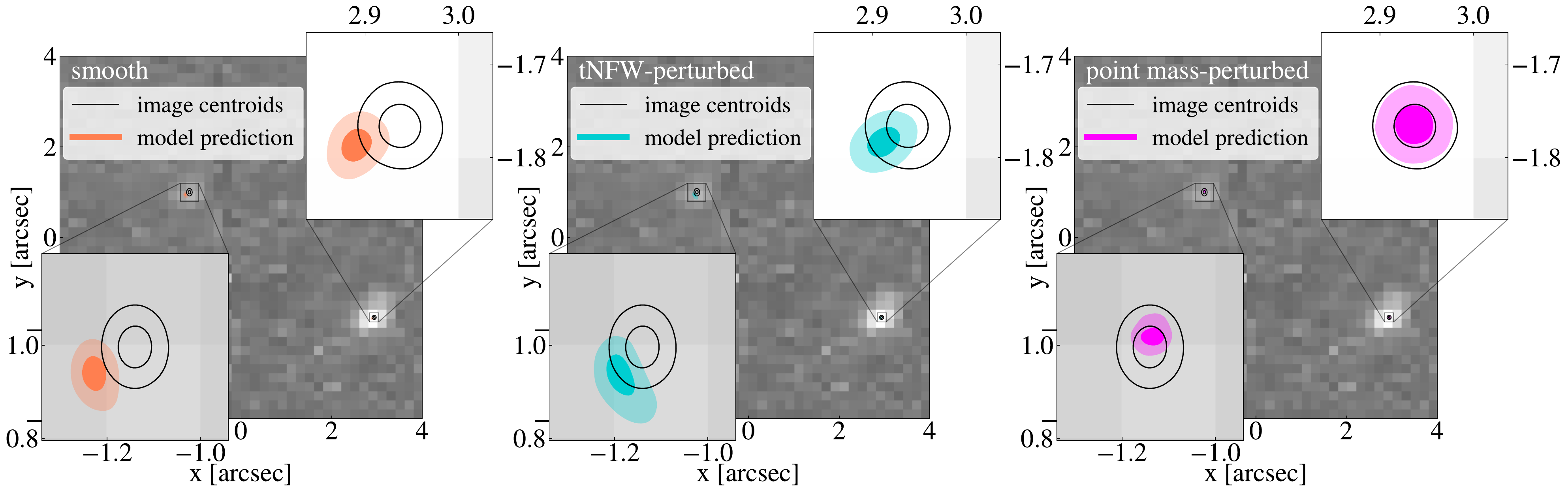}
    \caption{The $1\sigma$ and $2\sigma$ astrometric uncertainties (black contours) on the two image plane positions from the MUSE data (background image) with our posterior of $s3$ centroids forward ray--traced through our posterior of lens models to give our predicted $1\sigma$ and $2\sigma$ uncertainties on the image plane positions of $s3$, for our smooth (orange), tNFW--perturbed (cyan) and point mass--perturbed (magenta) models.}
    \label{fig:s3_delensing} 
\end{figure*}
\begin{figure}
    \centering
    \includegraphics[width=0.35\textwidth]{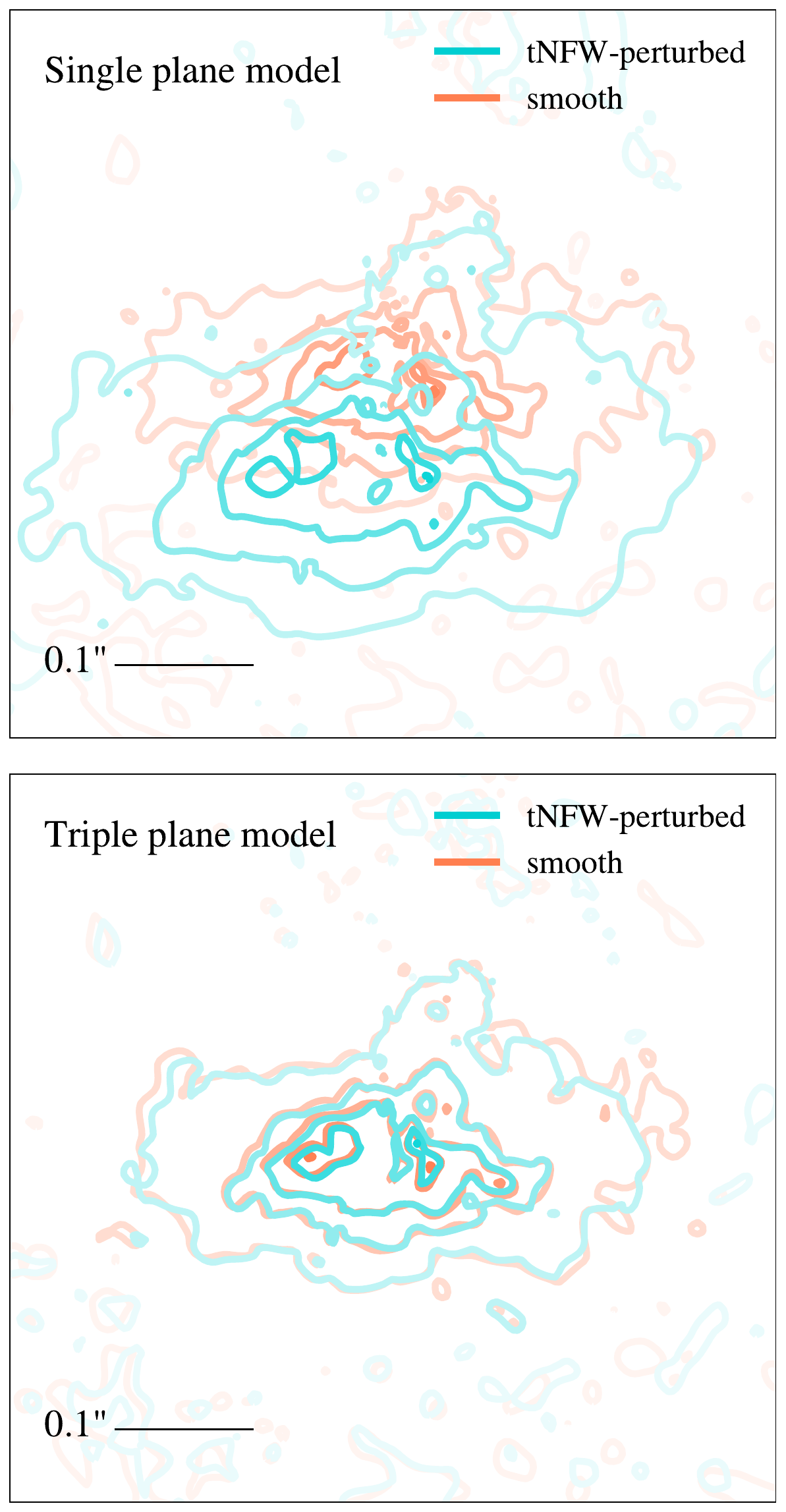}
    \caption{Isophotes of the I--band $s1$ reconstruction given the best tNFW--perturbed and smooth results from (top) the single plane modelling and (bottom) triple plane modelling. The alignment of the two source reconstructions in the latter case is indicative of a broken mass--sheet degeneracy.}
    \label{fig:broken_mass_sheet_degeneracy}
\end{figure}
\begin{figure*}
    \centering
    \includegraphics[width=0.95\textwidth]{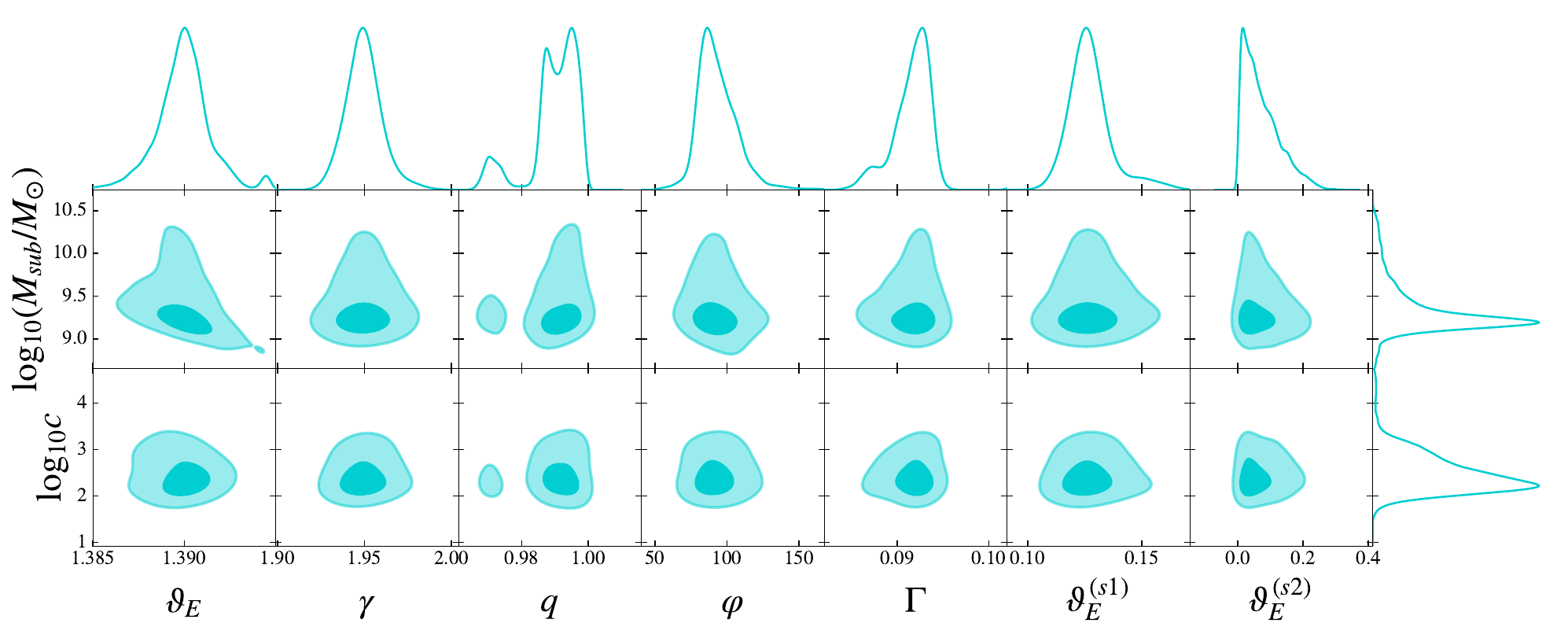}
    \caption{2D posterior distributions for the total mass, $\log_{10}(M_\mathrm{sub}/M_{\odot})$, and concentration, $\log_{10}c$, of the substructure, against a selection of other lens model parameters: (from left to right) the Einstein radius, $\vartheta_{E}$, power law slope, $\gamma$, axis ratio, $q$, and position angle, $\varphi$, of the main deflector, external shear strength, $\Gamma$, and Einstein radii of $s1$ and $s2$, $\vartheta_{E}^{(s1)}$ and $\vartheta_{E}^{(s2)}$, respectively.}
    \label{fig:m-c-degeneracies}
\end{figure*}

In this section, we present the results from our triple source plane (henceforth `fiducial') models, where we reconstruct $s1$ and $s2$ both in the I-- and U--band simultaneously, whilst also delensing $s3$ by mapping its two images to a common source plane position, with and without a tNFW perturbation. 

We use the same mass profiles and priors for the foreground lens as in our single--plane modelling, but we add an SIS at the centre of the delensed position of $s1$, allowing for a small offset between the centroids of the mass and light. We similarly add an SIS at $s2$ but enforce zero offset between the centroids of its mass and light, since \citetCS{} showed that this assumption has negligible impact on $s3$.

We find that we are able to simultaneously reproduce the I-- and U--band arcs of $s1$ and $s2$, and delens $s3$. Our source reconstructions and residuals are shown in Figure \ref{fig:sources_residuals_triple_plane}. The positions of the third source  are shown in Figure \ref{fig:s3_delensing}.

The extra data afforded from the outer set of arcs give much tighter constraints on the macro--model. We find that the super--isothermal results of \citetV{}, \citetM{}, and our single plane tNFW--perturbed models, do a comparatively poorer job of reconstructing $s2$. With our fiducial models, a near isothermal result is favoured for both the smooth and tNFW--perturbed cases, where $\gamma=1.956^{+0.009}_{-0.010}$ and $1.949^{+0.011}_{-0.010}$ respectively. The similarities between the recovered slopes and the reconstructed sources (as shown in Figure \ref{fig:broken_mass_sheet_degeneracy}) are clear demonstrations that the source position transformation of \citet{schneider_source-position_2014} has been broken by our multi--plane modelling. The $1\sigma$ and $2\sigma$ posterior distribution contours for these models -- as well as for the single plane dual I--band and U--band models -- can be found in Appendix \ref{sec:macromodelposteriors}.

We find that the existence of the tNFW perturbation is preferred with an evidence ratio $\Delta\ln\mathcal{Z}=19.64\pm0.03$ over the smooth model, corresponding to a $5.9\sigma$ detection. The preferred tNFW profile has a total mass $\log_{10}(M_\mathrm{sub}/M_{\odot})= 9.3^{+0.4}_{-0.1}$, 
with a virial mass $\log_{10}(M_{200}/M_{\odot})=10.3^{+1.2}_{-0.6}$ and concentration  $\log_{10}c=2.4^{+0.5}_{-0.3}$.

\begin{figure}
    \centering
    \includegraphics[width=0.40\textwidth]{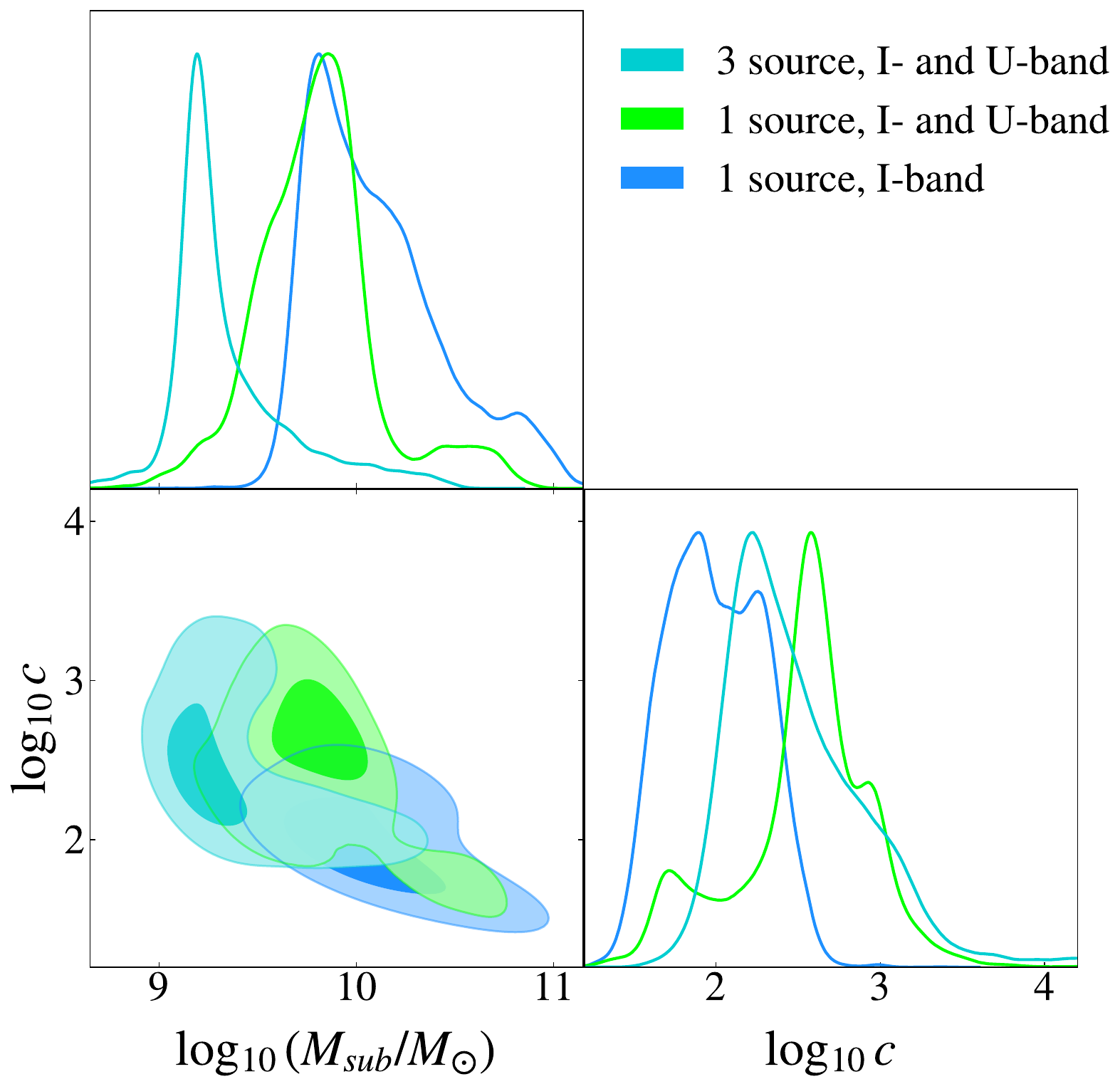}\\
    \includegraphics[width=0.40\textwidth]{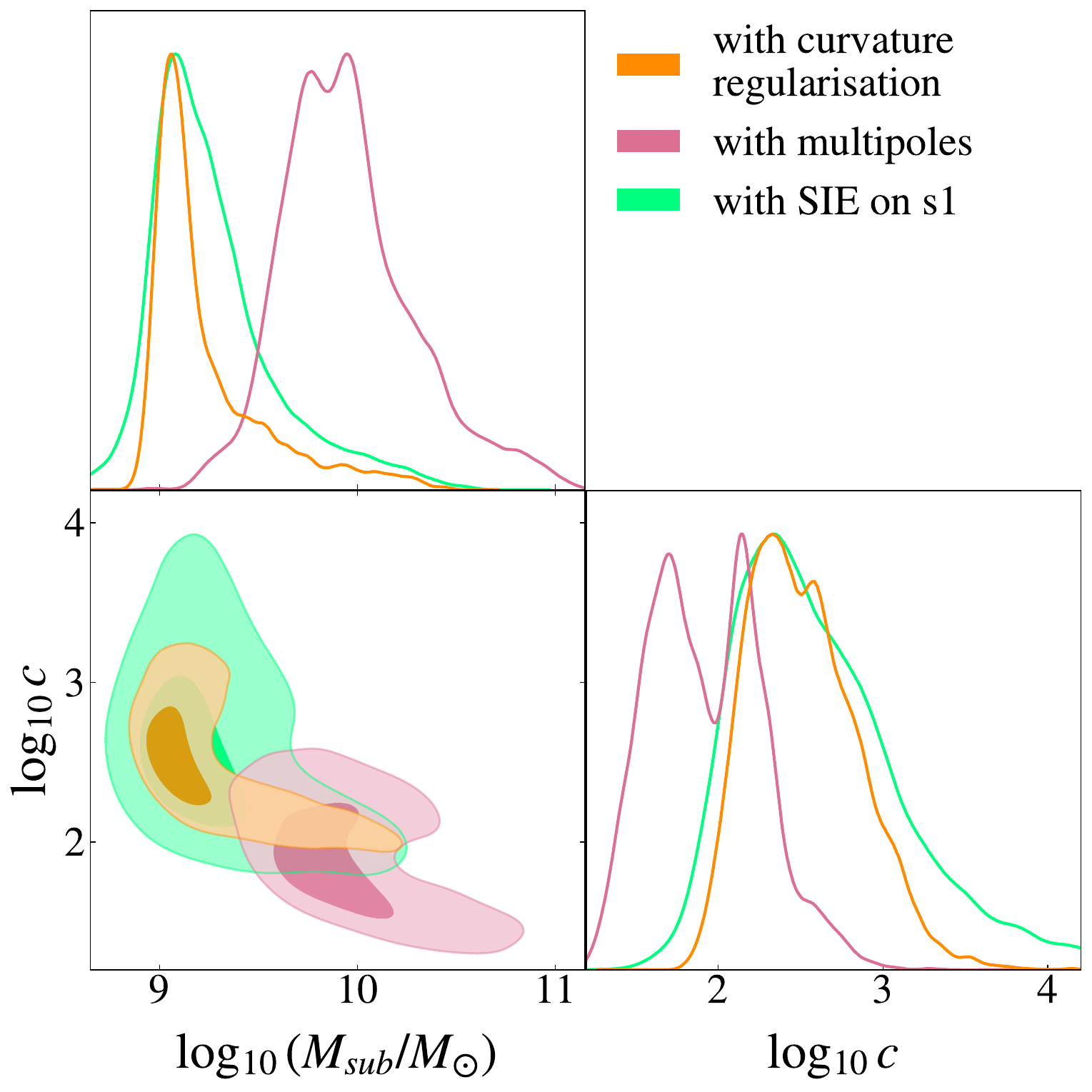}\\
    \hspace{0.15cm}\includegraphics[width=0.40\textwidth]{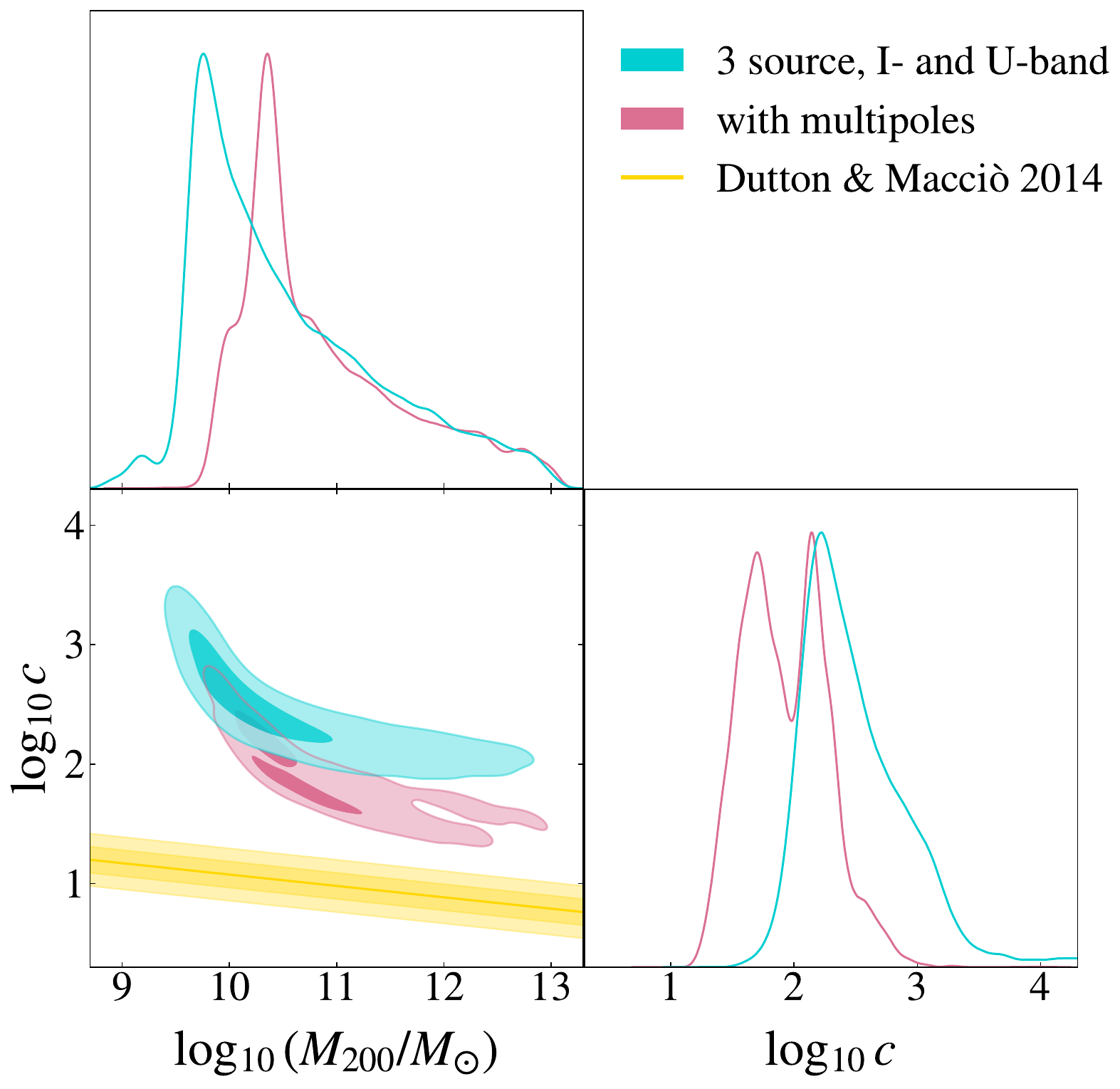} 
    \caption{The $1\sigma$ and $2\sigma$ $M_\mathrm{sub}$--$c$ posterior for our single and triple plane model fits utilising gradient regularisation (top), as well as for alternative source reconstruction and mass model assumptions (middle). The $M_{200}$--$c$ posterior for our highest evidence models from each of these two panels (fiducial and multipoles) are plotted against an $M_{200}$--$c$ relation for CDM halos from \citet{dutton_cold_2014}, with $1\sigma$ and $2\sigma$ uncertainty (bottom).}
    \label{fig:m_c_posteriors} 
\end{figure}

We show 2D posterior distributions of $M_\mathrm{sub}$ and $c$ against a selection of macro--model parameters, for the fiducial tNFW--perturbed model result in Figure \ref{fig:m-c-degeneracies}, wherein we observe a notable degeneracy between the Einstein radius of the main deflector and the mass of its substructure, since the total mass within the Einstein ring is well--constrained. Otherwise, there are no strong degeneracies. The 2D $M_\mathrm{sub}$--$c$ posterior distribution for our fiducial result is shown separately on the upper panel of Figure \ref{fig:m_c_posteriors}, overlaid with the single source plane results. Our fiducial $M_{200}$--$c$ posterior appears on the bottom panel of Figure \ref{fig:m_c_posteriors}, which also shows the $M_{200}$--$c$ relation of \citet{dutton_cold_2014}. The shape of this posterior distribution is similar to the results of \citetM{}, though our $\sigma_{c}$ is greater than theirs primarily because of our more flexible source model.  We find that our results differ from \citet{dutton_cold_2014} and the other aforementioned mass--concentration relations by 2.6--3.3 $\sigma_{c}$.

Assuming the stellar mass--subhalo mass relation in \citet{rodriguez_stellar_2012}, our virial mass implies a stellar mass $M_{\ast}\sim10^{7.5}M_{\odot}$. For a plausible stellar mass--to--light ratio of $\sim2M_{\odot}/L_{\odot}$ \citep[appropriate to a passive dwarf galaxy -- see e.g.][]{martin_comprehensive_2008}, this corresponds to an absolute magnitude $M_{I}\approx-15.4$, typical of dwarf elliptical populations in nearby galaxy groups. At this luminosity, such objects have typical sizes $\sim1\mathrm{kpc}$ \citep[][]{venhola_fornax_2019}. Introducing a simulated galaxy of these properties scaled to $z=0.222$ into the I--band image, we find that although such a galaxy would be detectable in isolation, it could not be unambiguously distinguished from other flux components if located at the position of the subhalo. Since the associated galaxy could easily be a factor of two fainter, or be more diffuse, than assumed here, we should not expect to see an easily--identified luminous galaxy hosted by the lensing substructure. The subhalo we have detected is therefore not unusually ``dark'', and appears compatible with being a dwarf satellite galaxy of the main deflector.

\section{Systematic tests}
\label{sec:systematics}

In this section, we examine several model assumptions that systematically could have influenced our ability to detect and measure a DM substructure. We perform tests on the choice of source regularisation and explore the effects of additional mass model complexity and an alternative hypothesis for the perturber. We explore all of these systematics for the triple source plane (I-- and U--band) case only.

\subsection{Degeneracy with source morphology}
One of the main systematic uncertainties is the degeneracy between the complexity of the mass and the source light distributions. While enforcing a smoother source could lead to a false positive detection of a lensing perturber, allowing too much freedom in the intrinsic structure of the source could lead to non--detections even in the presence of DM substructures.

In our fiducial model, we chose a gradient regularization scheme for the source galaxies, which allows for small--scale source structure. Alternatively, we can suppress these small--scale source features by regularising over curvature. This is the regularisation choice of \citetV. In this case, the substructure is detected with much higher significance: $\Delta\ln\mathcal{Z}$ = $67.00\pm0.02$, or $11.3\sigma$. Such a detection claim would be overconfident in our analysis since the evidence actually prefers gradient regularisation at $\sim$$20\sigma$ confidence. This result is true for both the smooth and perturbed models.

It is concerning that the significance of the detection changes hugely between the two regularisation schemes since neither 
is astrophysically motivated. It remains an open question whether alternative regularisation schemes or source reconstruction techniques could raise or lower the evidence for a substructure. We leave this exploration to future work.

The mass--concentration posterior for the substructure under the curvature regularisation scheme is shown in the centre panel of Figure \ref{fig:m_c_posteriors}. Whilst the detection significance has changed, the inferred subhalo parameters and their uncertainties have not changed significantly. The substructure would, therefore, remain a modest outlier given either regularization scheme.

\subsection{Mass model complexity}
\subsubsection{Angular structure in the main deflector}
\label{sec:multipoles}
Previous works have shown that lensing substructure inference can be sensitive to the flexibility of the main deflector mass model \citep[see e.g.][]{nightingale_scanning_2022,minor_unexpected_2021}. Therefore, we explore additional complexity in the foreground lens model by combining our EPL with the modes $m$ of a multipole expansion:
\begin{equation}
    \kappa(x,y)=\kappa_\mathrm{EPL}(x,y)\times [ ~1+k_{m}\cos{(m(\varphi-\varphi_{m})}) ~]
\end{equation}
where $\varphi=\arctan{(x/y)}$ and $0\leq k_{m}\leq 1$ is the amplitude of the $m^{th}$ mode with phase $\varphi_{m}$ \footnote{See \citet{chu_multipole_2013} and appendix B of \citet{xu_how_2015} for more details on multipoles}. Such an expansion can account for boxiness or diskiness of the lens galaxy. As in \citetM, we model multipole terms $m=3$ and $m=4$. We therefore add four non--linear parameters to the model: $k_{3}$, $k_{4}$, $\varphi_{3}$ and $\varphi_{4}$. The best fit source reconstructions and normalised image plane residuals are plotted in Appendix \ref{sec:sources_and_resiuals}.

Multipoles perform comparably well at reconstructing the data as the tNFW perturbation. In fact, a smooth model with added multipoles performs marginally better in reconstructing J0946 than a tNFW--perturbed model, with the data preferring the presence of multipoles over the presence of the tNFW profile with $1.5\sigma$ confidence. This is not solely due to there being fewer degrees of freedom in the multipoles case, since the best fit log--likelihood is also improved, with $\Delta\ln\mathcal{L}=3.74$.
The preference for non--zero multipole terms is unsurprising given detailed examination of the light profile, which reveals some disturbance in the shapes of the isophotes that can be absorbed by these extra parameters \citep{sonnenfeld_evidence_2012}.

Modelling the multipole terms and a tNFW--perturbation simultaneously provides the best reconstruction, where the substructure is detected with $6.2\sigma$ confidence. The inferred substructure in this case is more massive, with $\log_{10}(M_{\rm 200}/M_\odot) = 10.6^{+1.1}_{-0.4}$, but less concentrated, with $\log_{10}(c) = 1.9^{+0.4}_{-0.3}$, than in our fiducial model. Differences to the compared mass--concentration relations go down to 2.0--2.9 $\sigma_{c}$. The $M_{200}$--$c$ posterior for this model is shown in the bottom panel of Figure \ref{fig:m_c_posteriors}. 

\subsubsection{Additional complexity on s1}
Our fiducial model assumes a spherically symmetric mass distribution for $s1$, though its light profile is noticeably elliptical (see e.g. the top panels of Figure \ref{fig:sources_residuals_triple_plane}). We therefore perform a systematic test where we assign a Singular Isothermal Ellipsoid (SIE) to $s1$ rather than an SIS. This adds two parameters to our fiducial models: the axis ratio, $q$, and position angle, $\varphi$, of $s1$.
Our test shows that a smooth model prefers the presence of ellipticity components on $s1$ over the presence of a substructure in the main deflector with $2.9\sigma$ confidence, where both scenarios have the same number of degrees of freedom. Modelling smooth and tNFW--perturbed models with an ellipsoidal $s1$ simultaneously yields a substructure of total mass $\log_{10}(M_\mathrm{sub}/M_{\odot})=9.20^{+0.35}_{-0.21}$, virial mass $\log_{10}(M_{200}/M_{\odot})=10.04^{+1.31}_{-0.52}$ and concentration $\log_{10}c=2.53^{+0.59}_{-0.40}$ detected at the $4.8\sigma$ confidence level; this is a lower evidence substructure result than the tNFW perturbation with multipoles. The difference to the $\Lambda$CDM predictions of the mass--concentration relation remain at a level of 2.5--3.1 $\sigma_{c}$.

\subsubsection{A wandering black hole?}
Since the dark halo in \citetM{} is hard to accommodate within $\Lambda$CDM and our results have only partially alleviated that tension, it is worth considering alternative hypotheses for the perturber in J0946. Given the anomalously high concentration, and the surprising lack of a galaxy hosted within the halo, we investigate whether the perturber could be a supermassive black hole \citep[see e.g.][]{ricarte_origins_2021}.

The non--zero multipoles of the lens mass and the disrupted morphology of the light profile of the lens galaxy are characteristics of a merger where the ejection of such a black hole may not be implausible, either through 3--body ejection \citep{hoffman_dynamics_2007} or gravitational radiation recoil \citep{campanelli_maximum_2007}.

To test this proposal, we fit a 3--source model with a EPL, external shear and a point mass at the main deflector redshift, and recover a point mass of $\log_{10}(M_\mathrm{BH}/M_{\odot})={8.94^{+0.19}_{-0.08}}$. Given J0946 has a velocity dispersion of $\sim$280 km s$^{-1}$ \citet{gavazzi_sloan_2008}, the $M$--$\sigma$ relation implies that there should be a black hole of a few times $10^9 M_\odot$ \citep{kormendy_coevolution_2013} at the centre of the lens. 
Thus, the proposed ``wandering'' black hole would need to be of comparable mass to the 
expected central black hole.

The point mass--perturbed model is formally preferred over the equivalent tNFW--perturbed model at $2.7\sigma$. This is not definitive evidence and does not account for any prior preference between the models. This result is also driven purely by Occam's razor: the point mass perturbed model has a slightly lower likelihood
than the tNFW model but has fewer parameters.

As the right panel of Figure \ref{fig:s3_delensing} shows, the $s3$ image positions are sensitive to the change in mass profile, and the MUSE data is better reproduced with a point mass perturber. The significance of this is marginal, given that in all three panels the predicted centroids are well within the brightest parts of the $s3$ images. A more sophisticated treatment of $s3$ with higher--resolution data would be necessary to discriminate between possible density profiles for the perturbation.

\subsection{Alternative Lens Light Subtraction}

Previous studies \citep[e.g.][]{nightingale_scanning_2022} have highlighted the importance of an accurate model of the light distribution of the foreground lens, since lens light residuals may lead to false positive dark substructure detections.

The bulk of this work used the lens light subtraction of \citetCA. That light subtraction followed the method of \citet{auger_compact_2011}. It assumes that the lensing mass is a singular isothermal ellipsoid with external shear, and both sources are single Sérsic profiles. The lens light was then described as the sum of 3 Sérsic profiles. The best fitting triple--Sérsic model was then subtracted to give our fiducial lens light--subtracted image.

To quantify the potential source of systematic error on our substructure detection confidence coming from inaccuracies in lens light subtraction, we repeat our triple plane analysis from Section \ref{sec:multiplane} after performing an alternative lens subtraction on the data, by way of a Multi-Gaussian Expansion (MGE) fitted to the lens light and subtracted from the original image. We simultaneously fit the lens light with a 15 component MGE and $s1$ with a pixelated grid lensed by an EPL plus external shear model. The MGEs were concentric, but their amplitude, scale radius, ellipticity and position angle were free parameters for each component. We then subtracted off the MGE decomposed light profile to leave us with an alternative lens light--subtracted image to analyse.

With this alternative lens light subtraction we find a preference for the substructure with $\Delta\ln\mathcal{Z}=11.84\pm{0.03}$, corresponding to a $4.9\sigma$ detection, this is lower than the $5.9\sigma$ found with the original lens light--subtracted image. This example shows that whilst different reasonable choices for modelling lens light can alter the confidence of a substructure detection, the impact is unlikely to fundamentally change conclusions about the presence, or absence, of a substructure in J0946.

\subsection{Impact of the Point Spread Function}

We also test the potential for systematic errors to be introduced by a poor choice of PSF. The PSF in all our modelling was a well--motivated choice, having come from a star within the instrument's field of view. To investigate the impact of a poor choice of PSF on the detection of the substructure, we rotate our PSF through 90 degrees. Repeating the triple plane analysis of Section \ref{sec:multiplane}, but with the rotated PSF we recover the substructure with $\Delta\ln\mathcal{Z}=12.65\pm0.03$, corresponding to a $4.7\sigma$ detection. Therefore, the substructure detection again remains robust to $\sim1\sigma$ accuracy, even assuming a blurring operator which mismatches the instrument's optics.

\section{Conclusions}
\label{sec:conclusions}
In this paper, we have presented a gravitational imaging case study of the compound lens SDSSJ0946+1006. Our model is remarkably successful in its ability to simultaneously reproduce the images of two background sources in this system in both the HST I and U bands and the image positions of a third source observed by MUSE.

By including multiple sources in our analysis, we were able to lift many of the lens modelling degeneracies that are present for a single source plane lens, whilst modelling multiple passbands simultaneously enabled us to probe different source structures, and possibly different regions in the source plane, thus disentangling structure in the lens from structures in the source\footnote{Additionally, differences between the I-- and U--band structures in the s1 arcs (and source reconstructions) strongly suggest the presence of dust in $s1$, in exactly the part of the source plane that is most sensitive to the lensing substructure, yet poorly probed by the strongly attenuated U--band data. 
Upcoming 400--GHz ALMA observations of J0946 may be able to recover any dust continuum emission from s1, providing another set of constraints on the perturbing structure.}.

By comparing the Bayesian evidence of a smooth halo model to that of a tNFW--perturbed model, we test the claims that a dark subhalo exists in J0946 (in agreement with e.g. \citetV{}, \citet{nightingale_scanning_2022}). Our model prefers the existence of a subhalo with an evidence ratio $\Delta\ln\mathcal{Z}=19.64\pm0.03$ over the smooth model, corresponding to a $5.9\sigma$ detection.

The virial mass of the halo is $\log_{10}(M_{200}/M_{\odot})=10.3^{+1.2}_{-0.6}$, and its concentration is $\log_{10}c=2.4^{+0.5}_{-0.3}$, which is $2.6$--$3.3\sigma_{c}$ higher than predicted by simulations. This is a much weaker tension than reported in \citetM{} due to the inclusion of more data, the use of wider priors, and our more flexible source model. Additionally, \citet{nadler2023} recently showed that gravothermal core collapse seen in some Self--Interacting Dark Matter (SIDM) models \citep{Despali2019} is a potential mechanism to produce the substructure reported by \citetM{}; our less concentrated result should therefore be even easier to accommodate in SIDM.

The stellar mass of the subhalo, $M_{\ast}\sim10^{7.5}M_{\odot}$, implied by its virial mass indicates that any luminous component to the subhalo would not be possible to detect in the data, given its proximity to the upper arc image of $s1$ or possible blending with residual flux from the subtracted light profile of the lens. It is therefore unsurprising that the lensing perturber is dark, and we cannot confidently distinguish between it being a dwarf satellite galaxy or a DM substructure of the main deflector.

We can alternatively model the data with a black hole of $\log_{10}(M_\mathrm{BH}/M_\odot)={8.94^{+0.19}_{-0.08}}$, which is preferred over the truncated NFW profile at $2.7\sigma$ due to having fewer degrees of freedom. This scenario represents a supermassive black hole being ejected from the lens galaxy as a consequence of a merger event. For the $M$--$\sigma$ relation to hold, our resultant wandering black hole has comparable mass to the black hole expected at the centre of the lens galaxy.

Our analysis confirms that the distant source $s$3 is especially sensitive to the properties of the lensing perturbation, but the results are currently limited by the relatively low angular resolution of the MUSE data. High--resolution imaging of this source would be extremely powerful to probe the profile of the dark substructure, but will require a substantial investment of telescope time.

We also tested changes to the shape of the mass distribution in the macro--model by fitting of third and fourth order multipoles, as well as fitting for the ellipticity of $s1$. Whilst our macro--model has moved somewhat under these changes, our highest evidence model (with multipoles in the main deflector) yields $\sim6\sigma$ preference for the presence of a substructure in J0946. Its substructure gives the best compatibility with CDM simulations that we have found, at $2.0\sigma_{c}$.

We demonstrated that we are able to recover the subhalo with much higher confidence ($11.3\sigma$ versus $5.9\sigma$) when regularising over the curvature of the sources rather than the gradient of the sources. Curvature regularisation makes the sources intrinsically smoother whilst the addition of a dark substructure counteracts this by adding small--scale perturbations to the arcs. However, the Bayesian evidence vastly prefers our fiducial gradient regularisation scheme. 

 Ultimately, we conclude that precision lens modelling is challenging. Alongside cosmography, gravitational imaging is perhaps the hardest lens modelling challenge of all. Even with the luxuries afforded by a compound lens in its ability to suppress the mass--sheet degeneracy, there are nuances in how complexity is afforded to the lensing mass models, and the reconstruction of light profiles in background sources, that make it difficult to draw conclusions about small--scale structures with certainty. Much care needs to be taken over the choices of the background source model before embarking on detailed lens modelling. In reality, random draws from the priors of curvature or gradient regularised sources look nothing like astrophysical galaxies: ultimately neither regularisation scheme is physical; much more work is needed to understand how to reconstruct sources, and the need for evidence calculations will make this work computationally expensive. The potential payoff for this work is huge: with hundreds of thousands of lenses to be discovered in the next decade \citep{collett_population_2015}, gravitational imaging should yet place stringent constraints on the small--scale validity of $\Lambda$CDM.

\section*{Acknowledgements}
\label{sec:acknowledgements}
We are grateful to James Nightingale and Qiuhan He for sharing their results on lens modelling with natural neighbour interpolation.  Adopting this approach allowed us to overcome the sampling issues inherent to linear interpolation shown in Figure \ref{fig:likelihood_discontinuities}.
We thank Quinn Minor for insightful discussions at the IAU strong lensing symposium in Otranto.
We are grateful to Karina Rojas for comments on the manuscript.
DJB is funded by a graduate studentship from UK Research and Innovation's STFC and the University of Portsmouth.
TEC is funded by a Royal Society University Research Fellowship.
DJB,  WJRE and TEC and this project have received funding from the European Research Council (ERC)
under the European Union’s Horizon 2020 research and innovation
programme (LensEra: grant agreement No 945536).
HCT is funded by a STFC studentship.
RJS is supported by the STFC through the Durham Astronomy Consolidated Grants 
(ST/T000244/1 and ST/X001075/1).
This work was made use of the SCIAMA computing cluster at Portsmouth
 For the purpose of open access, the authors have applied a Creative Commons Attribution (CC BY) licence to any Author Accepted Manuscript version arising. 
The authors also acknowledge seedcorn funding from the DiRAC HPC Facility (project dp285).
This work was performed using the Cambridge Service for Data Driven Discovery (CSD3), part of which is operated by the University of Cambridge Research Computing on behalf of the STFC DiRAC HPC Facility (www.dirac.ac.uk). The DiRAC component of CSD3 was funded by BEIS capital funding via STFC capital grants ST/P002307/1 and ST/R002452/1 and STFC operations grant ST/R00689X/1. 
This work further used the DiRAC@Durham facility managed by the Institute for Computational Cosmology on behalf of the STFC DiRAC HPC Facility (www.dirac.ac.uk). The equipment was funded by BEIS capital funding via STFC capital grants ST/P002293/1 and ST/R002371/1, Durham University and STFC operations grant ST/R000832/1. DiRAC is part of the National e--Infrastructure.
\section*{Data Availability}
\label{sec:dataavailability}
Supporting research data are available on request from the corresponding author and from the HST and VLT archives.
 



\bibliographystyle{mnras}
\bibliography{bibliography} 



\appendix
\section{Scaling convention of our Einstein radii}
\label{sec:scale_factor}
For multi--plane lensing, the Einstein radius of a lens is not a well defined quantity since it changes with the source redshift. The physical deflection angle is source redshift independent, but it is not a convenient quantity for strong lensing, since it would require angular diameter distances to appear throughout the lens equation.

\citet{schneider_gravitational_1992}  adopt a convention for scaling deflection angles, such that the Einstein radius, $\vartheta_{E}$, of any deflector is defined with respect to light originating on the final source plane. This produces the following, recursive multi--plane lens equation, which relates $\boldsymbol{\theta}_{\nu}$, the angular position on plane $\nu$, to the image plane position, $\boldsymbol{\theta}_{1}$:
\begin{equation}
\label{eq:multiplane_schneider}
    \boldsymbol{\theta}_{\nu} = \boldsymbol{\theta}_{1} - \sum_{\mu=1}^{\nu-1}{\beta_{\mu\nu}\boldsymbol{\alpha}_{\mu}(\boldsymbol{\theta}_{\mu}}) ,
\end{equation}
where beta is defined as
\begin{equation}
    \beta_{\mu\nu}=\frac{D_{\mu\nu}D_{s}}{D_{\nu}D_{\mu s}} ,
\end{equation}
with $s$ representing the most distant source.

As pointed out by \citetCS, Equation \ref{eq:multiplane_schneider} is inflexible to the discovery of additional, more distant source planes. One option to circumvent this is to define Einstein radii acting on sources coming from infinity, but such sources are never observable so these Einstein radii would be highly degenerate with other lens model parameters. Instead, we derive a modified scaling relation that defines Einstein radii as acting on the source most immediately behind each lens.

For a lens at redshift $z_{l}$, we compute the same projected mass density, $\Sigma(\boldsymbol{\theta})$, regardless of whether we measure the Einstein radius from a lensed source at redshift $z_{i}$ or $z_{j}$. Recalling that $\Sigma(\boldsymbol{\theta})=\kappa(\boldsymbol{\theta})\Sigma_\mathrm{crit}$, we must therefore always satisfy the condition
\begin{equation}
    \kappa(\boldsymbol{\theta}, z_{i})\Sigma_\mathrm{crit}(z_{l}, z_{i})=\kappa(\boldsymbol{\theta}, z_{j})\Sigma_\mathrm{crit}(z_{l}, z_{j}) .
\end{equation}
We therefore define the scaling relation
\begin{equation}
    \eta_{ij}=\frac{\kappa(\boldsymbol{\theta}, z_{j})}{\kappa(\boldsymbol{\theta}, z_{i})}=\frac{\Sigma_\mathrm{crit}(z_{l}, z_{i})}{\Sigma_\mathrm{crit}(z_{l}, z_{j})} .
\end{equation}
Since $\Sigma_\mathrm{crit}(z_{A}, z_{B})\propto\frac{D_{B}}{D_{AB}D_{A}}$, this relation can be rewritten as
\begin{equation}
    \eta_{ij}=\frac{D_{i}D_{lj}}{D_{li}D_{j}}.
\end{equation}
This is mathematically identical to the scale factor $\beta_{\mu\nu}$ in \citet{schneider_gravitational_1992}, though the subscripts are defined differently. Their scaling parameter, $\beta_{\mu\nu}$, is defined such that when $\nu$ is the redshift of the furthest source, $\beta_{\mu\nu}=1$. Our equivalent condition, $\eta_{ij}=1$, is achieved when $j=i$. Physically, this corresponds to the source being on the plane most immediately behind the lens.

Throughout this work, we opt to use $\eta_{ij}$, such that the Einstein radius of the main deflector is quoted according to the Einstein ring produced by $s1$ observed in the image plane. The Einstein radius we quote for $s1$ would correspond to an Einstein ring being produced by $s2$ observed at $s1$, and so on. This is an intuitive convention as it preserves the value of Einstein radius for the main deflector from previous single source plane studies of J0946, which is approximately measurable by visual inspection given the physical scale of a pixel.

Equation \ref{eq:multiplanelensequation} describes the ray tracing through multiple source planes using this new scaling parameter. Explicitly, the single, double and triple source plane cases become
\begin{equation}
    \begin{split}
        \boldsymbol{\theta}_{1}&=\boldsymbol{\theta}_{0}-\eta_{11}\boldsymbol{\alpha}_{0}(\boldsymbol{\theta}_{0})&\\
        &=\boldsymbol{\theta}_{0}-\boldsymbol{\alpha}_{0}(\boldsymbol{\theta}_{0}) ,
    \end{split}
\end{equation}

\begin{equation}
    \begin{split}
        \boldsymbol{\theta}_{2}&=\boldsymbol{\theta}_{0}-\eta_{12}\boldsymbol{\alpha}_{0}(\boldsymbol{\theta}_{0})-\eta_{22}\boldsymbol{\alpha}_{1}(\boldsymbol{\theta}_{1})&\\
        &\begin{split}
            \hspace{2.5pt}=\boldsymbol{\theta}_{0}&-\eta_{12}\boldsymbol{\alpha}_{0}(\boldsymbol{\theta}_{0})&\\
                                                  &-\eta_{22}\boldsymbol{\alpha}_{1}(\boldsymbol{\theta}_{0}-\eta_{11}\boldsymbol{\alpha}_{0}(\boldsymbol{\theta}_{0}))
        \end{split}
        &\\
        &\begin{split}
            \hspace{2.5pt}=\boldsymbol{\theta}_{0}&-\eta_{12}\boldsymbol{\alpha}_{0}(\boldsymbol{\theta}_{0})&\\
                                                  &-\boldsymbol{\alpha}_{1}(\boldsymbol{\theta}_{0}-\boldsymbol{\alpha}_{0}(\boldsymbol{\theta}_{0})) ,
        \end{split}
    \end{split}
\end{equation}
and
\begin{equation}
    \begin{split}
        \boldsymbol{\theta}_{3}&=\boldsymbol{\theta}_{0}-\eta_{13}\boldsymbol{\alpha}_{0}(\boldsymbol{\theta}_{0})-\eta_{23}\boldsymbol{\alpha}_{1}(\boldsymbol{\theta}_{1})-\eta_{33}\boldsymbol{\alpha}_{2}(\boldsymbol{\theta}_{2})&\\
        &\begin{split}
            \hspace{2.5pt}=\boldsymbol{\theta}_{0}&-\eta_{13}\boldsymbol{\alpha}_{0}(\boldsymbol{\theta}_{0})&\\
            &-\eta_{23}\boldsymbol{\alpha}_{1}(\boldsymbol{\theta}_{0}-\eta_{11}\boldsymbol{\alpha}_{0}(\boldsymbol{\theta}_{0}))&\\
            &-\eta_{33}\boldsymbol{\alpha}_{2}(\boldsymbol{\theta}_{0}-\eta_{12}\boldsymbol{\alpha}_{0}(\boldsymbol{\theta}_{0})-\eta_{22}\boldsymbol{\alpha}_{1}(\boldsymbol{\theta}_{0}-\eta_{11}\boldsymbol{\alpha}_{0}(\boldsymbol{\theta}_{0})))
        \end{split}
        &\\
        &\begin{split}
            \hspace{2.5pt}=\boldsymbol{\theta}_{0}&-\eta_{13}\boldsymbol{\alpha}_{0}(\boldsymbol{\theta}_{0})&\\
            &-\eta_{23}\boldsymbol{\alpha}_{1}(\boldsymbol{\theta}_{0}-\boldsymbol{\alpha}_{0}(\boldsymbol{\theta}_{0}))&\\
            &-\boldsymbol{\alpha}_{2}(\boldsymbol{\theta}_{0}-\eta_{12}\boldsymbol{\alpha}_{0}(\boldsymbol{\theta}_{0})-\boldsymbol{\alpha}_{1}(\boldsymbol{\theta}_{0}-\boldsymbol{\alpha}_{0}(\boldsymbol{\theta}_{0}))) ,
        \end{split}
    \end{split}
\end{equation}
respectively.
\section{The Likelihood function of our model}
\label{sec:likelihood}
\begin{figure*}
    \centering
    \includegraphics[width=0.8\textwidth]{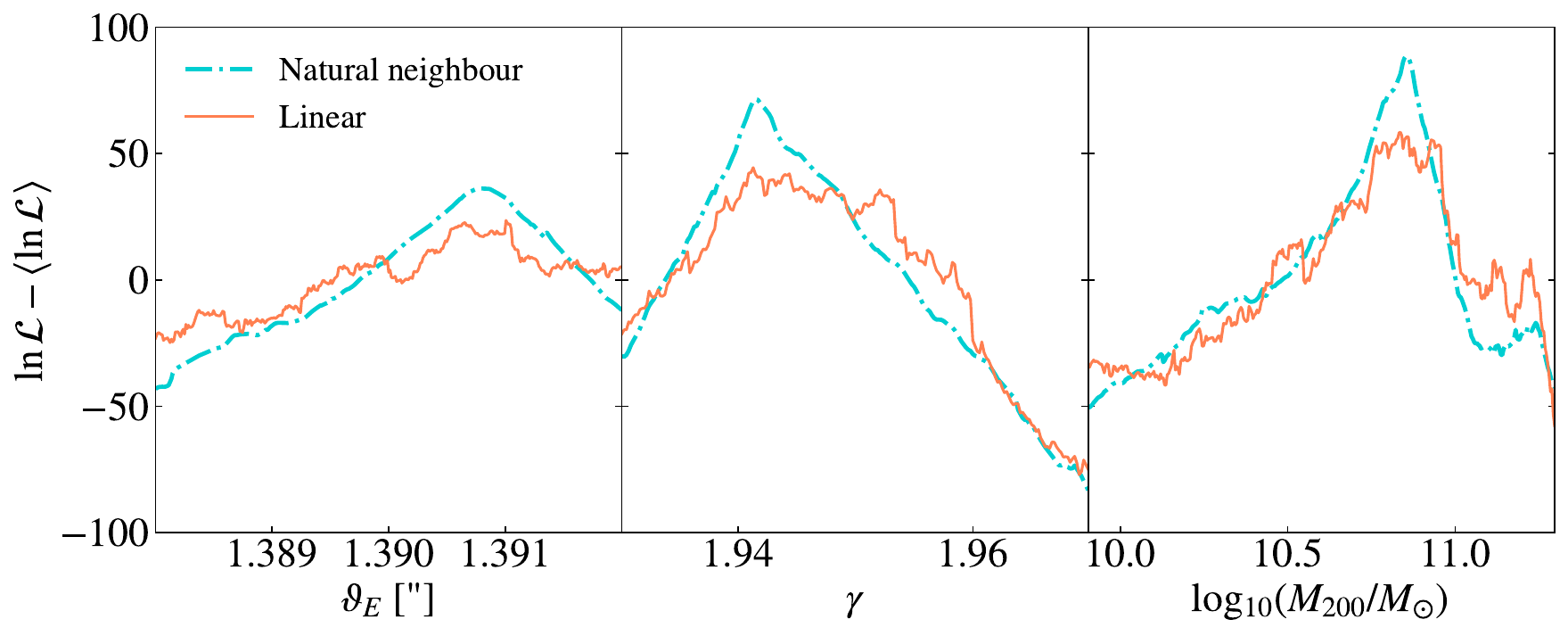}
    \caption{Slices through the logarithmic likelihood in (from left to right) the Einstein radius, $\vartheta_{E}$, and slope, $\gamma$, of the main deflector, and the virial mass, $\log_{10}(M_{200}/M_{\odot})$, of the tNFW substructure, shown for natural neighbour and linear interpolation schemes. Both cases have their mean likelihood subtracted for easier comparison. While the linear interpolation scheme shows many local discrete jumps, the natural neighboring interpolation renders the likelihood smooth with the highest peak much easier to identify.}
    \label{fig:likelihood_discontinuities}
\end{figure*}
We construct our likelihood similarly to other gravitational imaging methods \citep[see e.g.][]{warren_semilinear_2003, vegetti_bayesian_2009, nightingale_adaptive_2015}. We assume that the data vector $\boldsymbol{d}$ is a linear function of the source brightness $\boldsymbol s$ and the instrumental noise $\boldsymbol n$:
\begin{equation}
\boldsymbol d = \boldsymbol{F}(\boldsymbol\xi) \boldsymbol s + \boldsymbol n
\end{equation}
The operator $\boldsymbol F(\boldsymbol \xi) =\boldsymbol M \boldsymbol P \boldsymbol L(\boldsymbol \xi) $ is the product of the mask operator, $\boldsymbol  M$, the blurring operator describing the effects of the band--specific PSF, $\boldsymbol P$, and the lensing operator mapping each source pixel to the corresponding positions of the lensed images according to the lens model parameters, $\boldsymbol 
 L(\boldsymbol\xi)$. This mapping follows the interpolation scheme described in Section \ref{sec:source}. Furthermore, the source $\boldsymbol s$ and the noise $\boldsymbol n$ follow zero--mean Gaussian priors with the covariance matrices $\boldsymbol R$ and $\boldsymbol C$, respectively. As a result, the source marginalised likelihood for one source becomes
\begin{equation}
\mathcal{L} (\boldsymbol\xi) = \int d \boldsymbol s~ \mathcal{G}( \boldsymbol d -  \boldsymbol  F(\boldsymbol\xi) \boldsymbol s  , \boldsymbol C )  \mathcal{G}( \boldsymbol s , \lambda^{-1} \boldsymbol R) .
\end{equation}

By choosing mutually exclusive masks for the two Einstein radii, we can decompose the likelihood into individual likelihoods. For source $i$ in band $j$, the source--marginalised likelihood is given by
\begin{equation}
\mathcal{L}_{ij}(\boldsymbol \xi) = \int d \boldsymbol s_{ij} ~ \mathcal{G}(  \boldsymbol d_j - \boldsymbol F_{ij}(\boldsymbol \xi) \boldsymbol s_{ij} , \boldsymbol C_j )  \mathcal{G}( \boldsymbol s_{ij} , \lambda_{ij}^{-1} \boldsymbol R_{ij}) ,
\end{equation}
with $\boldsymbol F_{ij}(\boldsymbol \xi)  = \boldsymbol  M_i \boldsymbol P_j \boldsymbol L_i( \boldsymbol \xi)$ applying the mask and lensing operator with respect to source $i$ and the point spread function of band $j$.
The product of the above likelihoods provides the corresponding joint likelihood:
\begin{equation}
\mathcal{L}_{\rm tot} (\boldsymbol \xi) = \prod_{ij} \mathcal{L}_{ij}( \boldsymbol \xi) .
\label{eq:likelihood_product}
\end{equation}

To remove over-- and under--focused solutions from our likelihood, we trace four of the brightest image plane pixels of $s1$ to their source plane, where they are expected to correspond to roughly the same location. Whenever a pixel from outside the first source mask falls within one standard deviation around the centre of these four points, we decrease our likelihood by a factor of $10^{-10}$. This modification prevents our posterior from containing very smooth but non--physical source reconstructions that resemble the rings of the original data rather than a realistic source galaxy, for example, the scenario where there is no lensing and the sources are intrinsically arcs. Such solutions would otherwise be allowed by our mass model and source model priors. 

Finally, we include the image positions of the third source in our likelihood by adding an additional Gaussian likelihood term to Equation \ref{eq:likelihood_product}. This term punishes the chi--squared difference between predicted image positions of the third source for given model parameters $\boldsymbol \xi$ and observed positions in the VLT--MUSE data, with the positions and uncertainties taken from \citetCS{}.
\subsection{Source interpolation and likelihood}
The choice of interpolation scheme affects the stability and speed of our analyses. While linear interpolation schemes on Delaunay meshes have been popular in previous studies (e.g. \citetV{}), they tend to create discontinuities in the likelihoods of lens modelling. In contrast, natural neighbour interpolation gives rise to much smoother likelihoods. For example, Figure \ref{fig:likelihood_discontinuities} illustrates this behaviour as a function for parameters close to a maximum a posteriori point. Although the parameter ranges plotted are smaller, the linear interpolation scheme using the adaptive Delaunay mesh shows an abundance of peaks and troughs.

While a single likelihood evaluation tends to be slower for the natural neighbour interpolation scheme, the discontinuities and abundance of local maxima in the linear interpolation require more evaluations overall. Furthermore, sampling algorithms may get stuck in local maxima and, as a result, may underestimate the uncertainty of various model parameters. The Bayesian evidence prefers the natural neighbour interpolation scheme for our fiducial model with $\Delta\ln\mathcal{L}\gtrsim10$. 
\section{Tables of Results}
\label{sec:prior_posterior_tables}
Here we show the assumed priors and inferred posteriors on all our model parameters for all our discussed results. Table \ref{tab:posteriors_grad} contains the results for the single plane cases (one band and two band) and triple plane fiducial result, all assuming gradient regularisation. Table \ref{tab:posteriors_curv} contains single and triple plane results similarly, but with curvature regularisation. Table \ref{tab:posteriors_systematics} shows results from the systematic tests on mass model complexity: including multipoles in the foreground lens, adding ellipticity to $s1$, and modelling the perturber as a point mass instead of a subhalo.
\begin{table*}
\caption{Chart showing all lens model results obtained using gradient--regularised source reconstructions. Posteriors are quoted as medians with $1\sigma$ error bars. The total mass of the substructure, $M_\mathrm{sub}$, is a derived parameter, obtained by integrating over the whole tNFW profile, and is shown here as well as the non--truncated virial mass, $M_{200}$, which is sampled. Note that $\Delta x$ and $\Delta y$ indicate where a centroid coordinate is quoted relative to the mean $s1$--plane location of four bright conjugate points from the inner set of image plane arcs.}
\label{tab:posteriors_grad}
\centering
\scalebox{0.85}{
\begin{tabular}{|l|l|l|l|l|l|l|l|}
\hline
\textbf{Parameter} & \textbf{Prior} & \multicolumn{6}{c}{\textbf{Posterior (median with $1\sigma$ uncertainties)}} \\ 
                    &                & \multicolumn{2}{c}{Single plane (I--band)} & \multicolumn{2}{c}{Single plane (I-- \& U--band)} & \multicolumn{2}{c}{Triple plane} \\  \hline         
EPL $\vartheta_{E}$ [$\arcsec$]              & $\mathcal{U}(0.9, 1.5)$            & $1.397^{+0.001}_{-0.001}$   & $1.386^{+0.003}_{-0.004}$   & $1.402^{+0.001}_{-0.001}$   & $1.382^{+0.003}_{-0.002}$ & $1.397^{+0.001}_{-0.001}$     & $1.390^{+0.001}_{-0.001}$ \\ 
EPL $\gamma$                          & $\mathcal{U}(1.5, 2.5)$            & $2.123^{+0.026}_{-0.065}$   & $2.265^{+0.046}_{-0.036}$   & $1.923^{+0.033}_{-0.019}$   & $2.227^{+0.022}_{-0.023}$ & $1.956^{+0.009}_{-0.010}$     & $1.949^{+0.011}_{-0.010}$ \\ 
EPL $q$                               & $\mathcal{U}(0.6, 1.0)$            & $0.960^{+0.006}_{-0.009}$   & $0.989^{+0.007}_{-0.010}$   & $0.971^{+0.006}_{-0.004}$   & $0.975^{+0.005}_{-0.004}$ & $0.972^{+0.002}_{-0.002}$     & $0.991^{+0.005}_{-0.006}$ \\ 
EPL $\varphi$ [$^{\circ}$]            & $\mathcal{U}(-180, 180)$           & $-70.113^{+4.199}_{-3.662}$ & $-6.226^{+36.628}_{-29.394}$ & $-86.355^{+8.415}_{-2.213}$ & $75.869^{+6.221}_{-5.439}$ & $-79.945^{+10.322}_{-4.532}$ & $-88.807^{+15.280}_{-10.115}$ \\ 
EPL $x$ [$\arcsec$]                          & $\mathcal{U}(-0.1, 0.1)$           & $0.031^{+0.002}_{-0.002}$   & $0.048^{+0.003}_{-0.002}$   & $0.022^{+0.001}_{-0.001}$ & $0.030^{+0.001}_{-0.002}$ & $0.018^{+0.002}_{-0.001}$       & $0.023^{+0.002}_{-0.001}$ \\ 
EPL $y$ [$\arcsec$]                          & $\mathcal{U}(-0.1, 0.1)$           & $0.047^{+0.003}_{-0.002}$   & $0.039^{+0.003}_{-0.003}$   & $0.065^{+0.001}_{-0.001}$ & $0.043^{+0.002}_{-0.002}$ & $0.059^{+0.001}_{-0.002}$       & $0.048^{+0.002}_{-0.002}$ \\  \hline
Shear $\Gamma$                        & $\mathcal{U}(0.0, 0.2)$            & $0.107^{+0.003}_{-0.006}$   & $0.135^{+0.005}_{-0.004}$   & $0.086^{+0.011}_{-0.002}$ & $0.115^{+0.003}_{-0.003}$ & $0.091^{+0.004}_{-0.002}$       & $0.092^{+0.002}_{-0.002}$ \\ 
Shear $\varphi_{\Gamma}$ [$^{\circ}$] & $\mathcal{U}(-180, 180)$           & $-23.243^{+0.693}_{-2.207}$ & $-21.938^{+0.647}_{-0.540}$ & $-22.848^{+0.894}_{-0.749}$ & $-22.656^{+0.459}_{-0.352}$ & $-24.692^{+0.389}_{-0.405}$ & $-22.934^{+0.442}_{-0.396}$ \\  \hline
tNFW $\log_{10}(M_{200}/M_{\odot})$          & $\ln\mathcal{U}(7.0, 13.0)$ & -                           & $10.834^{+1.253}_{-0.562}$  & -                           & $9.954^{+0.582}_{-0.107}$ & -                             & $10.302^{+1.197}_{-0.573}$ \\ 
tNFW $\log_{10}(M_\mathrm{sub}/M_{\odot})$   & -                           & -                           & $10.033^{+0.413}_{-0.262}$  & -                           & $9.797^{+0.225}_{-0.278}$ & -                             & $9.249^{+0.373}_{-0.118}$ \\ 
tNFW $\log_{10}c$                     & $\ln\mathcal{U}(-4.0, 4.0)$        & -                           & $1.979^{+0.329}_{-0.291}$   & -                           & $2.574^{+0.336}_{-0.479}$ & -                            & $2.409^{+0.522}_{-0.299}$ \\ 
tNFW $\log_{10}(r_{t}/\arcsec)$                 & $\ln\mathcal{U}(-4.0, 4.0)$        & -                           & $-0.446^{+1.351}_{-0.353}$  & -                           & $0.719^{+0.865}_{-1.358}$ & -                            & $-0.891^{+0.558}_{-0.420}$ \\ 
tNFW $x$ [$\arcsec$]                         & $\mathcal{U}(-1.2, -0.3)$          & -                           & $-0.672^{+0.067}_{-0.042}$  & -                           & $-0.689^{+0.032}_{-0.015}$ & -                            & $-0.685^{+0.044}_{-0.017}$ \\ 
tNFW $y$ [$\arcsec$]                         & $\mathcal{U}(0.4, 1.3)$            & -                           & $1.107^{+0.031}_{-0.020}$   & -                           & $0.942^{+0.042}_{-0.045}$ & -                             & $0.933^{+0.029}_{-0.018}$ \\  \hline
SIS ($s1$) $\vartheta_{E}$ [$\arcsec$]       & $\mathcal{U}(0.0, 1.0)$            & - & - & - & - & $0.133^{+0.007}_{-0.008}$     & $0.126^{+0.009}_{-0.007}$ \\ 
SIS ($s1$) $\Delta x$ [$\arcsec$]                   & $\mathcal{G}(0.0, 0.1)$            & - & - & - & - & $0.020^{+0.006}_{-0.006}$    & $0.010^{+0.015}_{-0.009}$ \\ 
SIS ($s1$) $\Delta y$ [$\arcsec$]                   & $\mathcal{G}(0.0, 0.1)$            & - & - & - & - & $-0.017^{+0.008}_{-0.008}$    & $-0.027^{+0.009}_{-0.009}$ \\ \hline
SIS ($s2$) $\vartheta_{E}$ [$\arcsec$]       & $\mathcal{U}(0.0, 1.0)$            & - & - & - & - & $0.045^{+0.058}_{-0.032}$     & $0.058^{+0.071}_{-0.041}$ \\ \hline
$s3$ $x$ [$\arcsec$]                         & $\mathcal{U}(-0.4, 1.0)$           & - & - & - & - & $0.568^{+0.028}_{-0.048}$     & $0.573^{+0.035}_{-0.059}$ \\
$s3$ $y$ [$\arcsec$]                         & $\mathcal{U}(-1.0, 0.0)$           & - & - & - & - & $-0.431^{+0.031}_{-0.019}$    & $-0.437^{+0.037}_{-0.023}$ \\ \hline
\end{tabular}
}
\end{table*}
\begin{table*}
\caption{Likewise to Table \ref{tab:posteriors_grad}, for results obtained using curvature--regularised source reconstructions.}
\label{tab:posteriors_curv}
\centering
\scalebox{0.85}{
\begin{tabular}{|l|l|l|l|l|l|l|l|}
\hline
\textbf{Parameter} & \textbf{Prior} & \multicolumn{6}{c}{\textbf{Posterior (median with $1\sigma$ uncertainties)}} \\ 
                   &                & \multicolumn{2}{c}{Single plane (I--band)} & \multicolumn{2}{c}{Single plane (I-- \& U--band)} & \multicolumn{2}{c}{Triple plane} \\  \hline         
EPL $\vartheta_{E}$ [$\arcsec$]              & $\mathcal{U}(0.9, 1.5)$           & $1.398^{+0.002}_{-0.002}$      & $1.383^{+0.004}_{-0.005}$   & $1.396^{+0.001}_{-0.001}$   & $1.389^{+0.002}_{-0.002}$   & $1.399^{+0.001}_{-0.001}$   & $1.395^{+0.001}_{-0.001}$   \\ 
EPL $\gamma$                          & $\mathcal{U}(1.5, 2.5)$           & $2.186^{+0.076}_{-0.074}$      & $2.395^{+0.062}_{-0.068}$   & $2.187^{+0.039}_{-0.037}$   & $2.260^{+0.043}_{-0.038}$   & $1.944^{+0.014}_{-0.013}$   & $2.003^{+0.020}_{-0.019}$   \\ 
EPL $q$                               & $\mathcal{U}(0.6, 1.0)$           & $0.922^{+0.020}_{-0.020}$      & $0.976^{+0.014}_{-0.014}$   & $0.927^{+0.007}_{-0.008}$   & $0.972^{+0.007}_{-0.008}$   & $0.950^{+0.003}_{-0.003}$   & $0.969^{+0.004}_{-0.004}$   \\ 
EPL $\varphi$ [$^{\circ}$]            & $\mathcal{U}(-180, 180)$            & $-41.353^{+6.024}_{-5.037}$    & $10.691^{+19.254}_{-21.067}$ & $-45.327^{+2.866}_{-3.408}$ & $-59.580^{+7.706}_{-8.130}$ & $-54.094^{+2.284}_{-2.537}$ & $-68.097^{+4.926}_{-4.753}$ \\ 
EPL $x$ [$\arcsec$]                          & $\mathcal{U}(-0.1, 0.1)$          & $0.017^{+0.004}_{-0.003}$      & $0.044^{+0.004}_{-0.003}$   & $0.013^{+0.001}_{-0.001}$   & $0.027^{+0.002}_{-0.002}$   & $0.015^{+0.001}_{-0.001}$   & $0.025^{+0.002}_{-0.002}$   \\ 
EPL $y$ [$\arcsec$]                          & $\mathcal{U}(-0.1, 0.1)$          & $0.044^{+0.003}_{-0.003}$      & $0.033^{+0.004}_{-0.004}$   & $0.051^{+0.002}_{-0.002}$   & $0.050^{+0.002}_{-0.002}$   & $0.054^{+0.002}_{-0.002}$   & $0.054^{+0.002}_{-0.002}$   \\ \hline
Shear $\Gamma$                        & $\mathcal{U}(0.0, 0.2)$           & $0.131^{+0.010}_{-0.009}$      & $0.146^{+0.007}_{-0.007}$   & $0.129^{+0.004}_{-0.005}$   & $0.126^{+0.005}_{-0.005}$   & $0.100^{+0.002}_{-0.002}$   & $0.099^{+0.003}_{-0.003}$   \\ 
Shear $\varphi_{\Gamma}$ [$^{\circ}$] & $\mathcal{U}(-180, 180)$          & $-24.900^{+1.270}_{-0.924}$    & $-21.442^{+0.748}_{-0.693}$ & $-25.196^{+0.683}_{-0.562}$ & $-22.982^{+0.528}_{-0.607}$ & $-26.667^{+0.361}_{-0.365}$ & $-25.002^{+0.460}_{-0.469}$ \\ \hline
tNFW $\log_{10}(M_{200}/M_{\odot})$   & $\ln\mathcal{U}(7.0, 13.0)$       & -                              & $10.286^{+1.083}_{-0.228}$   & -                           & $9.849^{+1.672}_{-0.166}$   & -                           & $10.239^{+1.212}_{-0.531}$   \\ 
tNFW $\log_{10}(M_\mathrm{sub}/M_{\odot})$   & -                          & -                              & $10.034^{+0.321}_{-0.301}$   & -                           & $9.515^{+0.280}_{-0.281}$   & -                           & $9.128^{+0.388}_{-0.118}$   \\ 
tNFW $\log_{10}c$                     & $\ln\mathcal{U}(-4.0, 4.0)$           & -                              & $2.301^{+0.271}_{-0.405}$   & -                           & $2.821^{+0.340}_{-0.569}$   & -                           & $2.481^{+0.378}_{-0.296}$   \\ 
tNFW $\log_{10}(r_{t}/\arcsec)$                 & $\ln\mathcal{U}(-4.0, 4.0)$           & -                              & $0.572^{+0.963}_{-1.283}$   & -                           & $-0.397^{+1.641}_{-0.968}$   & -                           & $-1.013^{+0.509}_{-0.400}$   \\ 
tNFW $x$ [$\arcsec$]                         & $\mathcal{U}(-1.2, -0.3)$         & -                              & $-0.643^{+0.039}_{-0.061}$  & -                           & $-0.619^{+0.015}_{-0.022}$  & -                           & $-0.647^{+0.029}_{-0.018}$  \\ 
tNFW $y$ [$\arcsec$]                         & $\mathcal{U}(0.4, 1.3)$           & -                              & $1.090^{+0.024}_{-0.042}$   & -                           & $1.012^{+0.011}_{-0.018}$   & -                           & $0.995^{+0.019}_{-0.026}$   \\ \hline
SIS ($s1$) $\vartheta_{E}$ [$\arcsec$]       & $\mathcal{U}(0.0, 1.0)$           & -                              & -                           & -                           & -                           & $0.116^{+0.011}_{-0.011}$   & $0.163^{+0.016}_{-0.015}$   \\ 
SIS ($s1$) $\Delta x$ [$\arcsec$]                   & $\mathcal{G}(0.0, 0.1)$           & -                              & -                           & -                           & -                           & $-0.002^{+0.010}_{-0.009}$  & $-0.017^{+0.009}_{-0.008}$  \\ 
SIS ($s1$) $\Delta y$ [$\arcsec$]                   & $\mathcal{G}(0.0, 0.1)$           & -                              & -                           & -                           & -                           & $-0.023^{+0.009}_{-0.009}$  & $-0.023^{+0.009}_{-0.008}$  \\ \hline
SIS ($s2$) $\vartheta_{E}$ [$\arcsec$]       & $\mathcal{U}(0.0, 1.0)$           & -                              & -                           & -                           & -                           & $0.044^{+0.062}_{-0.031}$   & $0.051^{+0.064}_{-0.036}$   \\ \hline
$s3$ $x$ [$\arcsec$]                         & $\mathcal{U}(-0.4, 1.0)$          & -                              & -                           & -                           & -                           & $0.563^{+0.028}_{-0.051}$   & $0.580^{+0.031}_{-0.051}$   \\
$s3$ $y$ [$\arcsec$]                         & $\mathcal{U}(-1.0, 0.0)$          & -                              & -                           & -                           & -                           & $-0.421^{+0.032}_{-0.020}$  & $-0.446^{+0.034}_{-0.020}$  \\ \hline
\end{tabular}
}
\end{table*}
\begin{table*}
\caption{All lens model results obtained from our systematic tests. Note that the SIE profile on $s1$ is fixed to $q=1$, $\varphi=0$ for models that do not explore ellipticity in the mass distribution of this galaxy.}
\label{tab:posteriors_systematics}
\centering
\scalebox{0.85}{
\begin{tabular}{|l|l|l|l|l|l|l|}
\hline
\textbf{Parameter} & \textbf{Prior} & \multicolumn{5}{c}{\textbf{Posterior (median with $1\sigma$ uncertainties)}} \\ 
                    &                & \multicolumn{2}{c}{multipoles} & \multicolumn{2}{c}{SIE on $s1$} & \multicolumn{1}{c}{wandering BH} \\  \hline         
EPL $\vartheta_{E}$ [$\arcsec$]              & $\mathcal{U}(0.9, 1.5)$            & $1.399^{+0.001}_{-0.001}$   & $1.377^{+0.006}_{-0.008}$   & $1.398^{+0.002}_{-0.001}$   & $1.392^{+0.002}_{-0.002}$     & $1.393^{+0.002}_{-0.001}$ \\ 
EPL $\gamma$                          & $\mathcal{U}(1.5, 2.5)$            & $1.935^{+0.010}_{-0.008}$   & $1.947^{+0.012}_{-0.011}$   & $1.986^{+0.007}_{-0.023}$   & $1.982^{+0.018}_{-0.034}$     & $2.114^{+0.015}_{-0.031}$ \\ 
EPL $q$                               & $\mathcal{U}(0.6, 1.0)$            & $0.955^{+0.004}_{-0.004}$   & $0.964^{+0.010}_{-0.007}$   & $0.974^{+0.005}_{-0.003}$   & $0.978^{+0.018}_{-0.010}$     & $0.955^{+0.019}_{-0.003}$ \\ 
EPL $\varphi$ [$^{\circ}$]            & $\mathcal{U}(-180, 180)$           & $-81.974^{+2.121}_{-1.398}$ & $54.005^{+5.969}_{-3.557}$  & $70.230^{+24.882}_{-1.725}$ & $68.797^{+12.656}_{-37.621}$  & $-85.436^{+13.388}_{-1.669}$ \\ 
EPL $x$ [$\arcsec$]                          & $\mathcal{U}(-0.1, 0.1)$           & $0.010^{+0.006}_{-0.001}$   & $0.033^{+0.006}_{-0.007}$   & $0.028^{+0.001}_{-0.004}$   & $0.031^{+0.002}_{-0.002}$     & $0.020^{+0.001}_{-0.004}$ \\ 
EPL $y$ [$\arcsec$]                          & $\mathcal{U}(-0.1, 0.1)$           & $0.061^{+0.001}_{-0.002}$   & $0.039^{+0.006}_{-0.005}$   & $0.057^{+0.001}_{-0.001}$   & $0.051^{+0.006}_{-0.004}$     & $0.056^{+0.001}_{-0.003}$ \\  \hline
Shear $\Gamma$                        & $\mathcal{U}(0.0, 0.2)$            & $0.089^{+0.001}_{-0.001}$   & $0.087^{+0.001}_{-0.001}$   & $0.091^{+0.001}_{-0.001}$   & $0.092^{+0.001}_{-0.001}$     & $0.103^{+0.004}_{-0.001}$ \\ 
Shear $\varphi_{\Gamma}$ [$^{\circ}$] & $\mathcal{U}(-180, 180)$           & $-27.033^{+0.537}_{-0.548}$ & $-21.797^{+0.673}_{-0.931}$ & $-20.270^{+0.243}_{-1.672}$ & $-20.703^{+0.697}_{-0.792}$   & $-25.207^{+0.825}_{-0.334}$ \\ \hline
MP $k_{3}$                            & $\mathcal{U}(0.0, 0.2)$            & $0.020^{+0.001}_{-0.005}$   & $0.017^{+0.002}_{-0.002}$   & -   & -     & - \\ 
MP $\varphi_{3}$ [$^{\circ}$]         & $\mathcal{U}(-180, 180)$           & $26.523^{+0.838}_{-0.864}$  & $53.415^{+1.596}_{-2.188}$  & -   & -     & - \\ 
MP $k_{4}$                            & $\mathcal{U}(0.0, 0.2)$            & $0.008^{+0.005}_{-0.001}$   & $0.012^{+0.002}_{-0.002}$   & -   & -     & - \\ 
MP $\varphi_{4}$ [$^{\circ}$]         & $\mathcal{U}(-180, 180)$           & $21.343^{+2.859}_{-2.954}$  & $71.515^{+3.482}_{-3.214}$  & -   & -     & - \\ \hline
tNFW $\log_{10}(M_{200}/M_{\odot})$   & $\ln\mathcal{U}(7.0, 13.0)$        & -                           & $10.615^{+1.133}_{-0.405}$  & -   & $10.035^{+1.313}_{-0.517}$  & - \\ 
tNFW $\log_{10}(M_\mathrm{sub}/M_{\odot})$ & -                             & -                           & $9.916^{+0.364}_{-0.277}$   & -   & $9.199^{+0.350}_{-0.205}$   & - \\ 
tNFW $\log_{10}c$                     & $\ln\mathcal{U}(-4.0, 4.0)$        & -                           & $1.903^{+0.366}_{-0.331}$   & -   & $2.533^{+0.594}_{-0.404}$   & - \\ 
tNFW $\log_{10}(r_{t}/\arcsec)$                 & $\ln\mathcal{U}(-4.0, 4.0)$        & -                           & $-0.196^{+0.706}_{-0.469}$  & -   & $-0.812^{+0.870}_{-0.536}$  & - \\ 
tNFW $x$ [$\arcsec$]                         & $\mathcal{U}(-1.2, -0.3)$          & -                           & $-0.628^{+0.018}_{-0.013}$  & -   & $-0.641^{+0.067}_{-0.055}$  & - \\ 
tNFW $y$ [$\arcsec$]                         & $\mathcal{U}(0.4, 1.3)$            & -                           & $0.969^{+0.050}_{-0.059}$   & -   & $0.944^{+0.077}_{-0.051}$   & - \\ \hline
BH $\log_{10}(M_{\mathrm{BH}}/M_{\odot})$  & $\ln\mathcal{U}(7.0, 13.0)$   & -                           & -                           & -   & -     & $8.944^{+0.191}_{-0.080}$ \\ 
BH $x$ [$\arcsec$]                           & $\mathcal{U}(-1.2, -0.3)$          & -                           & -                           & -   & -     & $-1.111^{+0.012}_{-0.027}$ \\ 
BH $y$ [$\arcsec$]                           & $\mathcal{U}(0.4, 1.3)$            & -                           & -                           & -   & -     & $1.022^{+0.024}_{-0.017}$ \\ \hline
SIE ($s1$) $\vartheta_{E}$ [$\arcsec$]       & $\mathcal{U}(0.0, 1.0)$            & $0.109^{+0.007}_{-0.007}$   & $0.124^{+0.009}_{-0.008}$   & $0.152^{+0.005}_{-0.021}$   & $0.147^{+0.014}_{-0.026}$     & $0.265^{+0.011}_{-0.037}$ \\ 
SIE ($s1$) $q$                        & $\mathcal{U}(0.7, 1.0)$            & $1.000$                     & $1.000$                     & $0.710^{+0.011}_{-0.007}$   & $0.720^{+0.026}_{-0.014}$     & $1.000$ \\ 
SIE ($s1$) $\varphi$ [$^{\circ}$]     & $\mathcal{U}(-180, 180)$           & $0.000$                     & $0.000$                     & $37.493^{+2.182}_{-4.389}$  & $34.157^{+4.381}_{-3.738}$    & $0.000$ \\ 
SIE ($s1$) $\Delta x$ [$\arcsec$]            & $\mathcal{G}(0.0, 0.1)$            & $0.007^{+0.009}_{-0.011}$   & $0.005^{+0.006}_{-0.007}$   & $0.015^{+0.007}_{-0.005}$   & $0.021^{+0.008}_{-0.015}$     & $0.004^{+0.006}_{-0.005}$ \\ 
SIE ($s1$) $\Delta y$ [$\arcsec$]            & $\mathcal{G}(0.0, 0.1)$            & $-0.021^{+0.008}_{-0.008}$  & $-0.027^{+0.008}_{-0.008}$  & $-0.000^{+0.006}_{-0.014}$  & $-0.015^{+0.007}_{-0.007}$    & $-0.007^{+0.006}_{-0.007}$ \\ \hline
SIS ($s2$) $\vartheta_{E}$ [$\arcsec$]       & $\mathcal{U}(0.0, 1.0)$            & $0.053^{+0.069}_{-0.037}$   & $0.062^{+0.077}_{-0.044}$   & $0.057^{+0.069}_{-0.040}$   & $0.063^{+0.075}_{-0.044}$     & $0.140^{+0.121}_{-0.096}$ \\ \hline
$s3$ $x$ [$\arcsec$]                         & $\mathcal{U}(-0.4, 1.0)$           & $0.560^{+0.032}_{-0.058}$   & $0.577^{+0.037}_{-0.065}$   & $0.577^{+0.034}_{-0.059}$   & $0.584^{+0.039}_{-0.064}$     & $0.574^{+0.082}_{-0.104}$ \\ 
$s3$ $y$ [$\arcsec$]                         & $\mathcal{U}(-1.0, 0.0)$           & $-0.412^{+0.036}_{-0.021}$  & $-0.441^{+0.041}_{-0.025}$  & $-0.442^{+0.039}_{-0.024}$  & $-0.443^{+0.042}_{-0.027}$    & $-0.441^{+0.065}_{-0.052}$ \\ \hline
\end{tabular}
}
\end{table*}
\section{Macro--Model Posteriors}
\label{sec:macromodelposteriors}
\begin{figure*} \centering
\begin{tikzpicture}[      
        every node/.style={anchor=south west,inner sep=0pt},
        x=1mm, y=1mm,
      ]   
     \node (fig1) at (0,0)
       {\includegraphics[scale=0.175]{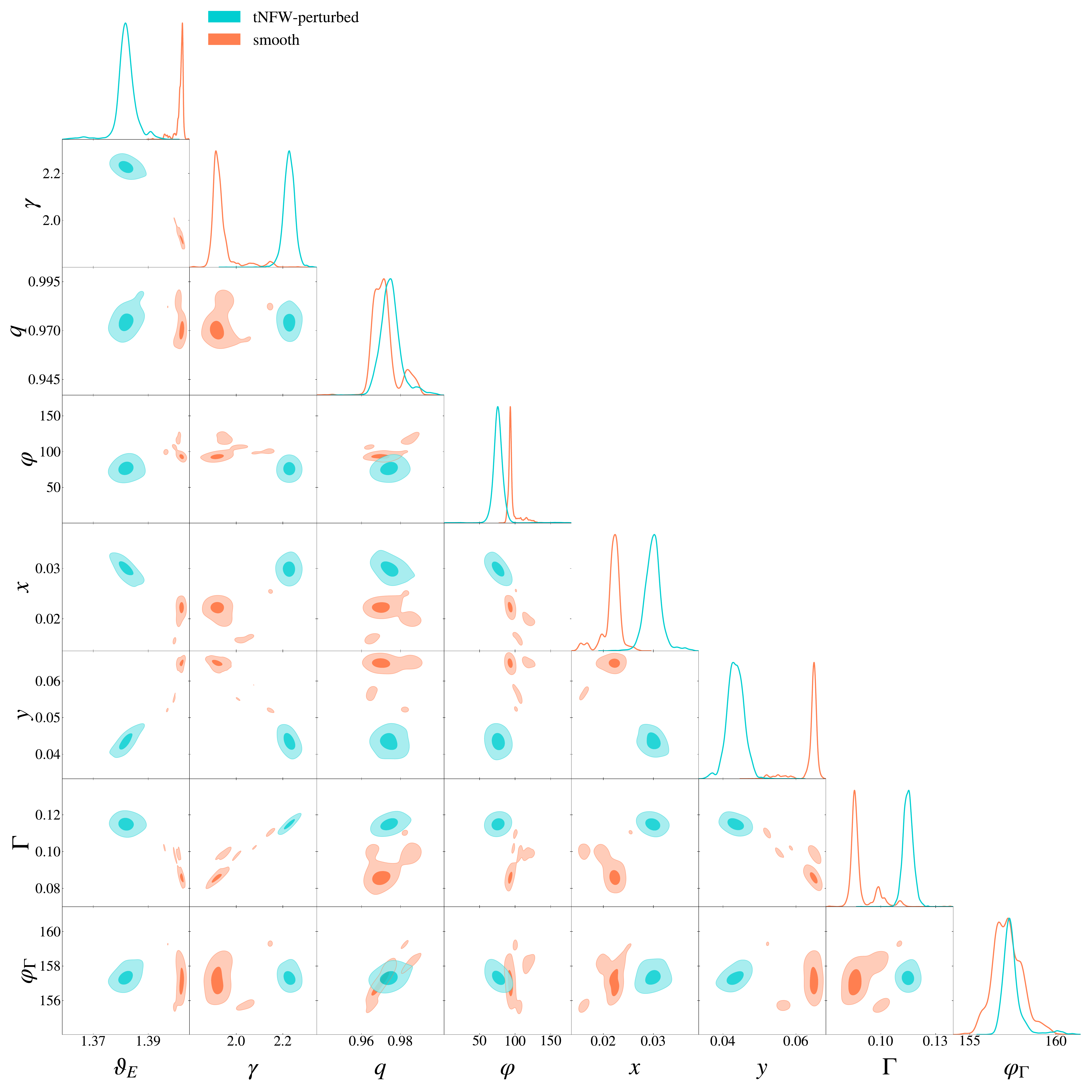}};
    \node (fig2) at (90,137.5)
       {\includegraphics[scale=0.275]{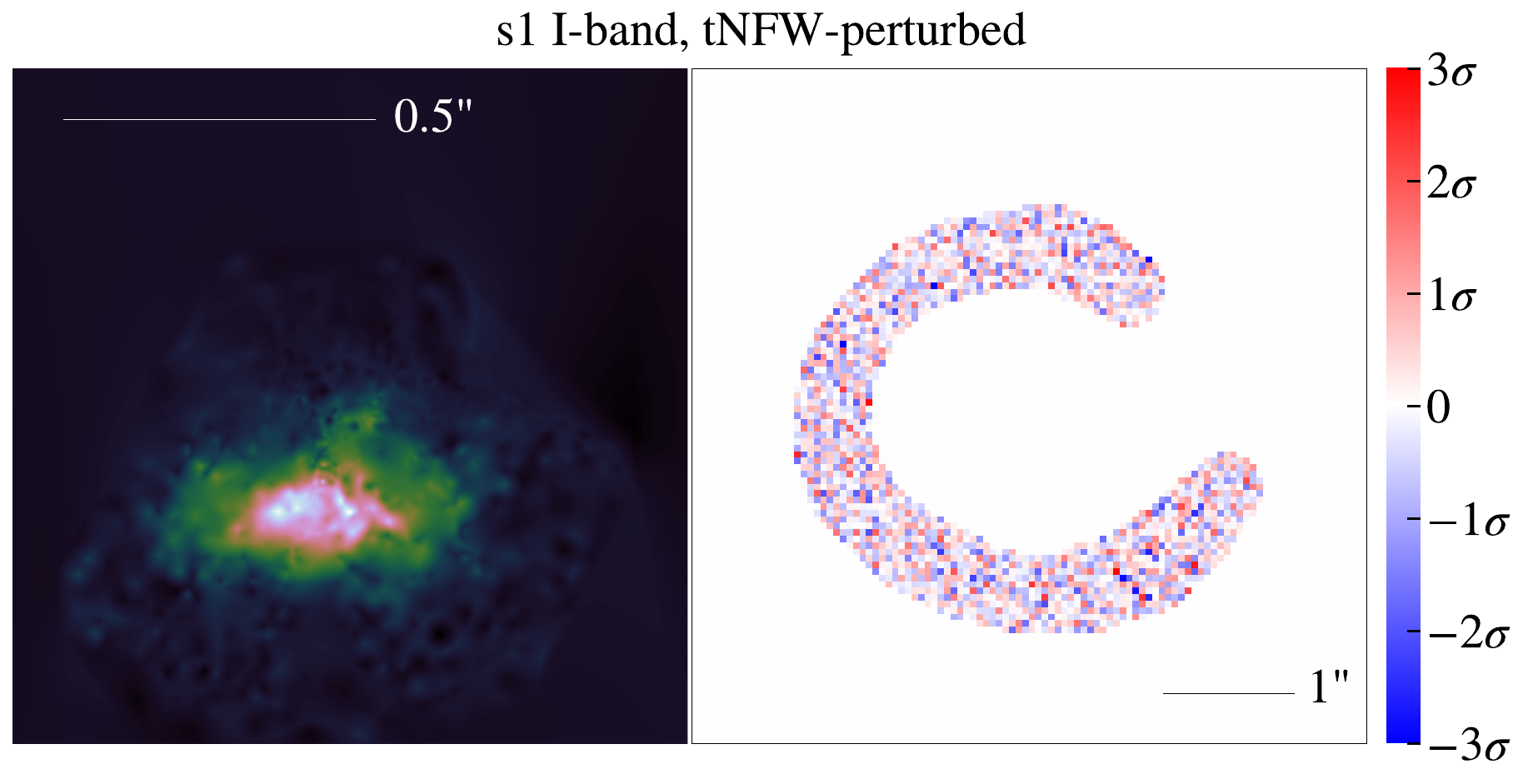}};
    \node (fig3) at (90,94)
       {\includegraphics[scale=0.275]{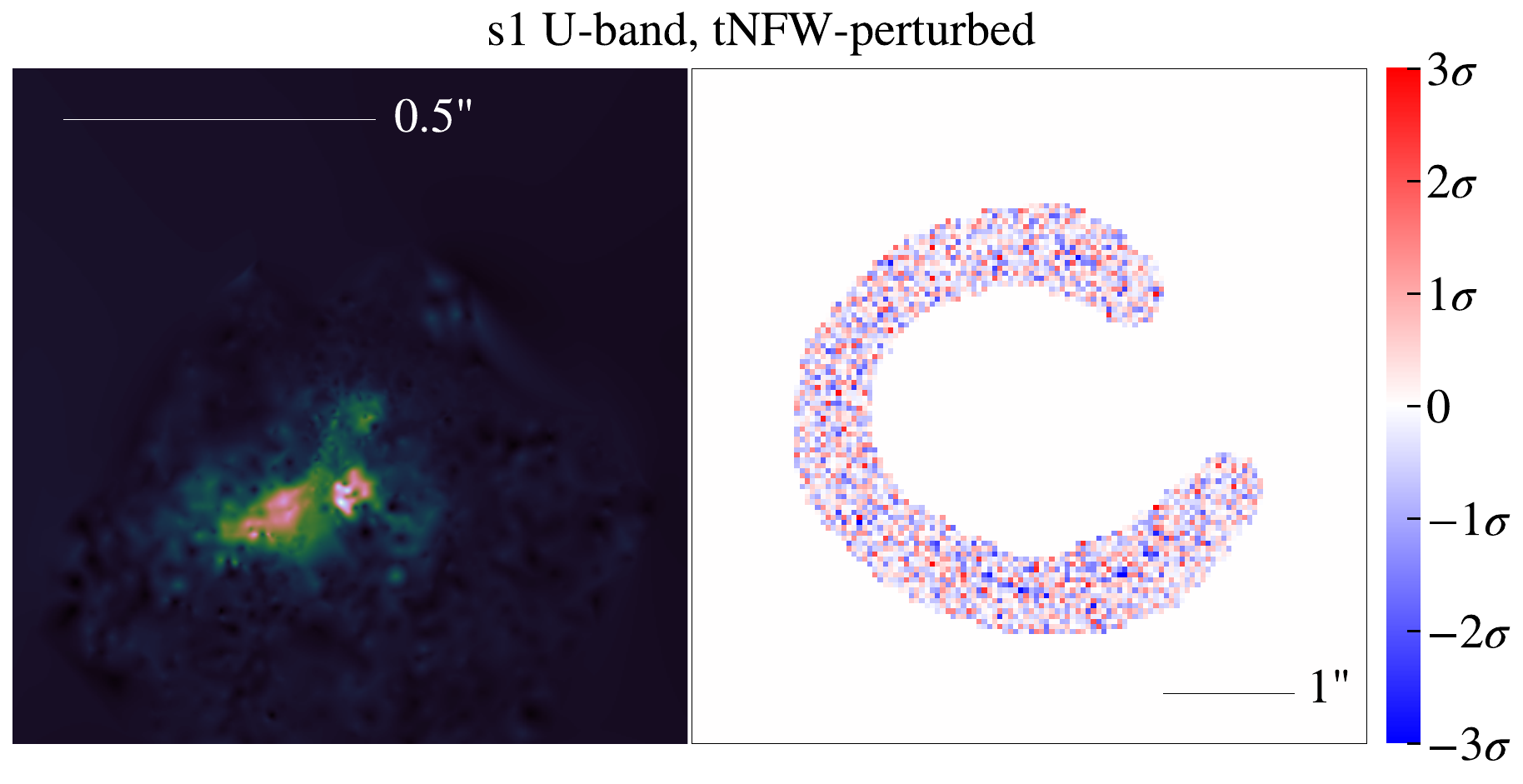}};
\end{tikzpicture}
\caption{Equivalent to Figure \ref{fig:1band_corner}, for our single source plane I-- and U--band models. The source reconstruction and image plane residual panels correspond to the tNFW--perturbed models, in (top) the I--band and (bottom) the U--band.}
\label{fig:2band_corner}
\end{figure*}

\begin{figure*}
    \centering
    \includegraphics[width=0.95\textwidth]{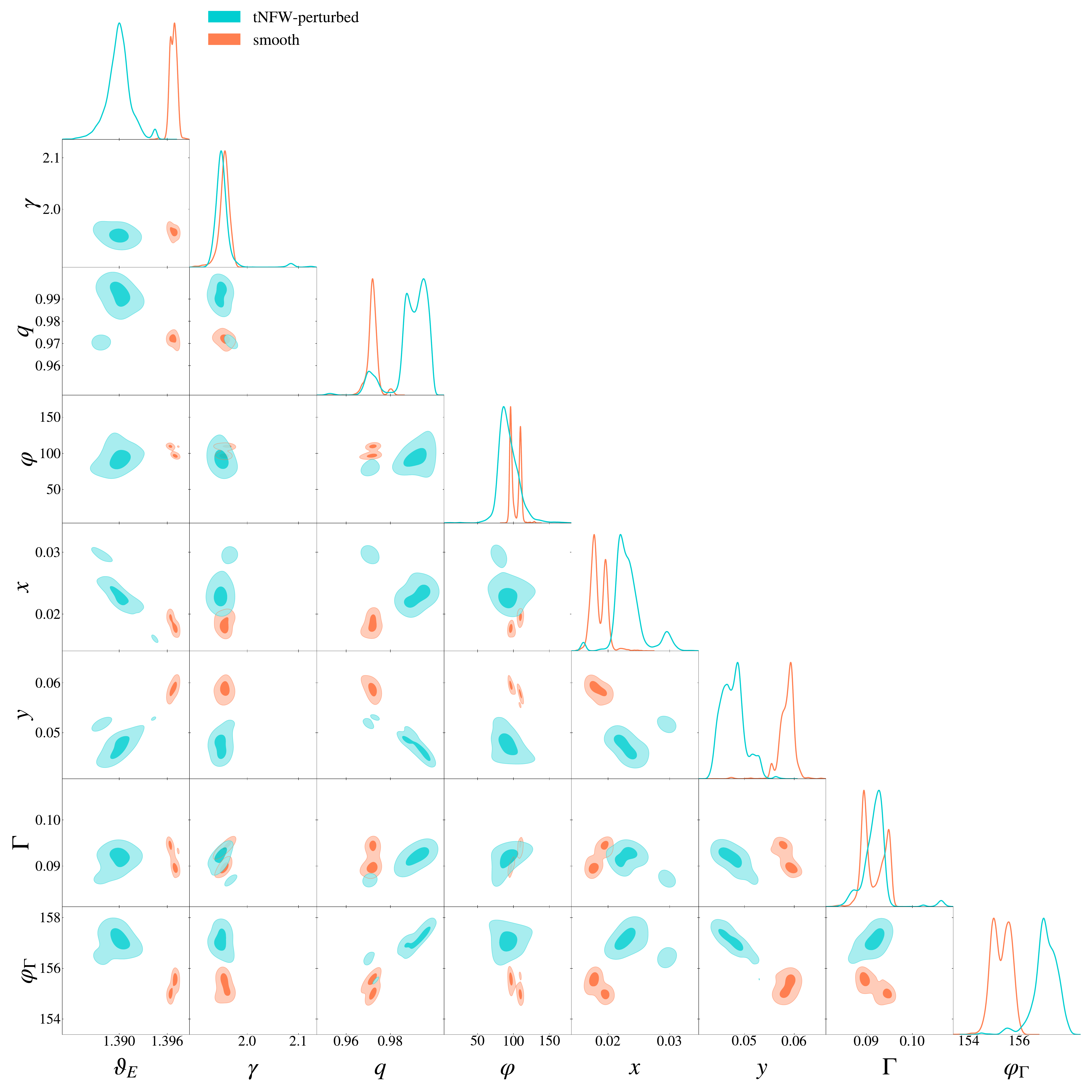}
    \caption{Equivalent to figures \ref{fig:1band_corner} and \ref{fig:2band_corner}, for our triple source plane I-- and U--band models. Source reconstructions and image plane residuals for the smooth and tNFW--perturbed cases are shown in Figure \ref{fig:sources_residuals_triple_plane}.}
    \label{fig:3source_corner}
\end{figure*}

Here we present how the macro model posterior distributions behave for the smooth and tNFW--perturbed cases of our simultaneous I-- and U--band single plane model (Figure \ref{fig:2band_corner}) and our fiducial triple plane model (Figure \ref{fig:3source_corner}), similarly to Figure \ref{fig:1band_corner}. 
\section{Sources and residuals from systematic tests}
\label{sec:sources_and_resiuals}
\begin{figure*} \centering
\begin{tikzpicture}[      
        every node/.style={anchor=south west,inner sep=0pt},
        x=1mm, y=1mm,
      ]   
     \node (fig1) at (-115,0)
       {\includegraphics[scale=0.28]{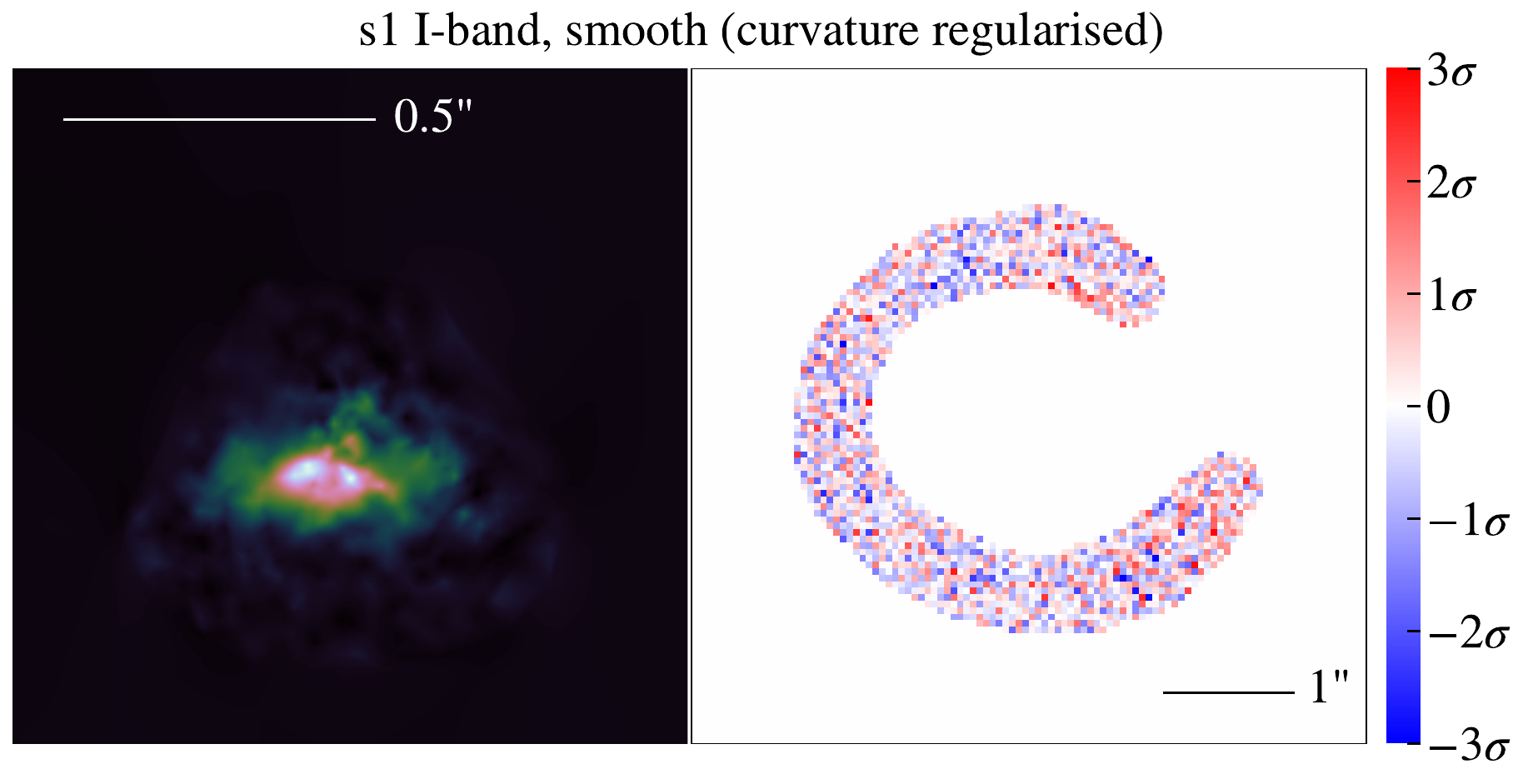}};
     \node (fig3) at (-25,0)
       {\includegraphics[scale=0.28]{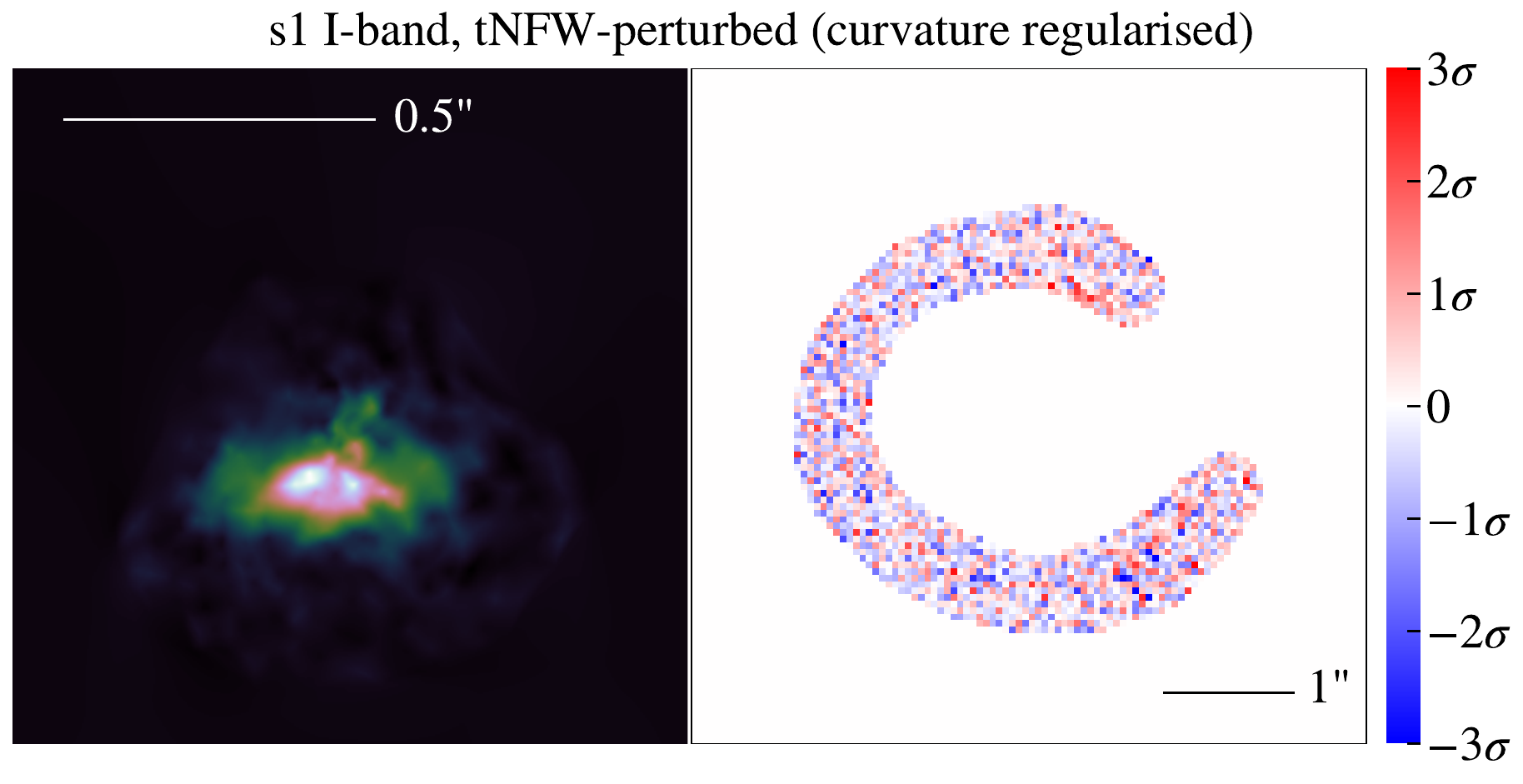}};
     \node (fig5) at (-115,-45)
       {\includegraphics[scale=0.28]{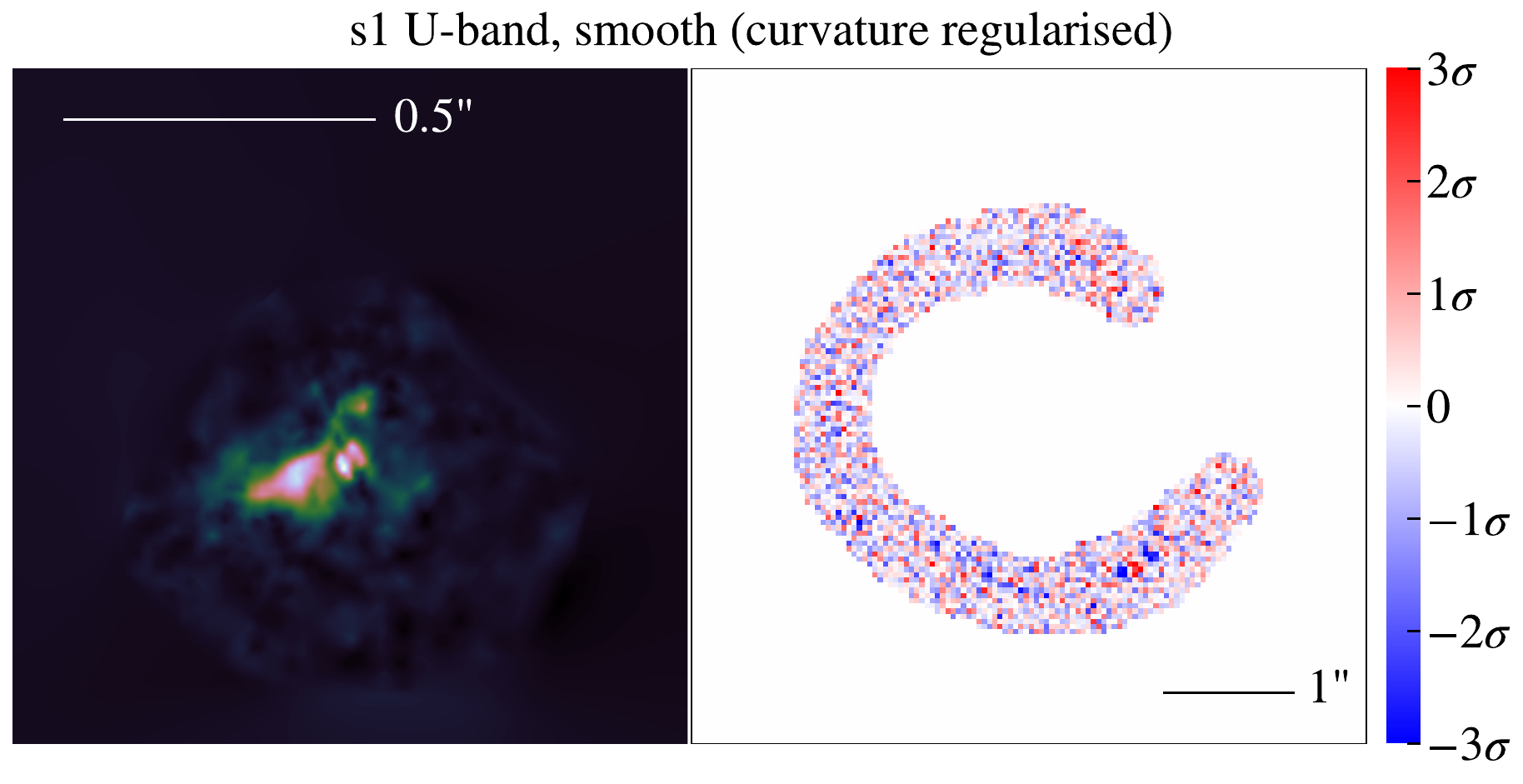}};
     \node (fig7) at (-25,-45)
       {\includegraphics[scale=0.28]{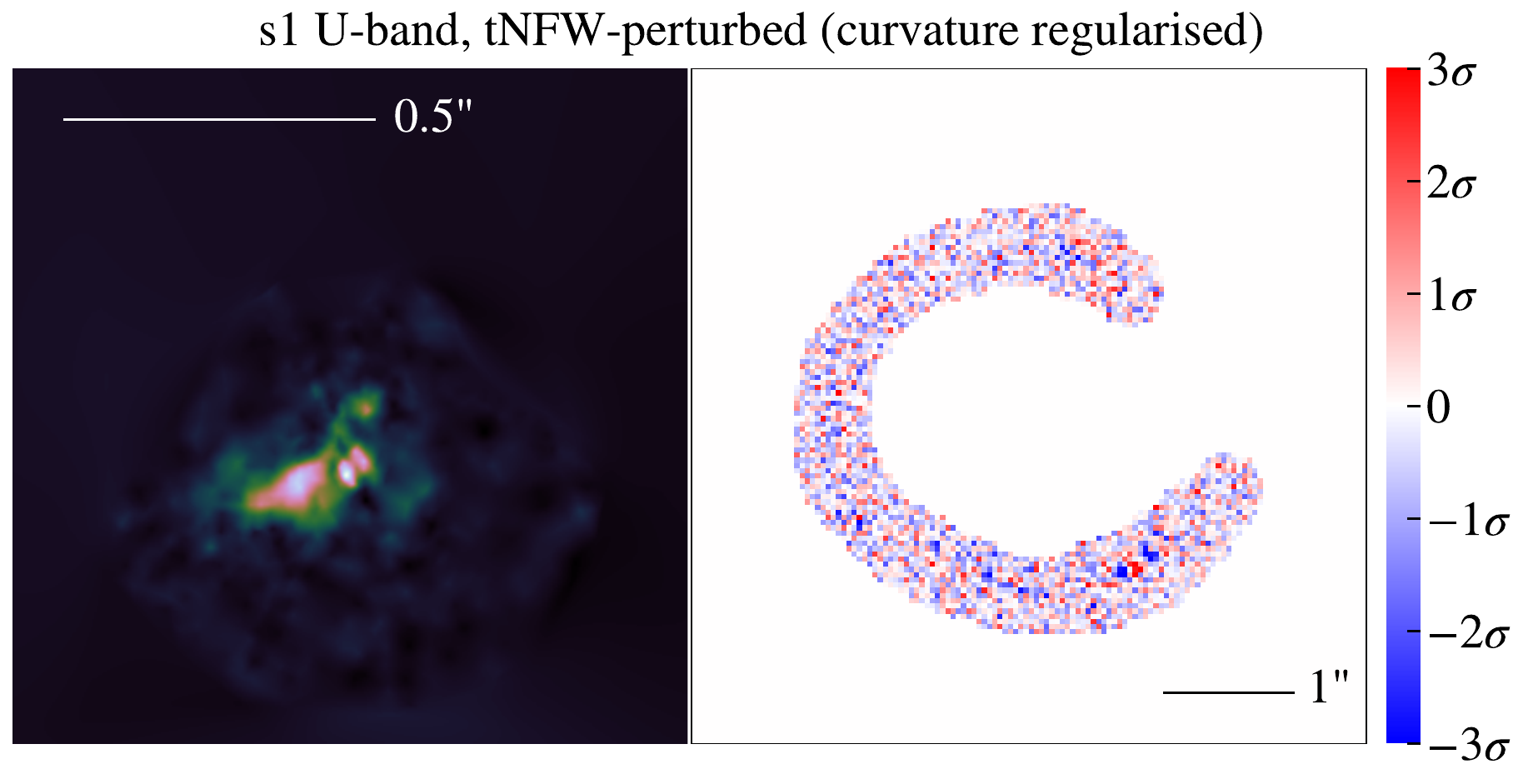}};
     \node (fig9) at (-115,-90)
       {\includegraphics[scale=0.28]{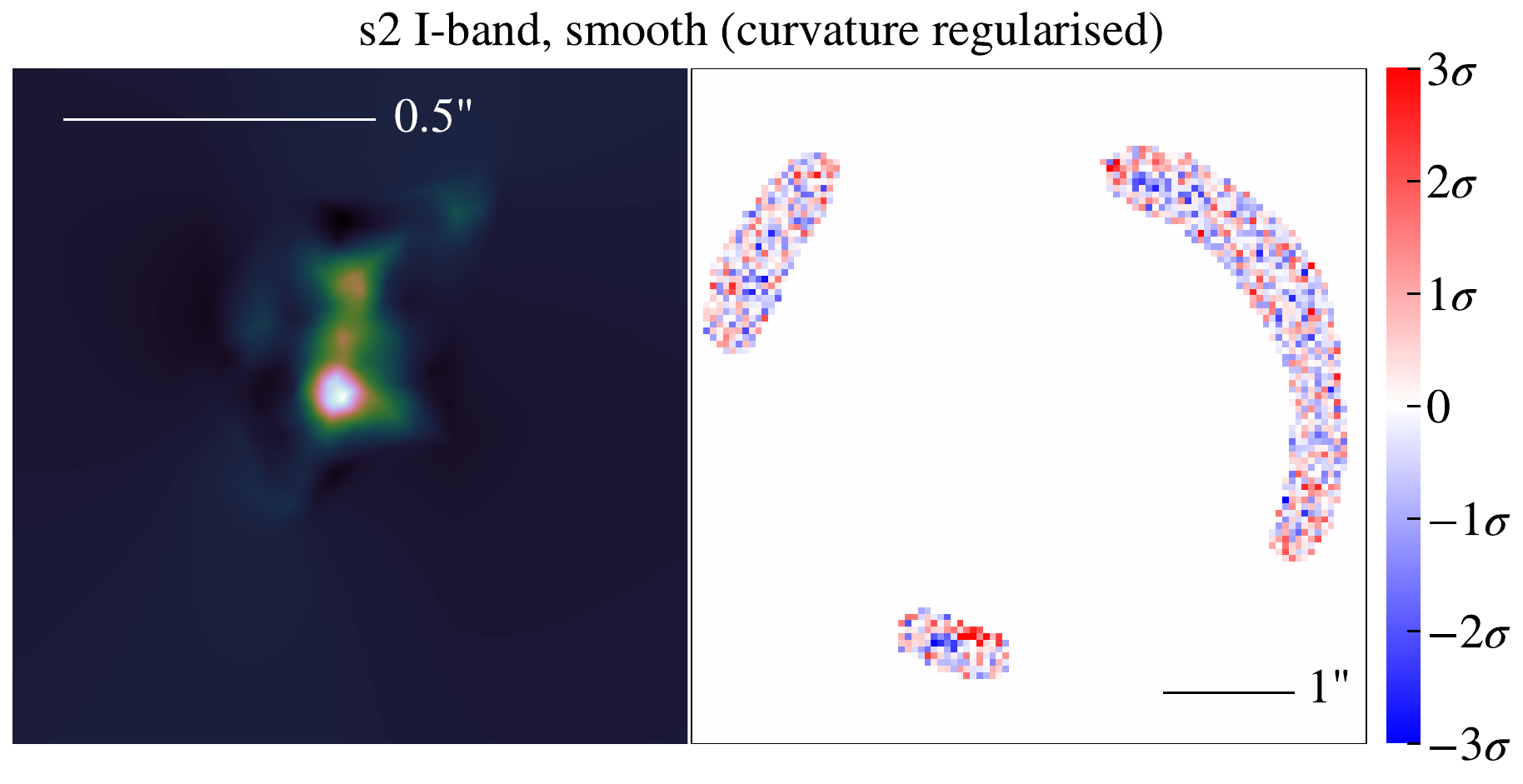}};
     \node (fig11) at (-25,-90)
       {\includegraphics[scale=0.28]{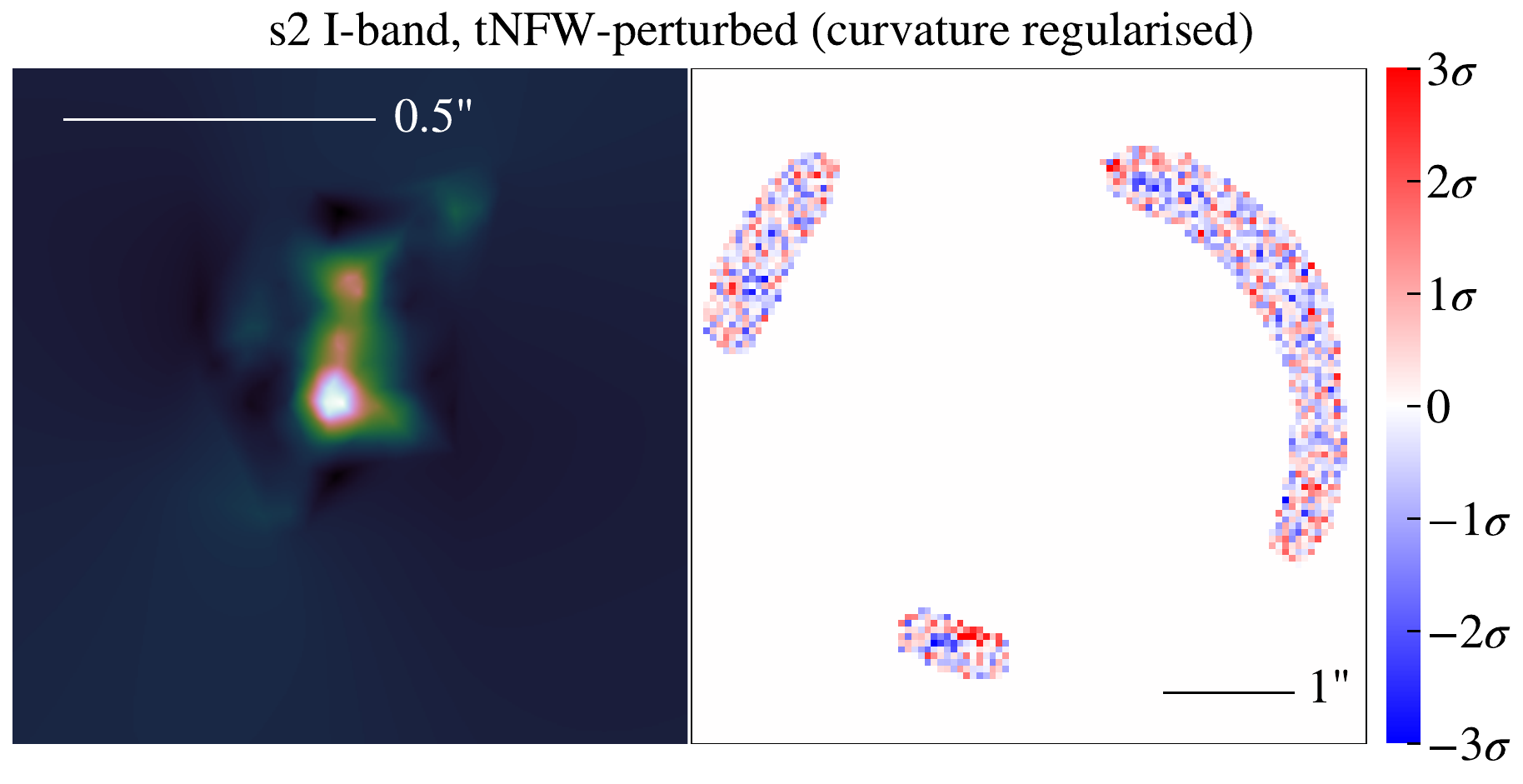}};
     \node (fig13) at (-115,-135)
       {\includegraphics[scale=0.28]{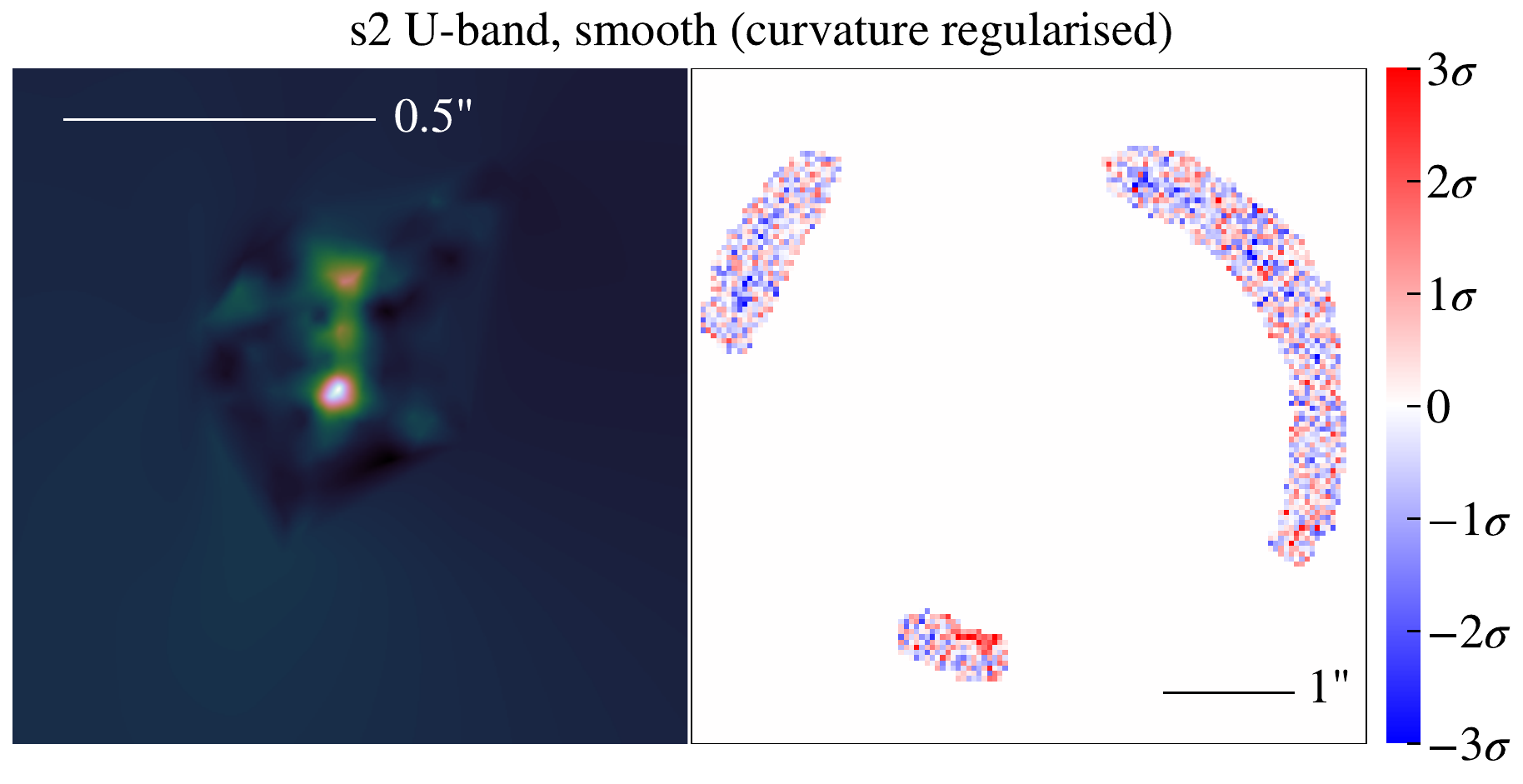}};
     \node (fig15) at (-25,-135)
       {\includegraphics[scale=0.28]{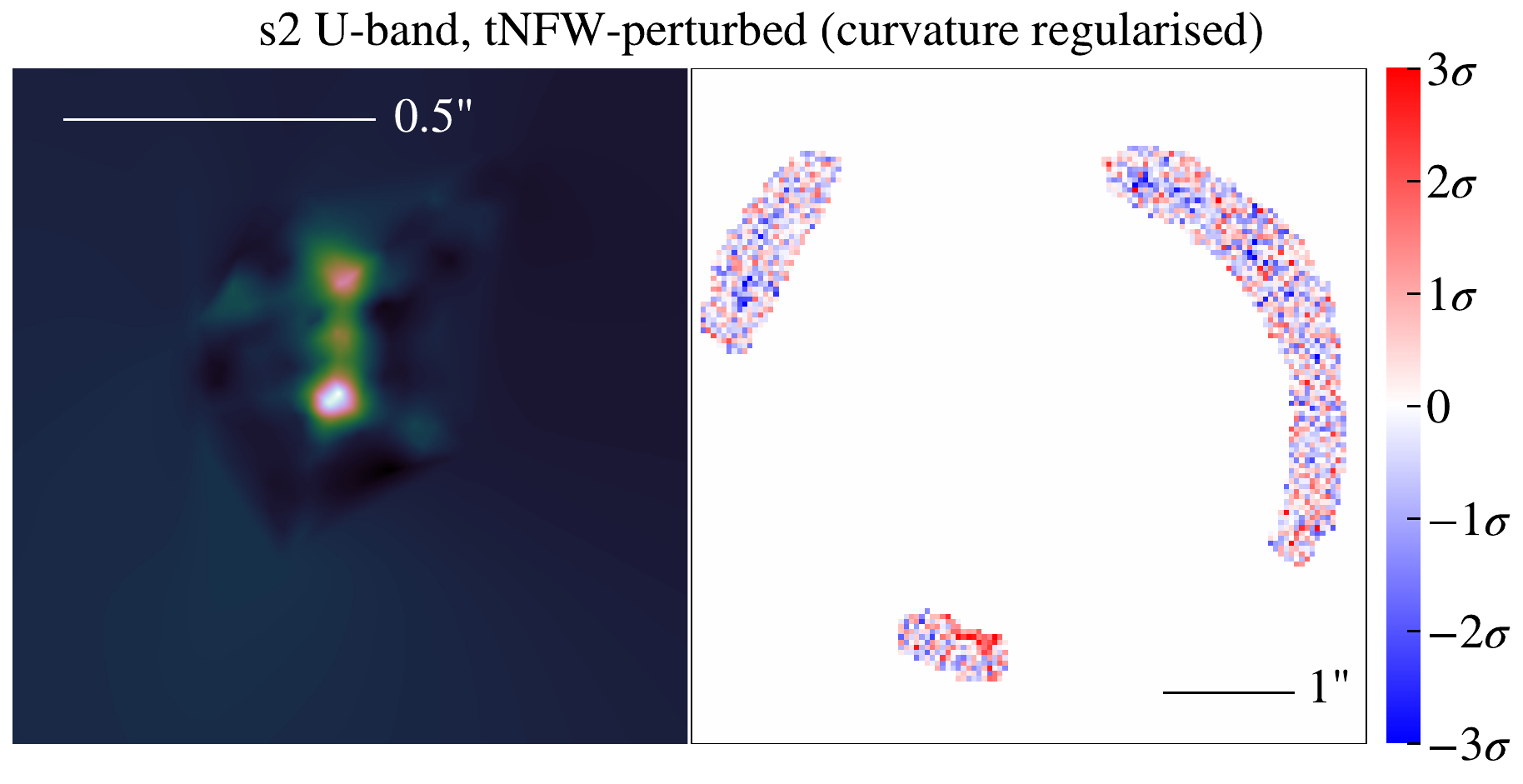}};
\end{tikzpicture}
\caption{Source plane reconstructions and normalised image plane residuals for our best fit smooth (left) and tNFW--perturbed (right) model, for (from top to bottom) $s1$ in I--band, $s1$ in U--band, $s2$ in I--band and $s2$ in U--band, where all sources are modelled with curvature regularisation.}
\label{fig:sources_residuals_triple_plane_curv}
\end{figure*}

\begin{figure*} \centering
\begin{tikzpicture}[      
        every node/.style={anchor=south west,inner sep=0pt},
        x=1mm, y=1mm,
      ]   
     \node (fig1) at (-115,0)
       {\includegraphics[scale=0.28]{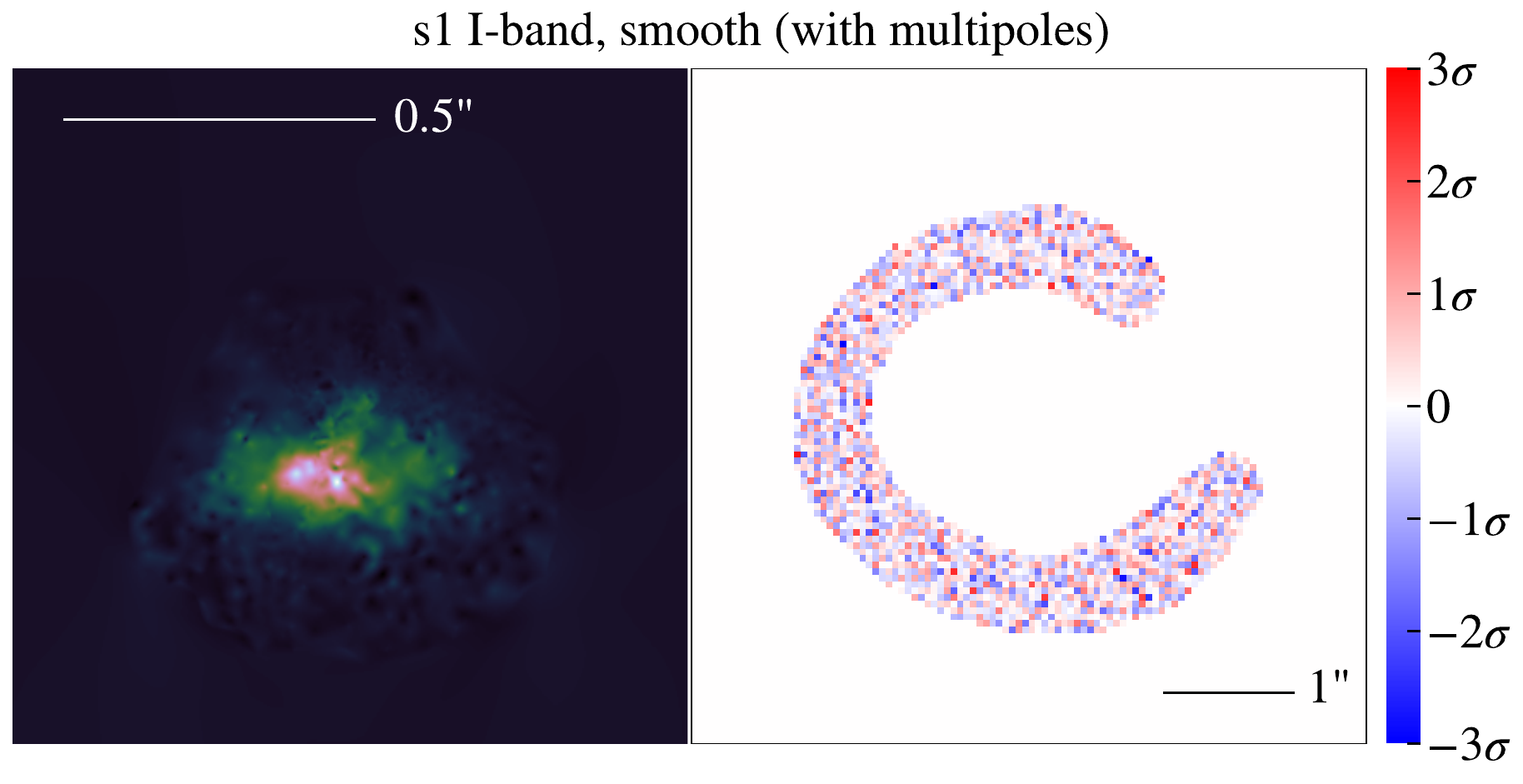}};
     \node (fig3) at (-25,0)
       {\includegraphics[scale=0.28]{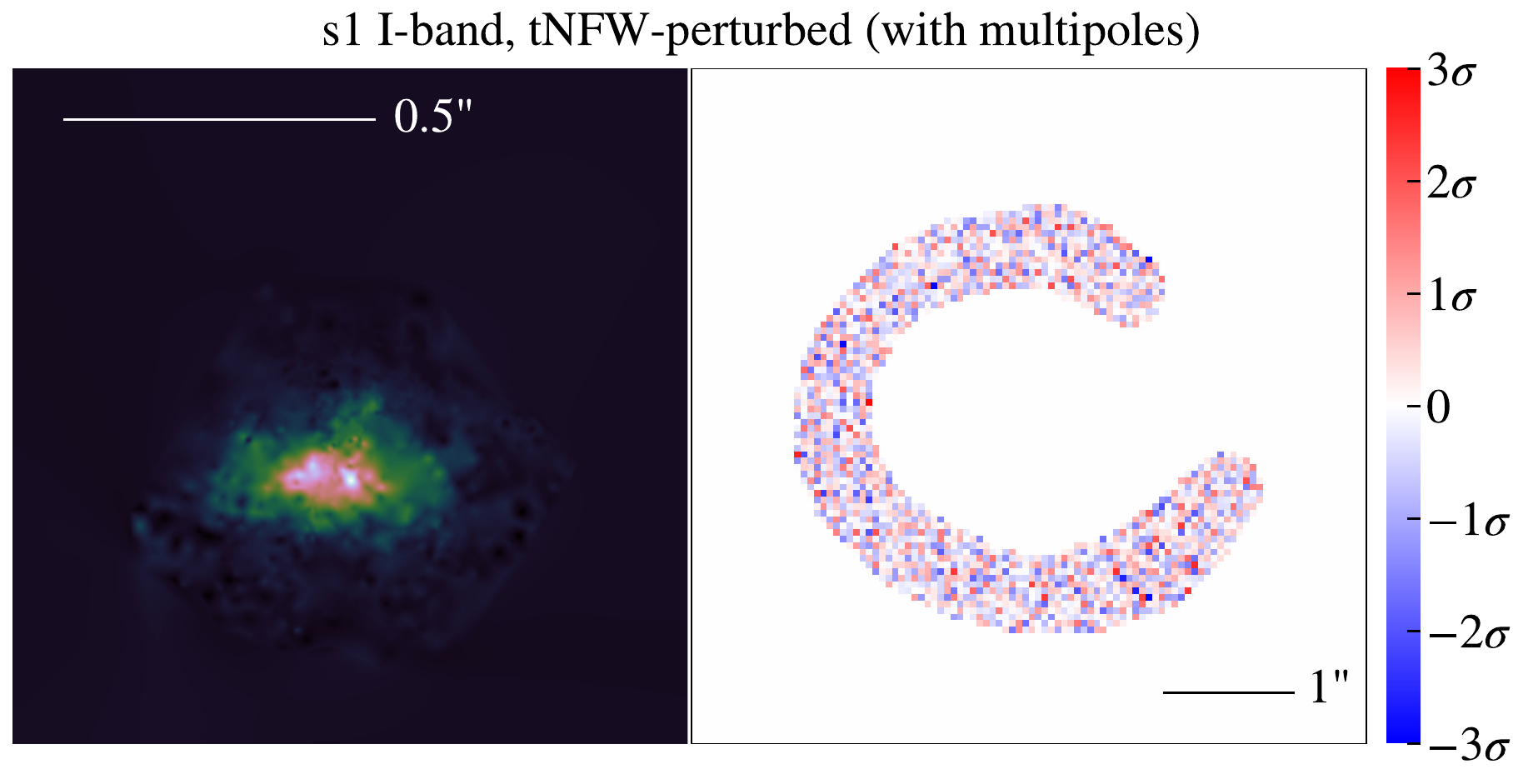}};
     \node (fig5) at (-115,-45)
       {\includegraphics[scale=0.28]{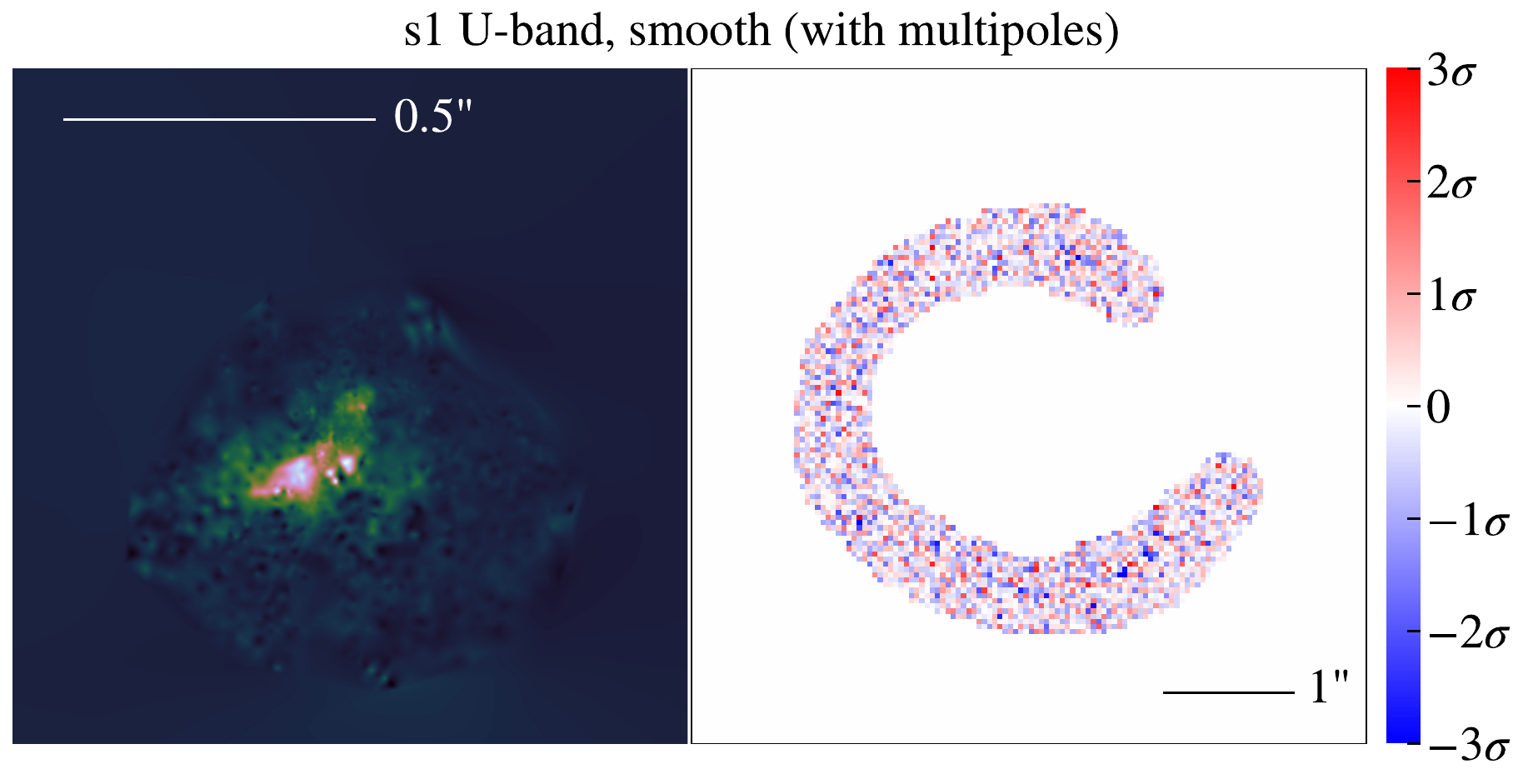}};
     \node (fig7) at (-25,-45)
       {\includegraphics[scale=0.28]{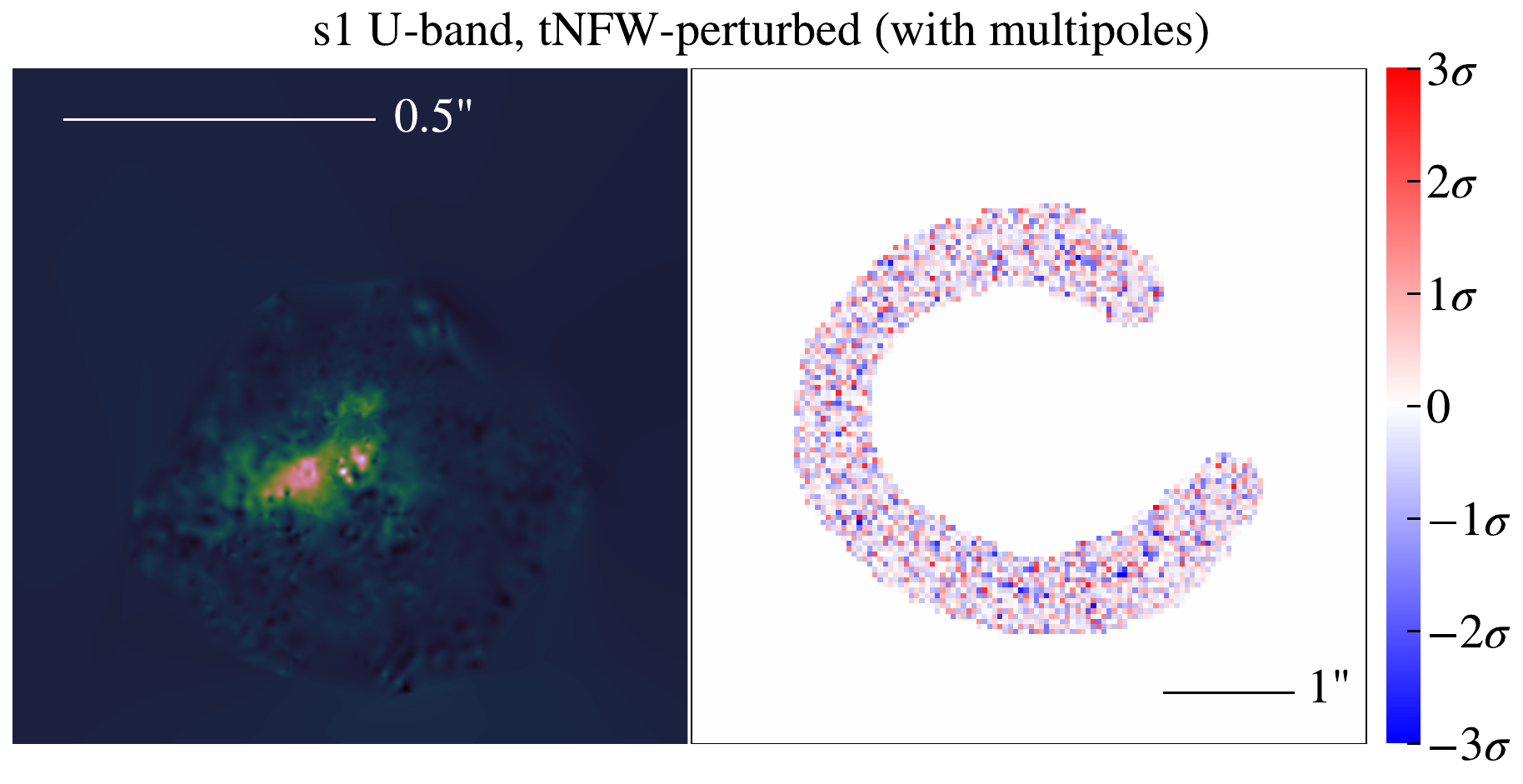}};
     \node (fig9) at (-115,-90)
       {\includegraphics[scale=0.28]{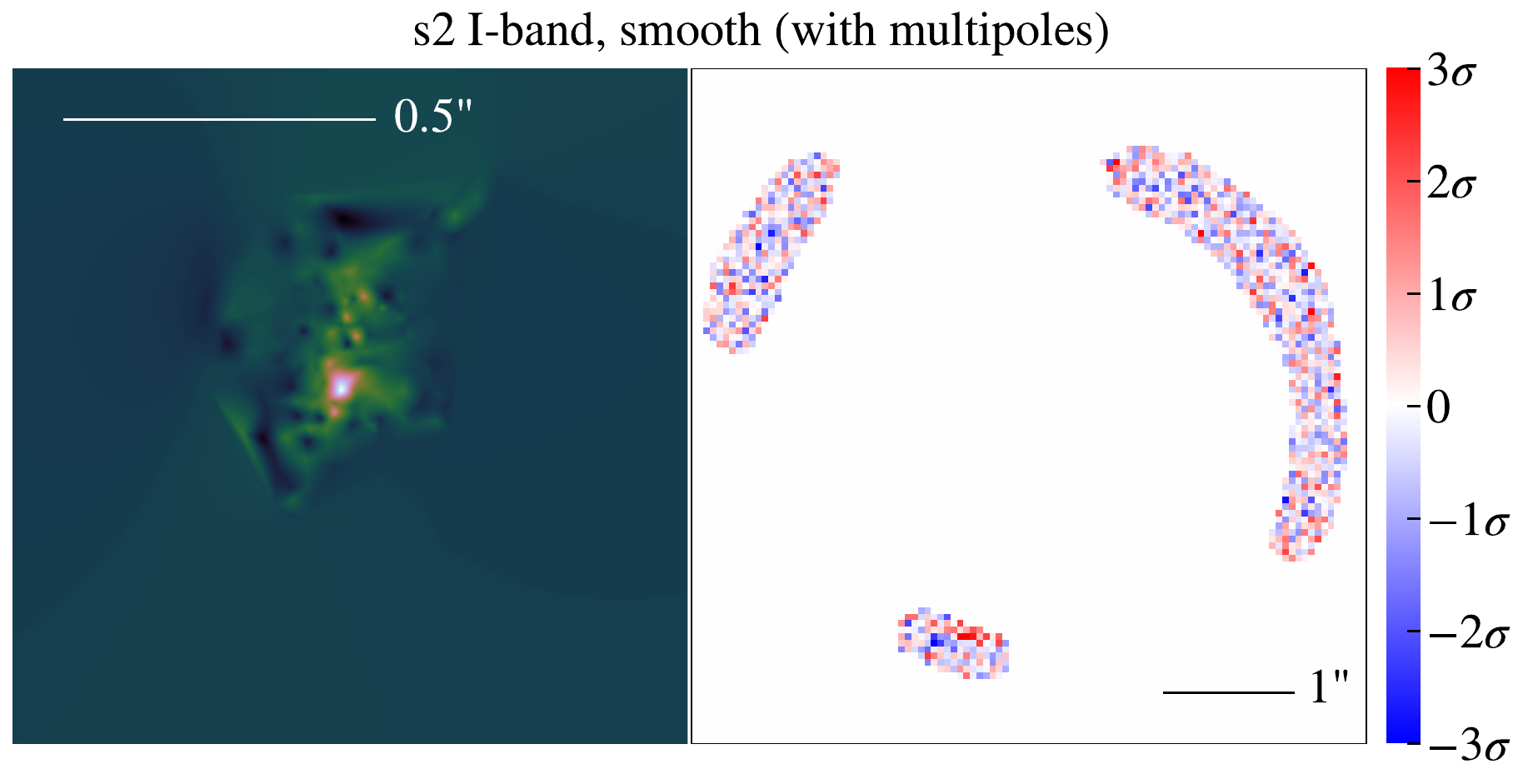}};
     \node (fig11) at (-25,-90)
       {\includegraphics[scale=0.28]{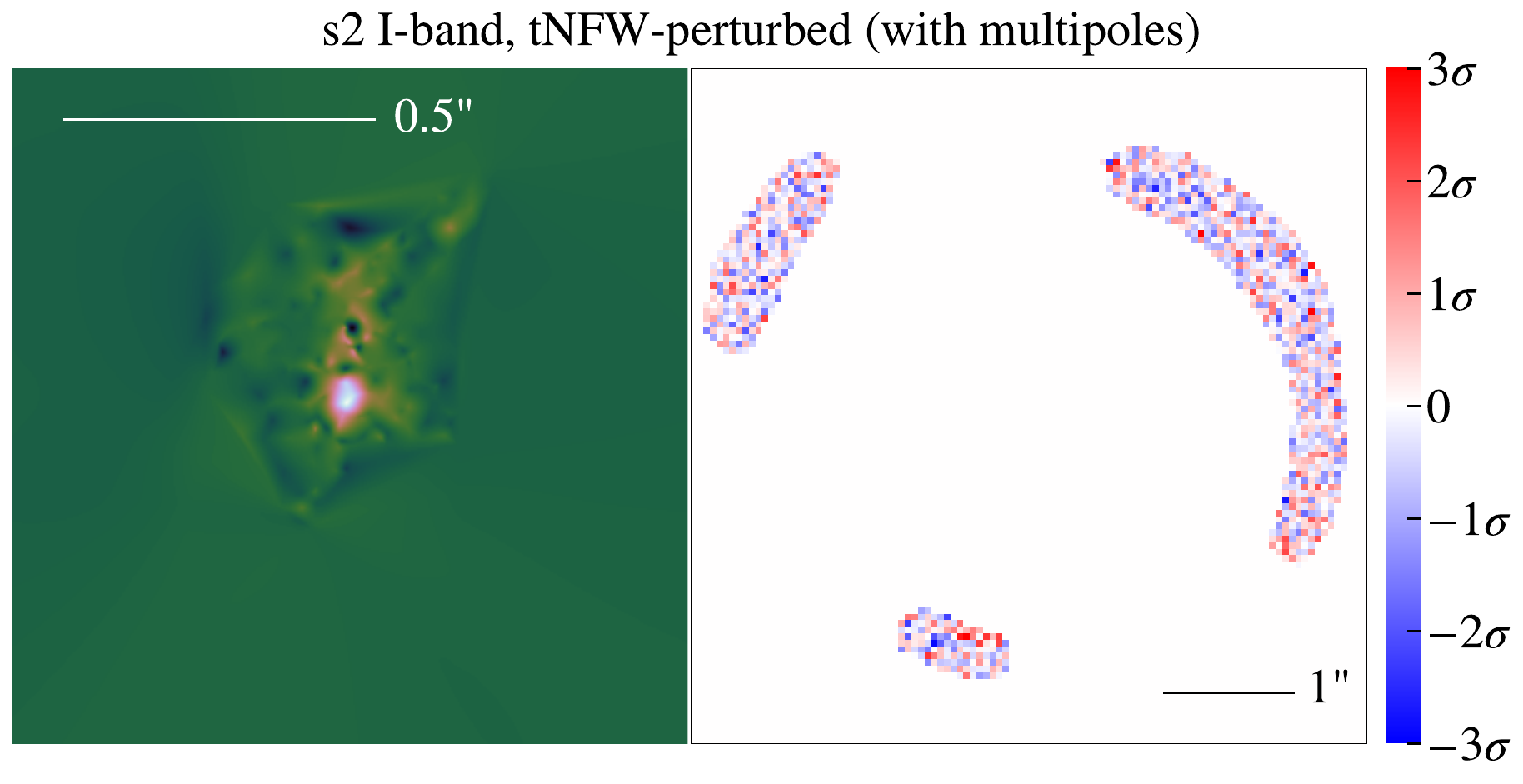}};
     \node (fig13) at (-115,-135)
       {\includegraphics[scale=0.28]{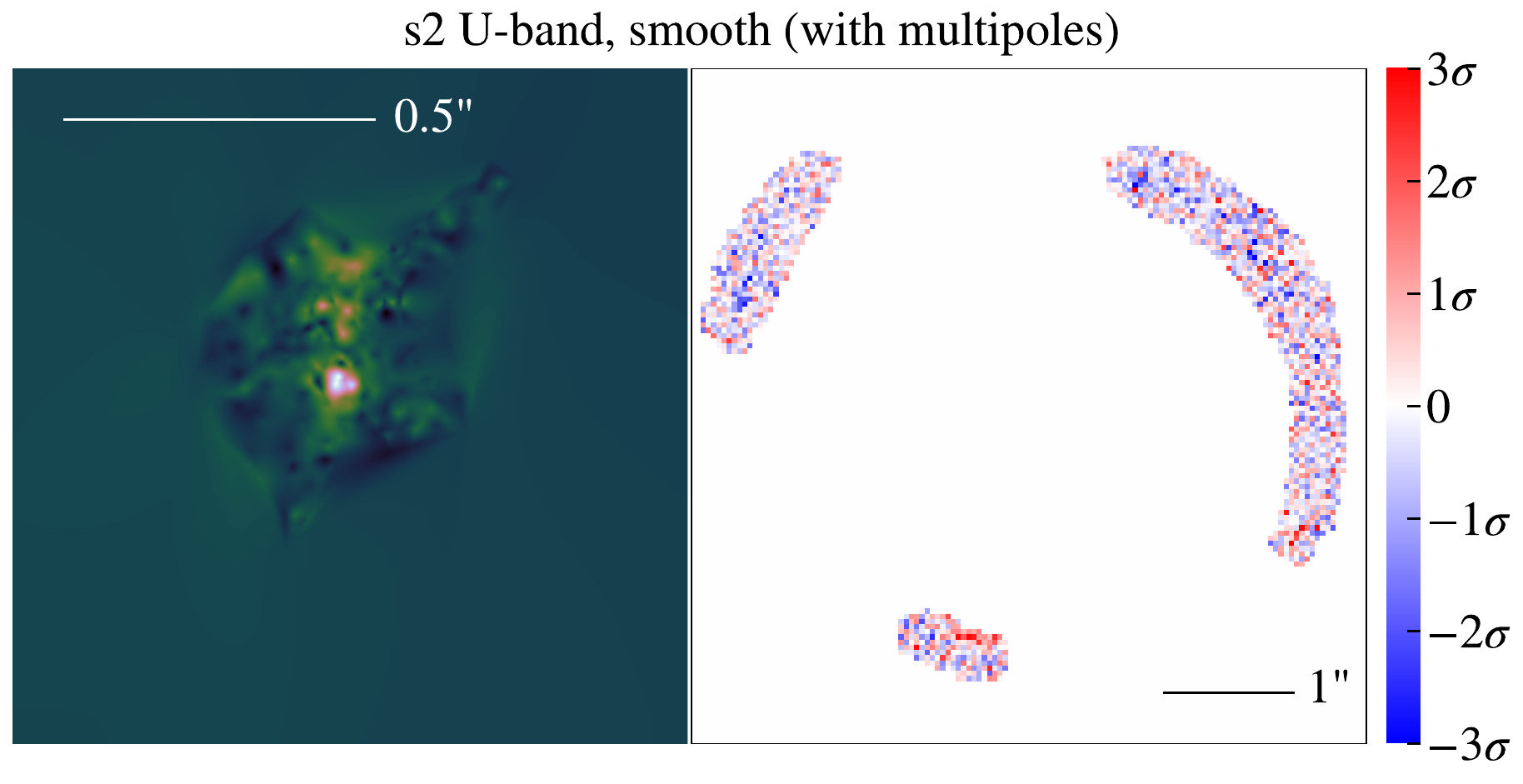}};
     \node (fig15) at (-25,-135)
       {\includegraphics[scale=0.28]{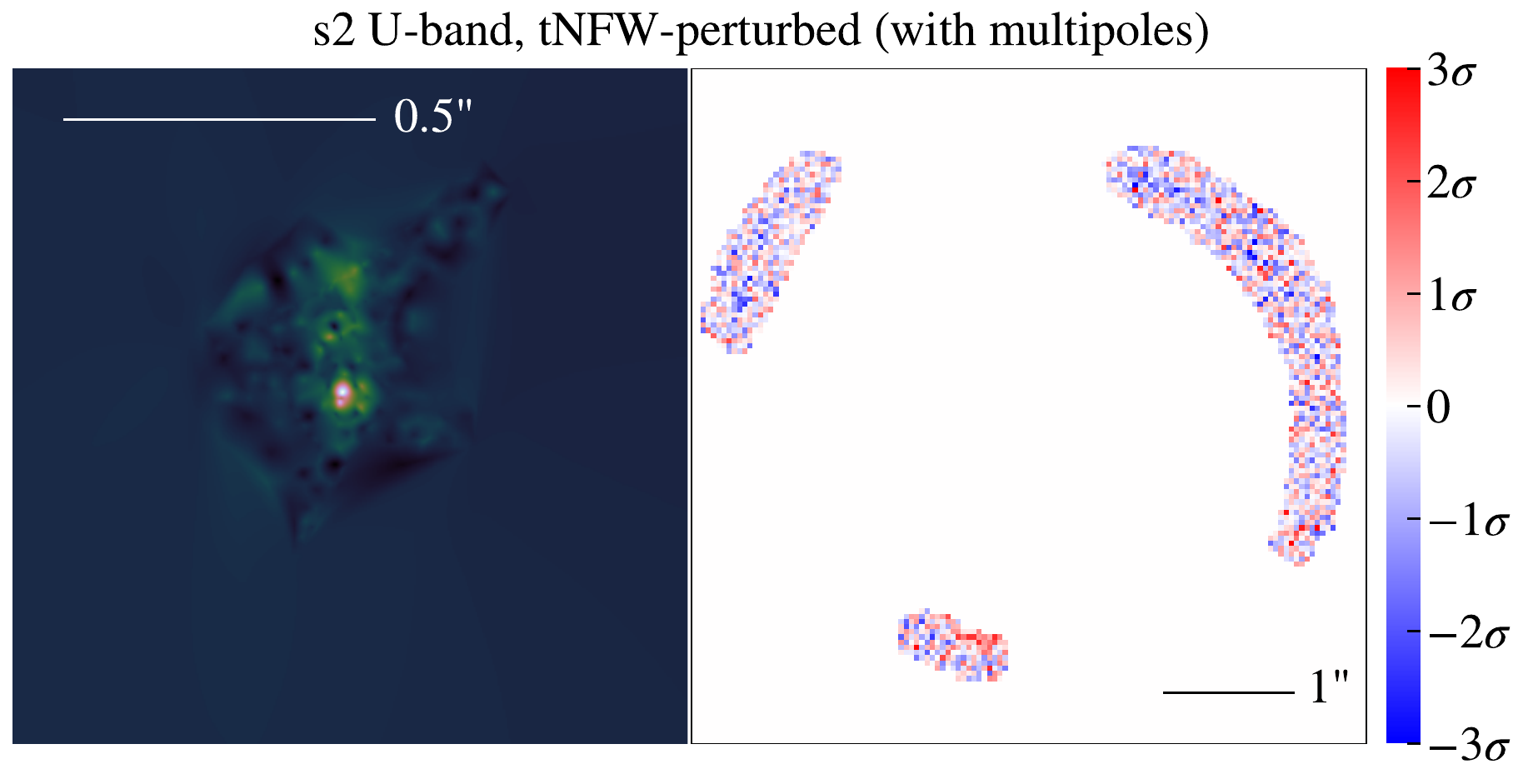}};
\end{tikzpicture}
\caption{Source plane reconstructions and normalised image plane residuals for our best fit smooth (left) and tNFW--perturbed (right) model, for (from top to bottom) $s1$ in I--band, $s1$ in U--band, $s2$ in I--band and $s2$ in U--band, where multipoles are included in the main deflector.}
\label{fig:sources_residuals_triple_plane_mp}
\end{figure*}

\begin{figure*} \centering
\begin{tikzpicture}[      
        every node/.style={anchor=south west,inner sep=0pt},
        x=1mm, y=1mm,
      ]   
     \node (fig1) at (-115,0)
       {\includegraphics[scale=0.28]{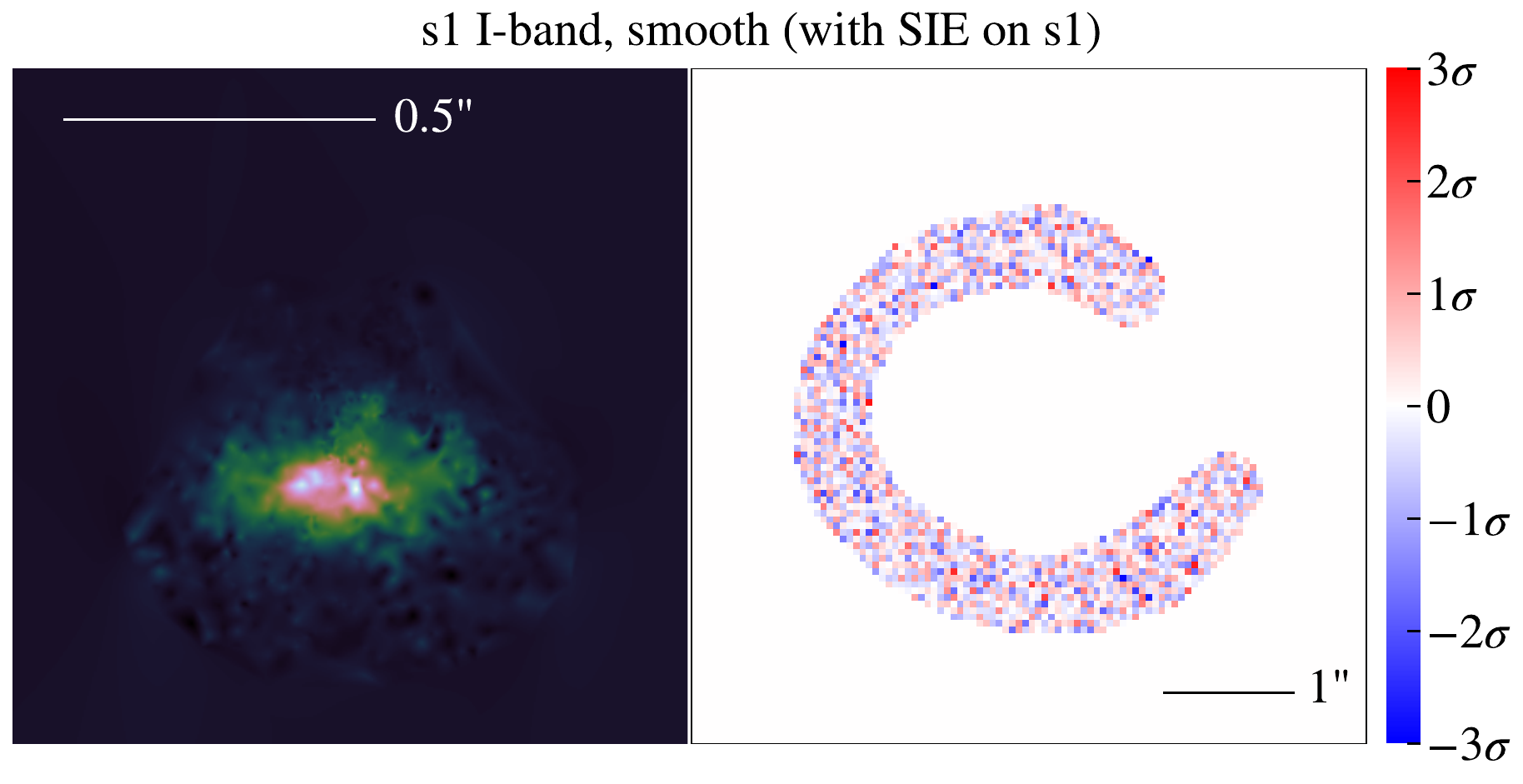}};
     \node (fig3) at (-25,0)
       {\includegraphics[scale=0.28]{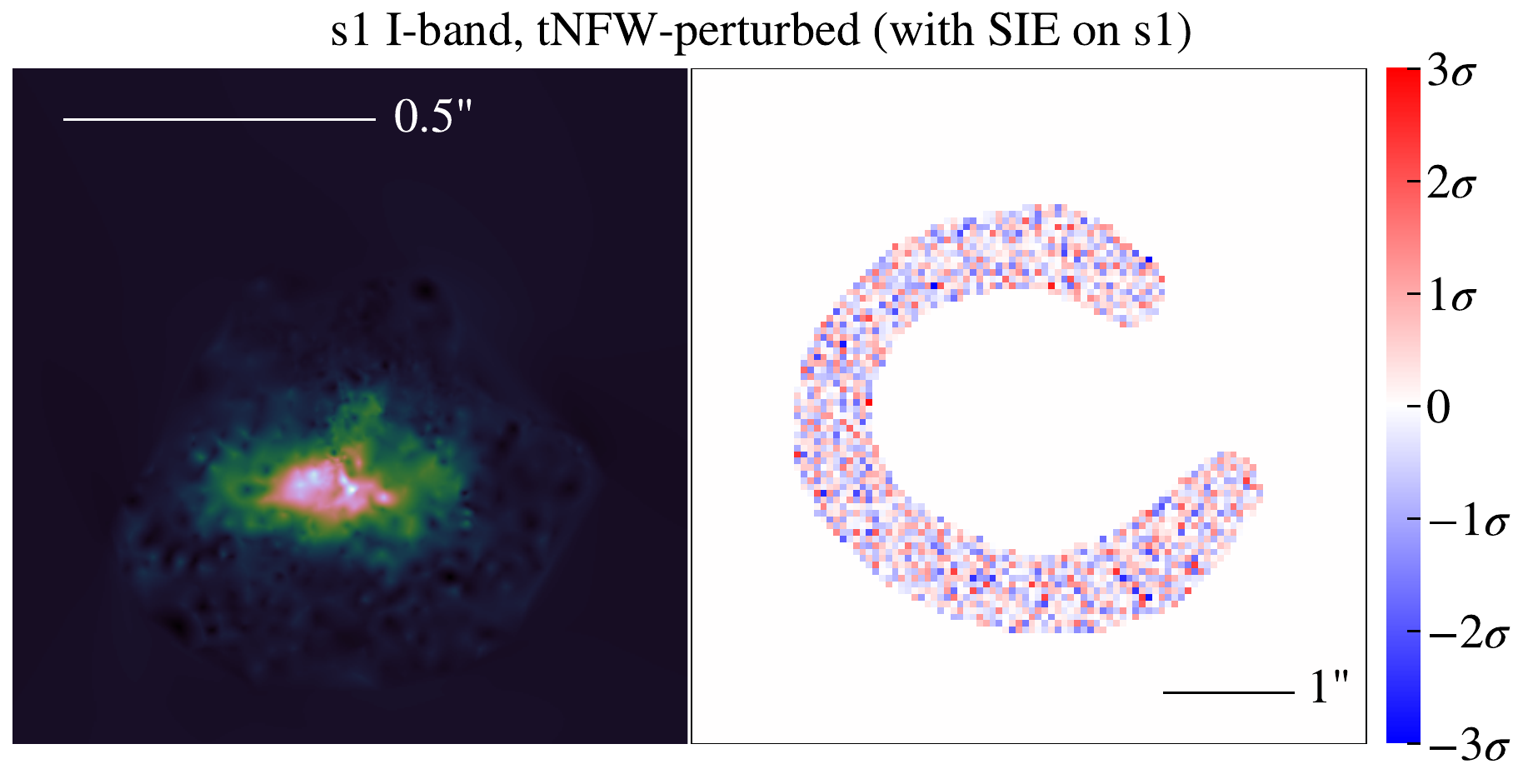}};
     \node (fig5) at (-115,-45)
       {\includegraphics[scale=0.28]{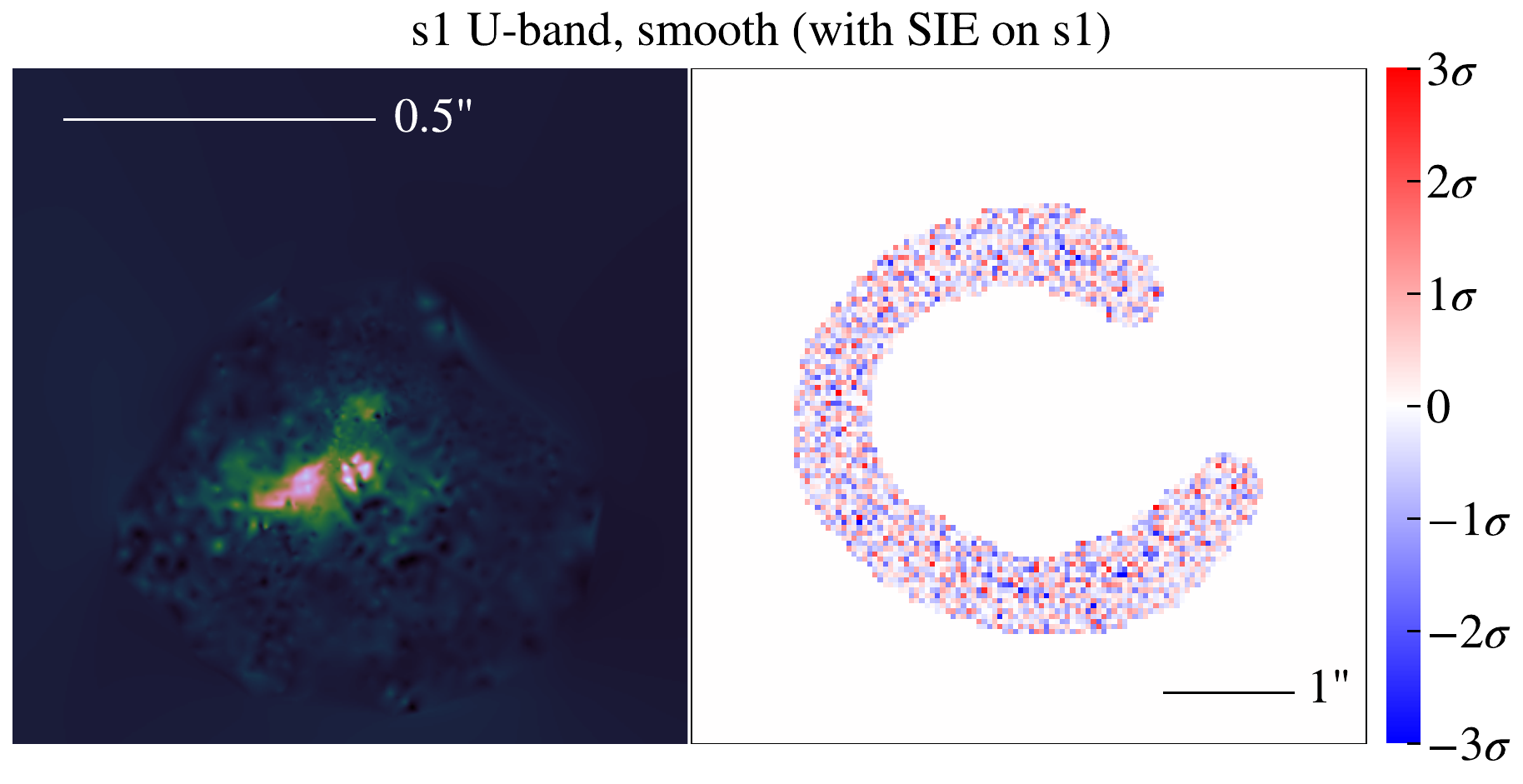}};
     \node (fig7) at (-25,-45)
       {\includegraphics[scale=0.28]{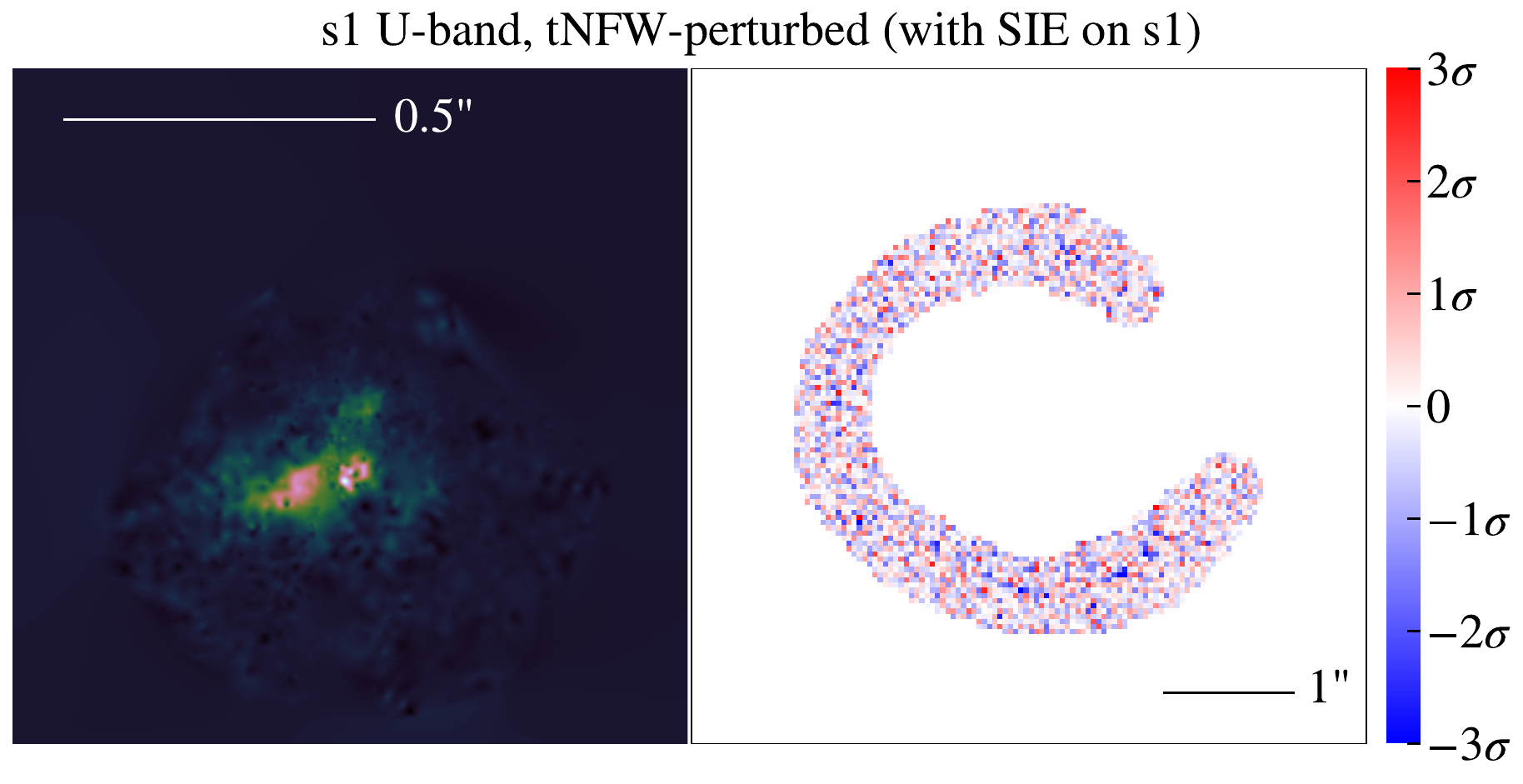}};
     \node (fig9) at (-115,-90)
       {\includegraphics[scale=0.28]{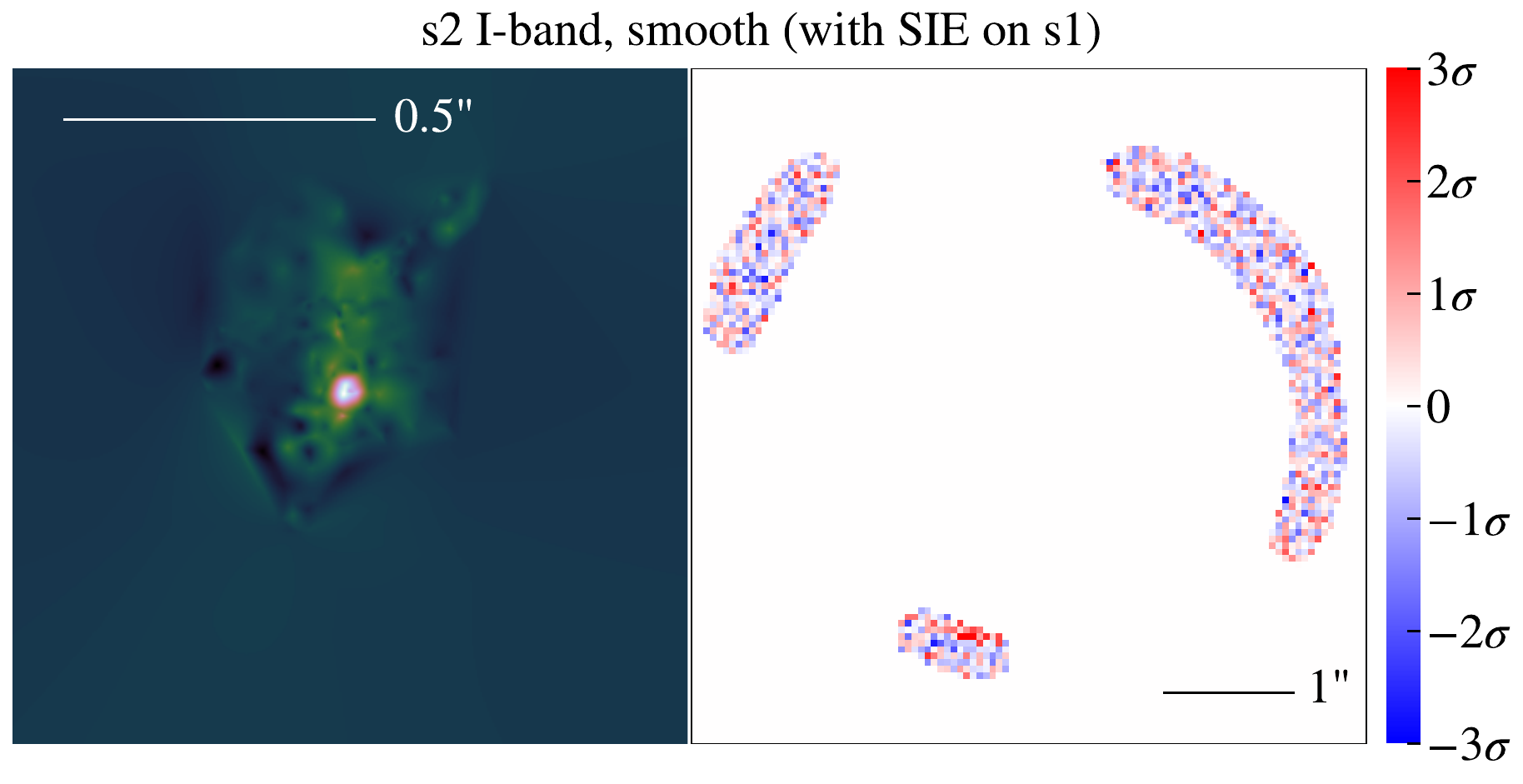}};
     \node (fig11) at (-25,-90)
       {\includegraphics[scale=0.28]{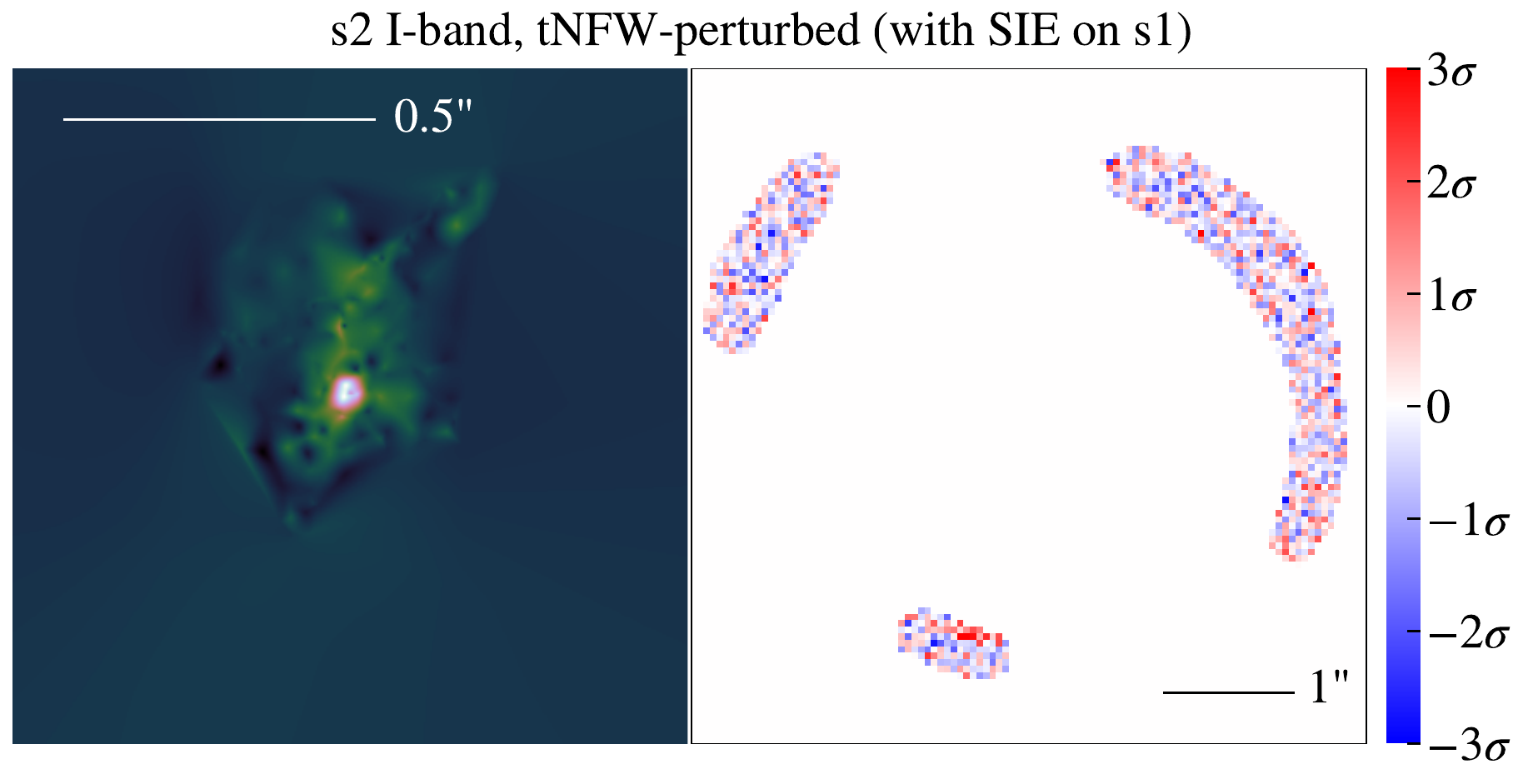}};
     \node (fig13) at (-115,-135)
       {\includegraphics[scale=0.28]{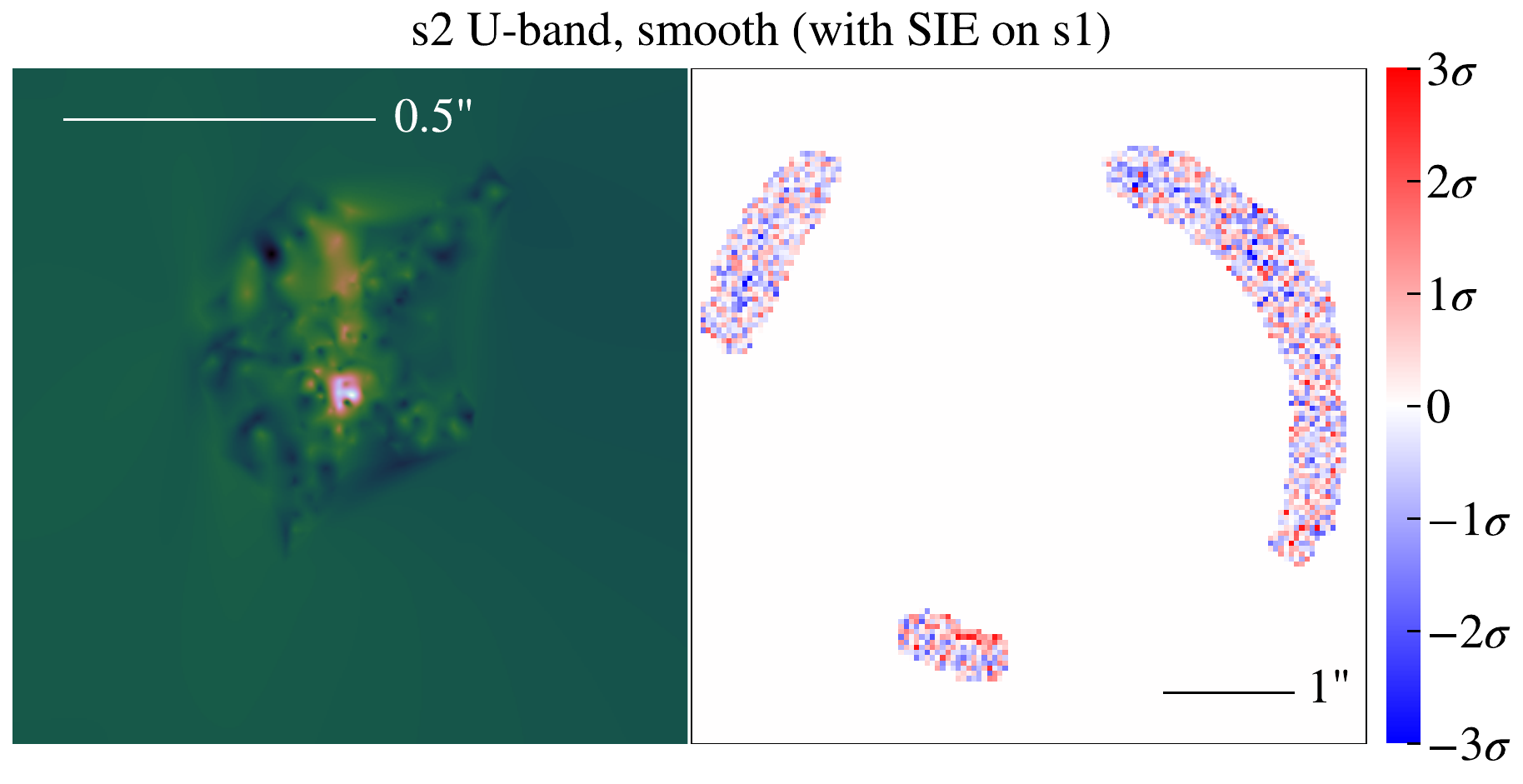}};
     \node (fig15) at (-25,-135)
       {\includegraphics[scale=0.28]{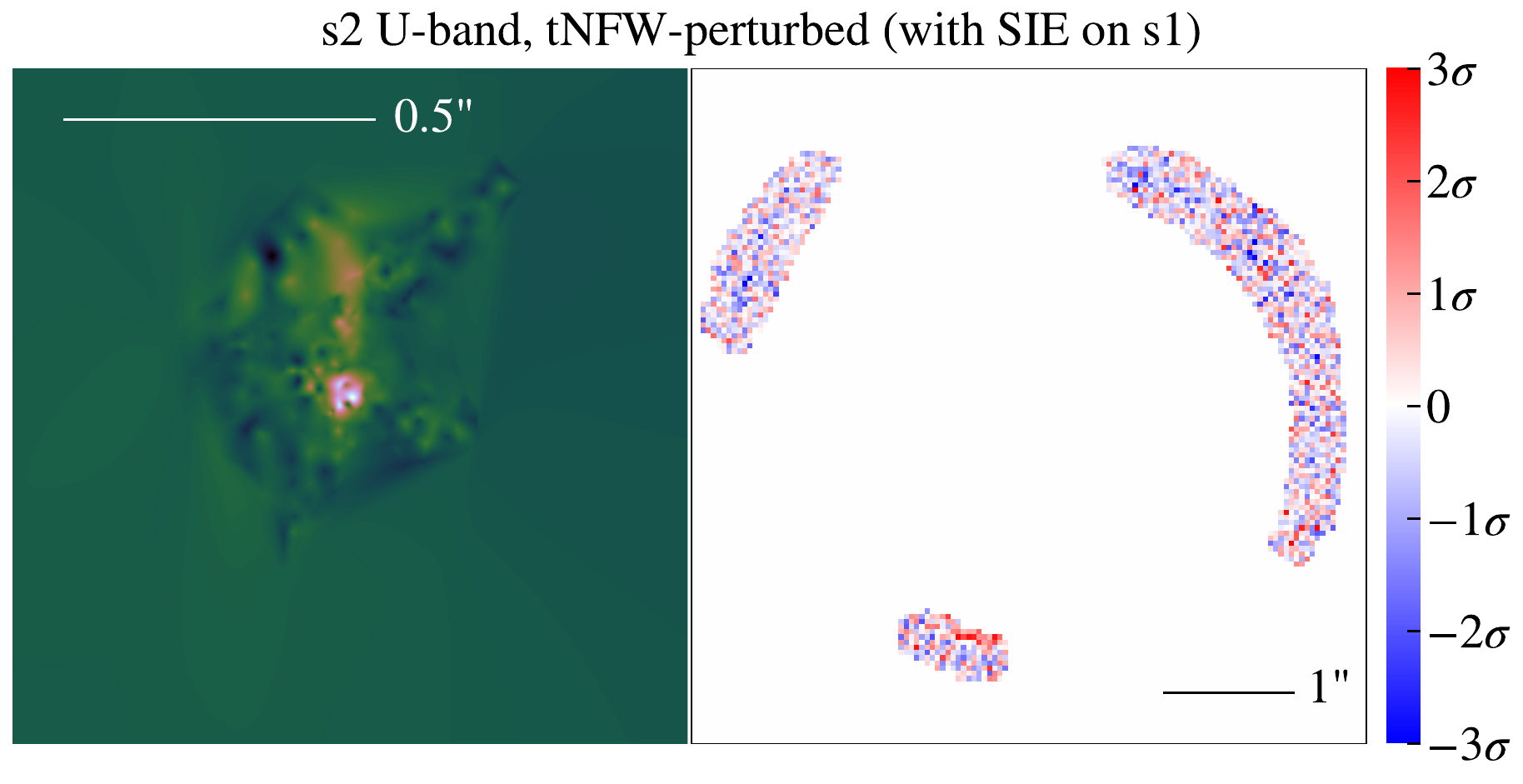}};
\end{tikzpicture}
\caption{Source plane reconstructions and normalised image plane residuals for our best fit smooth (left) and tNFW--perturbed (right) model, for (from top to bottom) $s1$ in I--band, $s1$ in U--band, $s2$ in I--band and $s2$ in U--band, where $s1$ is modelled as an SIE.}
\label{fig:sources_residuals_triple_plane_sie}
\end{figure*}

\begin{figure*} \centering
\begin{tikzpicture}[      
        every node/.style={anchor=south west,inner sep=0pt},
        x=1mm, y=1mm,
      ]   
     \node (fig1) at (-115,0)
       {\includegraphics[scale=0.28]{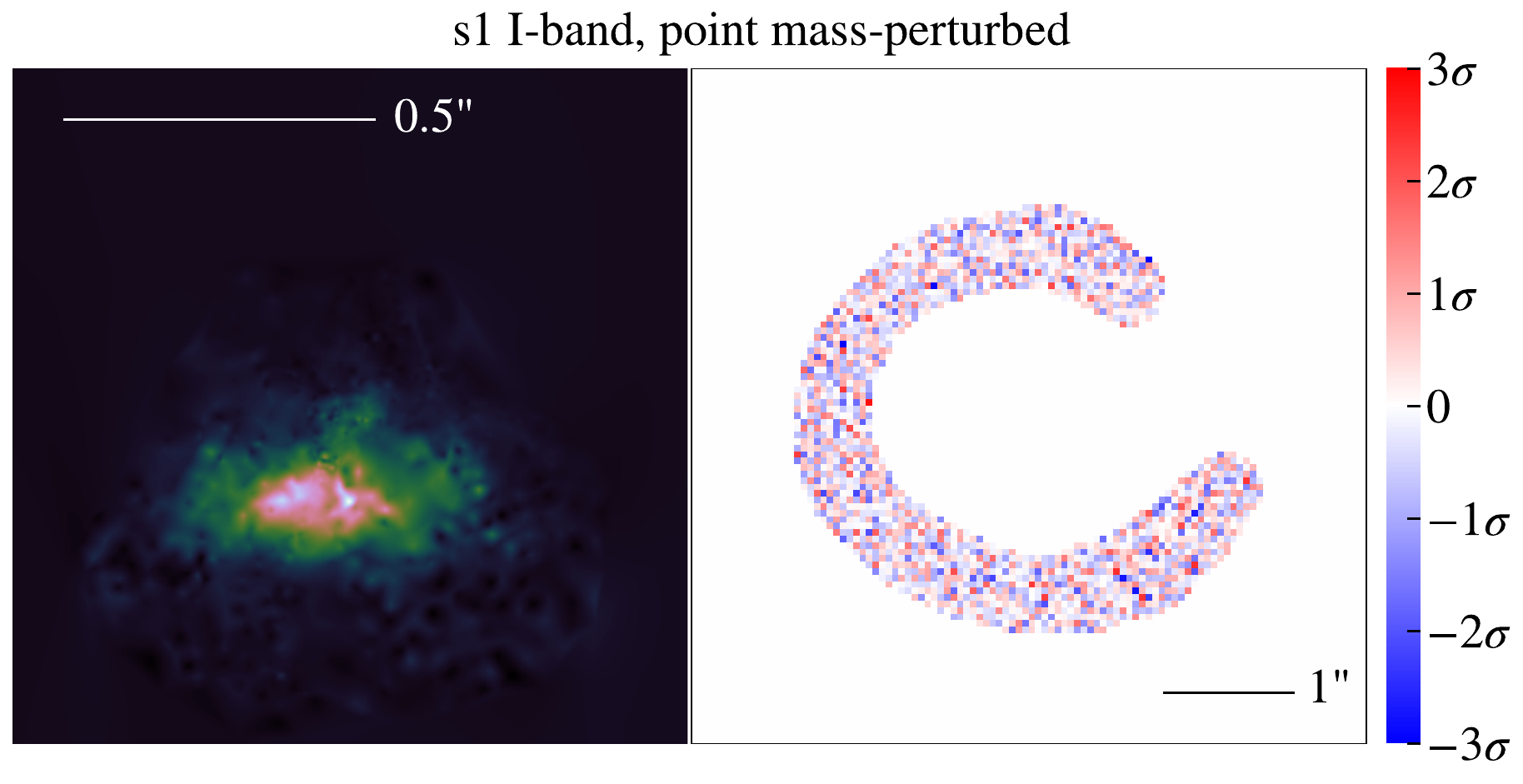}};
     \node (fig5) at (-115,-45)
       {\includegraphics[scale=0.28]{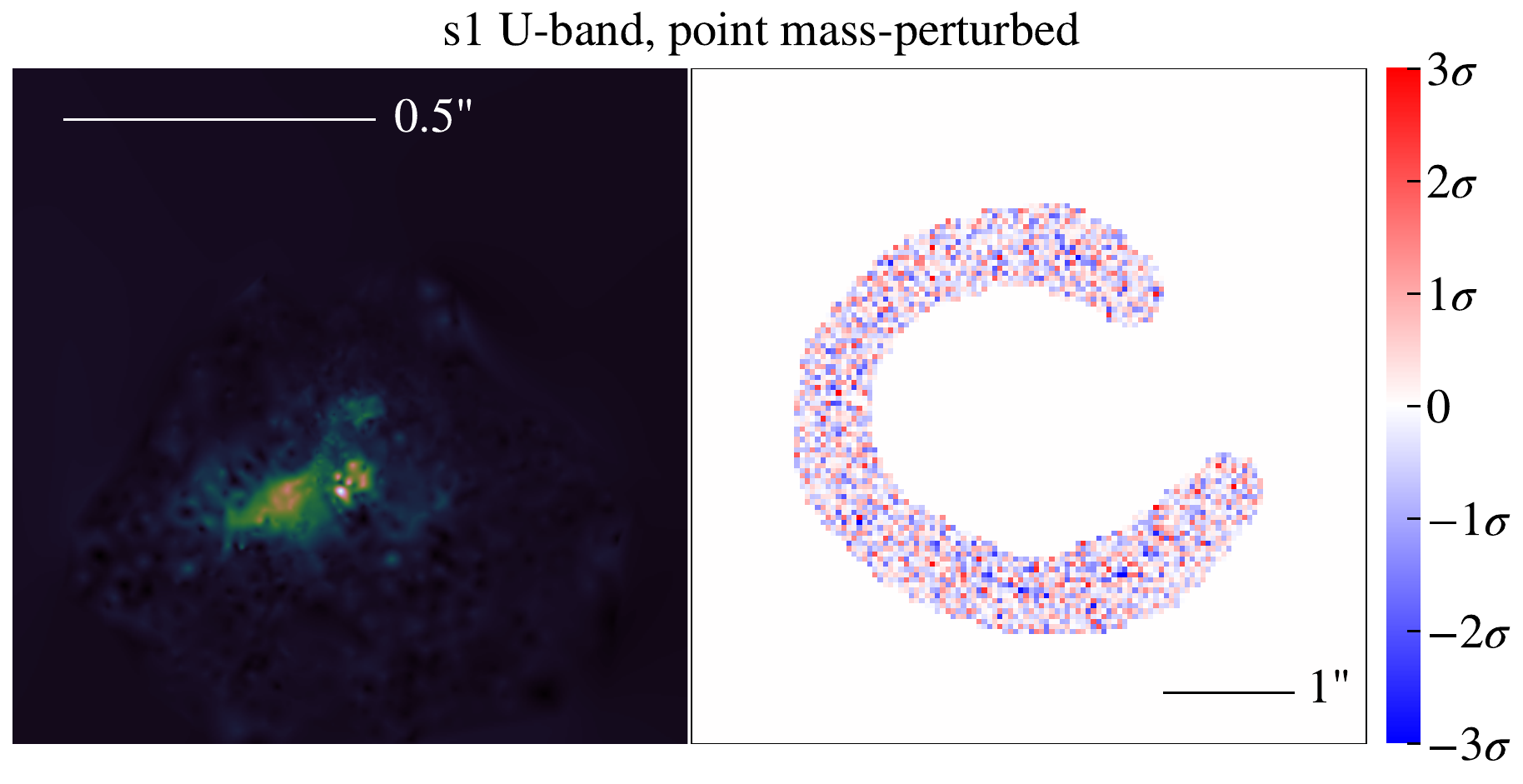}};
     \node (fig9) at (-115,-90)
       {\includegraphics[scale=0.28]{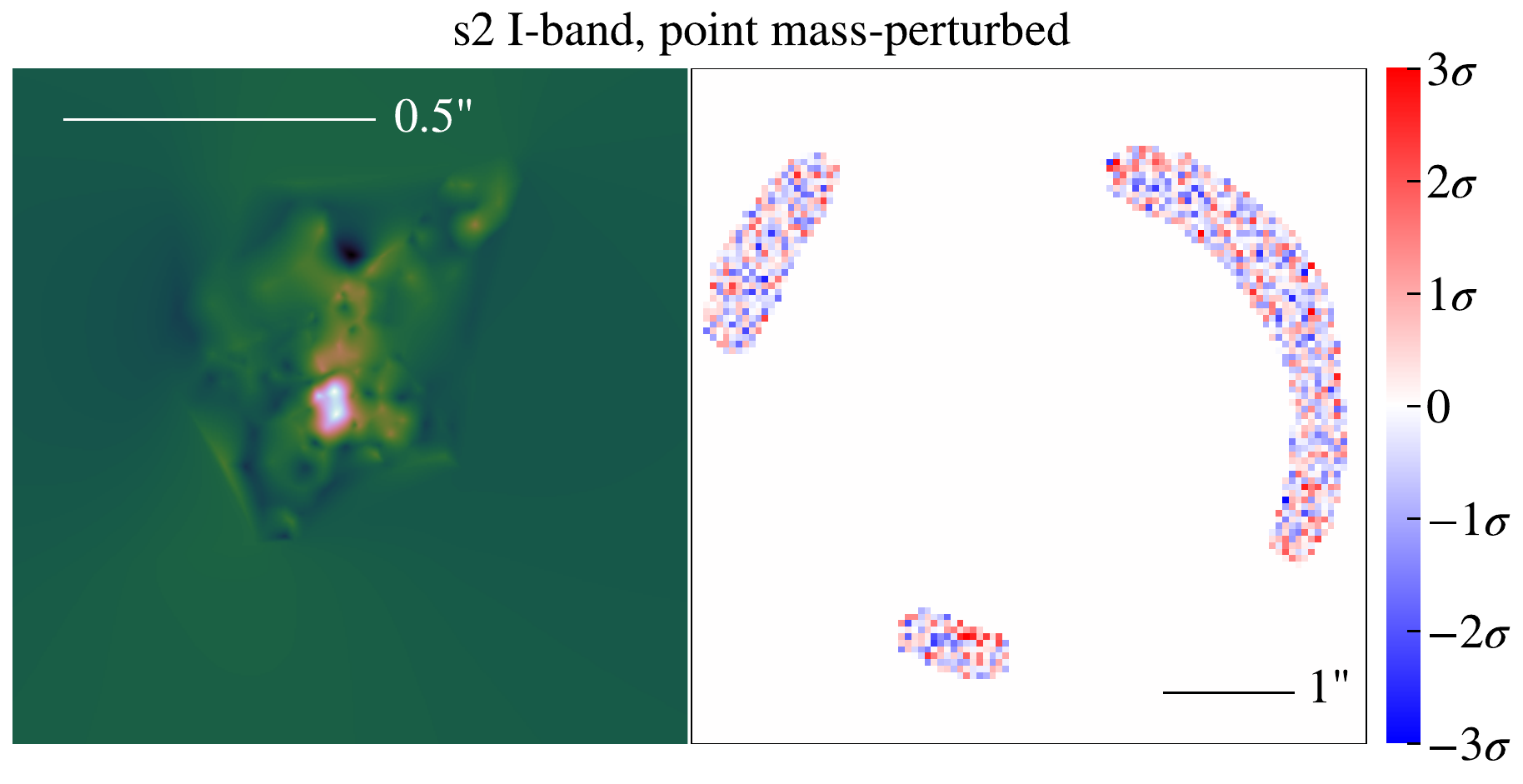}};
     \node (fig13) at (-115,-135)
       {\includegraphics[scale=0.28]{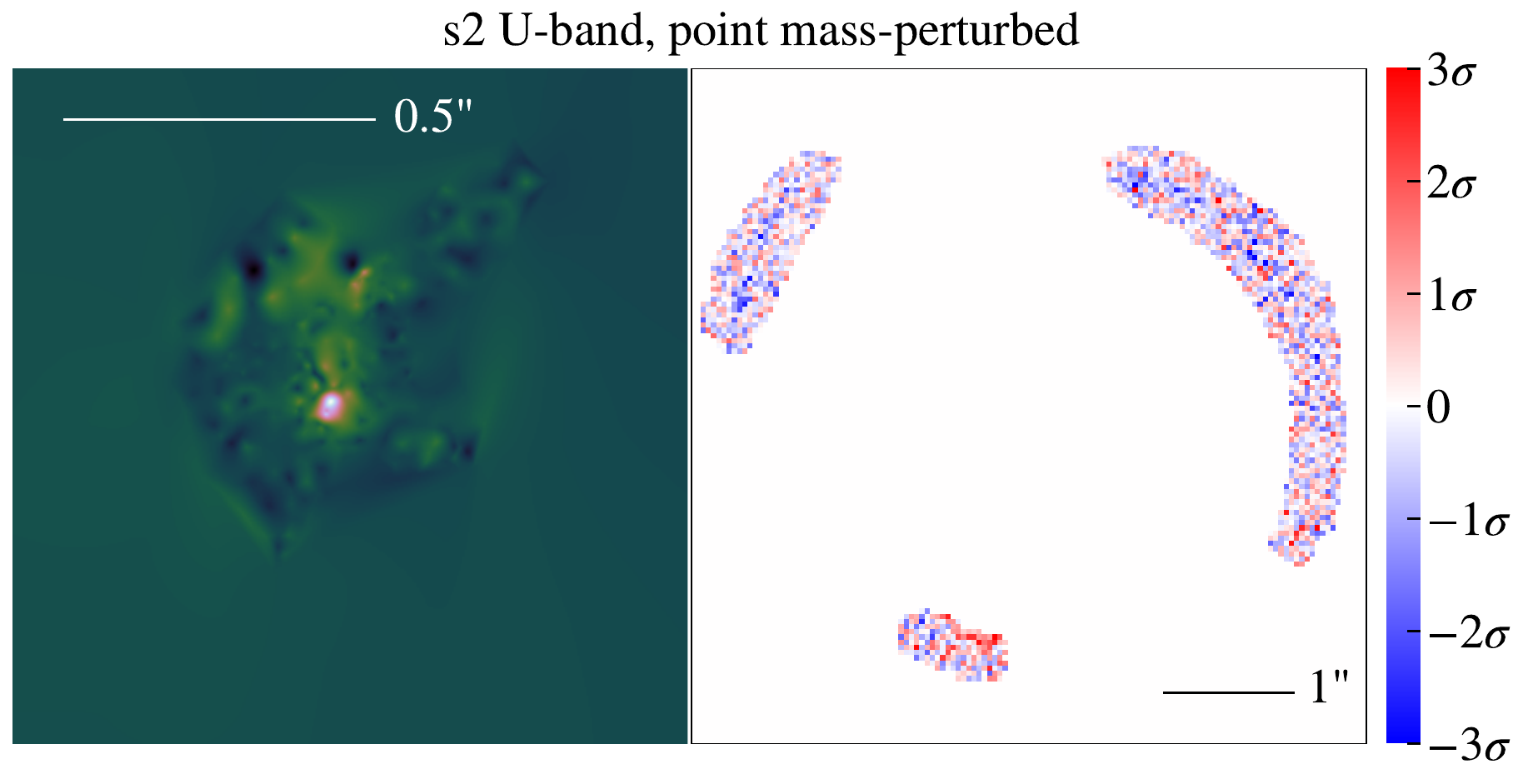}};
\end{tikzpicture}
\caption{Source plane reconstructions and normalised image plane residuals for our best fit model, for (from top to bottom) $s1$ in I--band, $s1$ in U--band, $s2$ in I--band and $s2$ in U--band, where the perturber is modelled as a point mass.}
\label{fig:sources_residuals_triple_plane_point}
\end{figure*}

Here we present all of the best fit sources and normalised image plane residuals corresponding to the systematic tests in Section \ref{sec:systematics}. Figure \ref{fig:sources_residuals_triple_plane_curv} shows the results of modelling with curvature--regularised sources; Figures \ref{fig:sources_residuals_triple_plane_mp}, \ref{fig:sources_residuals_triple_plane_sie} and \ref{fig:sources_residuals_triple_plane_point} show the results with multipoles in the foreground lens, an SIE on $s1$, and a point mass as the perturber, respectively.
%



\bsp	
\label{lastpage}
\end{document}